\documentclass[aps,prx,twocolumn,superscriptaddress]{revtex4-2}
\usepackage{bbm}
\usepackage{mathrsfs}
\usepackage{amsmath}
\usepackage{amsfonts}
\usepackage[colorlinks=true,linkcolor=blue,urlcolor=blue,citecolor=blue,anchorcolor=blue]{hyperref}
\usepackage{graphicx,epstopdf}
\usepackage{subfigure}
\usepackage{epsfig}
\usepackage{dcolumn}
\usepackage{bm}
\usepackage{color}
\usepackage{natbib}
\usepackage{amssymb}
\usepackage{xcolor}
\usepackage{braket}
\usepackage{ulem}
\usepackage{float}
\usepackage{lipsum}
\usepackage{pifont}

\usepackage{array}
\usepackage{booktabs} 
\definecolor{color1}{rgb}{0,0,1.0}

\definecolor{color2}{rgb}{1,0,0.0}

\begin{document}

\title{Asymptotic Entanglement across Dynamical Regimes of a Dissipative Bosonic Dimer}   

\author{Chenghe Yu}
\affiliation{State Key Laboratory of Artificial Microstructure and Mesoscopic Physics, School of Physics, Frontiers Science Center for Nano-optoelectronics, $\&$ Collaborative Innovation Center of Quantum Matter, Peking University, Beijing 100871, China}
\affiliation{Hefei National Laboratory, Hefei 230088, China}

\author{Mingsheng Tian}
\affiliation{Department of Physics, The Pennsylvania State University, University Park, Pennsylvania, 16802, USA}

\author{Feng-Xiao Sun}
\email{sunfengxiao@bupt.edu.cn}
\affiliation{State Key Laboratory of Information Photonics and Optical Communications, Beijing Key Laboratory of Information Metamaterials, $\&$ School of Physical Science and Technology, Beijing University of Posts and Telecommunications, Beijing 100876, China}
\affiliation{State Key Laboratory of Artificial Microstructure and Mesoscopic Physics, School of Physics, Frontiers Science Center for Nano-optoelectronics, $\&$ Collaborative Innovation Center of Quantum Matter, Peking University, Beijing 100871, China}

\author{Qiongyi He}
\affiliation{State Key Laboratory of Artificial Microstructure and Mesoscopic Physics, School of Physics, Frontiers Science Center for Nano-optoelectronics, $\&$ Collaborative Innovation Center of Quantum Matter, Peking University, Beijing 100871, China}
\affiliation{Hefei National Laboratory, Hefei 230088, China}
\affiliation{Collaborative Innovation Center of Extreme Optics, Shanxi University, Taiyuan, Shanxi 030006, China}

\begin{abstract}
Dynamical instability in parametrically driven bosonic systems is generally associated with diverging occupations and the absence of a stationary state, but its implications for long-time entanglement remain unclear. We investigate this question in a dissipative bosonic dimer with beam-splitter and two-mode-squeezing interactions, local loss, and a tunable common bath. The non-Hermitian dynamical spectrum governing the Gaussian moments partitions the parameter space into distinct regimes and provides a unified description of population and entanglement dynamics across both stable and unstable regions. We find that dynamical instability does not imply unbounded entanglement growth: although the bosonic population diverges, the logarithmic negativity remains bounded and approaches a finite asymptotic value. We further characterize the finite asymptotic entanglement and show that asymmetric local dissipation shifts both the stability boundary and the threshold for nonzero asymptotic entanglement. For identical local dissipation, the presence of a common bath leaves the dynamical-region boundaries unchanged while qualitatively reshaping the long-time entanglement landscape, producing broad enhancement regions within the stable regimes and a narrow enhancement region within the unstable regime. In the common-bath-only limit, this enhancement is attributed to an exact dark-mode (bound-state-in-the-continuum, BIC) protection mechanism, which reduces to a quasi‑BIC remnant under independent dissipation. Our results establish the non-Hermitian dynamical spectrum as a framework for connecting stability, nonequilibrium dynamics, and asymptotic entanglement in open quadratic bosonic systems.

\end{abstract}
	
\maketitle
	
\section{Introduction}
Continuous-variable (CV) entanglement is a fundamental resource for quantum information science~\cite{Braunstein2005Quantum,Weedbrook2012Gaussian}, underpinning applications ranging from quantum communication~\cite{Furusawa1998teleportation,Garc2009key,madsen2012continuous} and quantum networking~\cite{yonezawa2004demonstration,armstrong2012programmable,zheng2026large} to distributed quantum sensing~\cite{Xia2020,Kwon2022,guo2020distributed} and measurement-based quantum computation~\cite{Lloyd1999Computation,Ukai2011Computation,Larsen2021Fault}. Gaussian bosonic systems provide one of the leading platforms for realizing these protocols owing to their deterministic state preparation, scalability, and compatibility with optical and microwave architectures~\cite{Su2007Experimental,yokoyama2013ultra,cai2017multimode,Warit2019Generation,Mikkel2019Deterministic,Wang2020Large,roh2025generation,wang2025large,jia2025continuous, Jolin2023,Lingua2025,Pogorzalek2019}. Consequently, understanding how entanglement is generated, manipulated, and preserved in realistic bosonic systems has become an important problem in quantum science~\cite{Adesso2007,Asavanant2024multipartite}.

Considerable effort has therefore been devoted to the generation, manipulation, and stabilization of CV entanglement in open quantum systems~\cite{Krauter2011Entanglement,Djorw2014Robustness,ockeloen2018stabilized,Mamaev2018dissipative,Zheng2019Manipulation,zheng2021enhanced,Abdi2021Continuous,Li2025Programmable}. 
Among coherent mechanisms, nondegenerate parametric interactions provide a well-established route to generating bipartite and multipartite CV entanglement~\cite{Reid1988Quantum,Villar2005Generation,Coelho2009three}. 
In idealized lossless treatments, these interactions can produce strong parametric amplification and sustained entanglement growth. In realistic systems, however, uncontrolled dissipation and thermal noise generally degrade the generated quantum correlations. This has motivated reservoir-engineering approaches, in which carefully designed dissipative couplings are used to prepare and stabilize entangled states~\cite{Wang2013Reservoir,Woolley2014squeezed,pedram2023overview,Zheng2024Optomechanical,Qiu2024Genuine}. Accordingly, most studies of long-time entanglement have focused on dynamically stable regimes in which a stationary entangled state exists.

While the steady-state regime has been extensively characterized, the long-time entanglement properties of the unstable parameter region remain largely unexplored. Beyond the stability boundary, parametrically driven bosonic systems generally exhibit exponentially growing occupations, and no stationary state exists. Such dynamically unstable regimes are therefore often excluded from analyses of asymptotic entanglement. Nevertheless, they are not merely pathological regions of parameter space. In related bosonic platforms, including cavity optomechanics~\cite{Aspelmeyer2014optomechanics}, superconducting parametric circuits~\cite{Wustmann2013Parametric}, and nonlinear photonic systems~\cite{dutt2024nonlinear}, operation close to or even beyond the instability threshold is employed to enhance parametric gain~\cite{Wustmann2017Nondegenerate}, squeezing~\cite{Han2021four,Kustura2022Mechanical,Xu2026Modulation}, and transient entanglement generation~\cite{Hofer2011optomechanics,palomaki2013entangling,sun2017phase}. 
A fundamental question therefore remains: Does dynamical instability necessarily cause unbounded entanglement, or can entanglement remain finite even as the bosonic population grows without bound?

Addressing this question requires a framework that remains meaningful even in the absence of a stationary state. A natural candidate is provided by the non-Hermitian dynamical spectrum governing open-system evolution~\cite{Ashida2020,El2018Non}. 
The interplay of coherent couplings, parametric driving, gain, and loss can give rise to exceptional points (EPs), stability transitions, and other spectral features that organize nonequilibrium dynamics~\cite{McDonald2018Phase,ozdemir2019parity,Miri2019Exceptional,Bergholtz2021Exceptional,Lee2024Entanglement}. 
For open quadratic bosonic systems governed by a Lindblad master equation, a more appropriate non-Hermitian object is the dynamical matrix underlying the equations of motion for first- and second-order moments~\cite{Arkhipov2021Generating}, rather than the effective non-Hermitian Hamiltonian associated with postselected no-jump trajectories~\cite{Minganti2019Quantum}.
Its spectrum directly determines dynamical stability and naturally partitions the parameter space into qualitatively distinct dynamical regimes, making it a suitable framework for investigating entanglement dynamics across the full parameter space.

In this work, we investigate how the non-Hermitian dynamical spectrum shapes CV entanglement in a dissipative bosonic dimer with beam-splitter coupling, two-mode squeezing, local dissipation, and an optional common bath. We show that the dynamical spectrum provides a unified description of entanglement dynamics across both stable and unstable regimes. This unification is achieved by classifying the parameter space into distinct dynamical regions, each characterized by specific population and entanglement evolution patterns. A striking finding is that population divergence does not signal a corresponding unbounded entanglement. Specifically, although the bosonic occupation diverges in the dynamically unstable regime, the logarithmic negativity remains bounded and converges to a finite asymptotic limit. Thus, in the present open-system setting, dynamical instability signals the absence of a stationary covariance matrix, but does not by itself preclude finite asymptotic entanglement.

We further show that dissipation provides an effective means of engineering the entanglement behaviors. Specifically, the asymmetric local loss shifts both the dynamical stability boundary and the transition between vanishing and nonzero asymptotic entanglement. Moreover, a common bath can qualitatively enhance asymptotic entanglement over broad stable parameter regions and even within part of the dynamically unstable regime. This enhancement becomes particularly transparent in the common-bath-only limit, where the underlying protection mechanism admits an interpretation in terms of a dark mode, or equivalently, a bound-state-in-the-continuum (BIC) picture~\cite{hsu2016bound,Lei2023}.

The paper is organized as follows. Section II introduces the dissipative bosonic dimer and its non-Hermitian dynamical matrix. Section III analyzes the dynamical regimes and the corresponding time evolution of population and entanglement. Section IV presents the finite asymptotic entanglement diagram. Section V shows how a common bath engineers entanglement dynamics and asymptotic entanglement. Section VI explains the BIC-type protection mechanism underlying the common-bath-induced entanglement enhancement. Finally, we conclude in Section VII.

\section{Model and Dynamical Matrix}  

\begin{figure}
    \centering
    \includegraphics[width=0.9\linewidth]{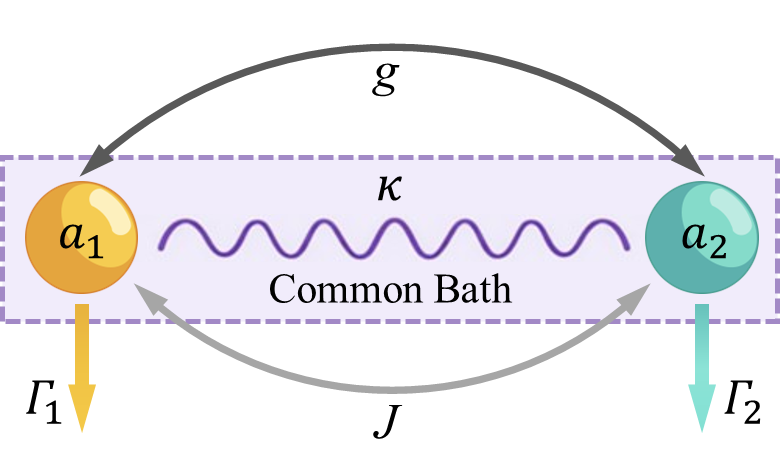}
    \caption{
    Schematic of the dissipative quadratic bosonic dimer. The two modes $a_1$ and $a_2$ are coupled via a beam-splitter interaction with strength $g$ and a two-mode-squeezing interaction with strength $J$. Each mode is subject to local dissipation at rates $\Gamma_{1,2}$, and both modes are additionally coupled to a common bath at rate $\kappa$.} 
    \label{fig1}
\end{figure}

    We consider a dissipative bosonic dimer, as illustrated in Fig.~\ref{fig1}. The two modes are directly coupled by beam-splitter and two-mode-squeezing interactions of strengths $g$ and $J$, respectively, with Hamiltonian
    \begin{equation}\label{eq1}
        \hat{H}
        =
        g \hat{a}_1^\dagger \hat{a}_2
        +
        J \hat{a}_1^\dagger \hat{a}_2^\dagger
        + \mathrm{h.c.}
    \end{equation}
    Here $\hat{a}_{1,2}$ and $\hat{a}_{1,2}^\dagger$ are the bosonic annihilation and creation operators for the two modes, respectively.
    
    In addition, each mode is subject to an independent dissipative channel with rates $\Gamma_1$ and $\Gamma_2$, and both modes are coupled to a common bath with the decay rate $\kappa$. The open-system dynamics are described by the Markovian master equation
    \begin{equation}
        \dot{\hat{\rho}}
        =
        -i[\hat{H},\hat{\rho}]
        + \Gamma_1 \mathcal{D}[\hat{a}_1]\hat{\rho}
        + \Gamma_2 \mathcal{D}[\hat{a}_2]\hat{\rho}
        + \kappa \mathcal{D}[\hat{a}_1+\hat{a}_2]\hat{\rho},
        \label{eq:ME}
    \end{equation}
    where $\mathcal{D}[\hat{o}]\hat{\rho}=\hat{o}\hat{\rho}\hat{o}^\dagger-\{\hat{o}^\dagger\hat{o},\hat{\rho}\}/2$.

    To characterize the entanglement dynamics, we consider the first- and second-order operator moments and define the operator vector
    \begin{equation}
        \Psi
        =
        \left(
        \hat{a}_1,
        \hat{a}_1^\dagger,
        \hat{a}_2,
        \hat{a}_2^\dagger
        \right)^{\mathrm T},
    \end{equation}
    together with the first-order moment vector
    \begin{equation}
        v_1 = \langle \Psi \rangle,
    \end{equation}
    and the second-order moment vector
    \begin{equation}
        v_2 = \mathrm{vec}\!\left(\langle \Psi \Psi^{\mathrm T} \rangle\right),
    \end{equation}
    where $\mathrm{vec}(\cdot)$ denotes vectorization.

    Starting from the master equation, we obtain linear evolution equations for these moments,
    \begin{equation}
        i\frac{d}{dt}v_1 = M_1 v_1,
        \label{eq:M1}
    \end{equation}
    and
    \begin{equation}
        i\frac{d}{dt}v_2 = M_2 v_2 + D,
        \label{eq:M2}
    \end{equation}
    where $D$ is a constant vector originating jointly from the Lindblad terms and bosonic commutation relations. The second-order dynamical matrix satisfies
    \begin{equation}
        M_2 = I \otimes M_1 + M_1 \otimes I,
        \label{eq:M2structure}
    \end{equation}
    with $I$ the $4\times4$ identity matrix. The detailed derivation of Eqs.~(\ref{eq:M1})-(\ref{eq:M2structure}), together with the explicit expressions of $M_1$ and $D$, is presented in Appendix~\ref{App-A1}. In the basis defined above, $M_1$ is related by a simple permutation to the standard bosonic Bogoliubov-de Gennes form~\cite{Ashida2020, Del2022, Wanjura2023, Yu2025Exceptional}, which is determined by the coherent couplings and the dissipative channels. Owing to the combined effects of squeezing and dissipation, the dynamical matrices $M_1$ and $M_2$ are generally non-Hermitian.

    To connect these moments with bipartite entanglement, we introduce the quadrature vector
    \begin{equation}
        R
        =
        \left(
        \hat{X}_1,\hat{P}_1,\hat{X}_2,\hat{P}_2
        \right)^{\mathrm T},
    \end{equation}
    where $\hat{X}_j=(\hat{a}_j+\hat{a}_j^\dagger)/\sqrt{2}$ and $\hat{P}_j=(\hat{a}_j-\hat{a}_j^\dagger)/(i\sqrt{2})$ denote the amplitude and phase quadratures of mode $j$. The covariance matrix (CM) is defined as
    \begin{equation}
        V_{mn}
        =
        \left\langle
        \hat{R}_m\hat{R}_n+\hat{R}_n\hat{R}_m
        \right\rangle
        -
        2\langle \hat{R}_m\rangle \langle \hat{R}_n\rangle .
    \end{equation}
    We quantify bipartite entanglement by the logarithmic negativity $E_N(t)=\max\{0,-\log \nu_-(t)\}$, where $\nu_-(t)$ is the smallest symplectic eigenvalue of the partially transposed CM~\cite{Simon2000,Adesso2007}. 
The first-order moments $v_1$ determine $\langle R\rangle$, while
$v_1$ and $v_2$ together determine the CM. Throughout the time-domain
analysis, we take the initial state to be the two-mode vacuum, for which
the first-order moments remain zero. Consequently, the mode populations
$n_i(t)=\langle \hat{a}_i^\dagger\hat{a}_i\rangle$, the CM, and the
logarithmic negativity $E_N(t)$ are fully determined by the second-order moments governed by $M_2$ (see Appendix~\ref{App-A2} for details).
Since $M_2=I\otimes M_1+M_1\otimes I$, its spectrum is constructed from
pairwise sums of the eigenvalues of $M_1$. The spectral boundaries relevant
to the dynamical regimes can thus already be identified from the simpler
first-order matrix $M_1$.

    The dynamical matrix also exhibits characteristic symmetry properties. In the absence of dissipation, $M_1$ satisfies chiral and anti-parity-time-reversal (anti-$\mathcal{PT}$) symmetry relations, which imply that the dynamical spectrum is symmetric with respect to both the real and imaginary axes, or equivalently centrosymmetric about the origin~\cite{Yu2025Exceptional, wakefield2024,yang2026spectral}. Once dissipation is included, the chiral symmetry is broken, and the spectral symmetry is reduced to a reflection symmetry about the imaginary axis. This change will be directly reflected in the dynamical spectra discussed below. The explicit symmetry operators and their derivations are given in Appendix~\ref{App-A3}.

\section{Dynamical Regimes and Entanglement Dynamics} \label{sec3} 
\begin{figure}
    	\centering
    	\includegraphics[width=\linewidth]{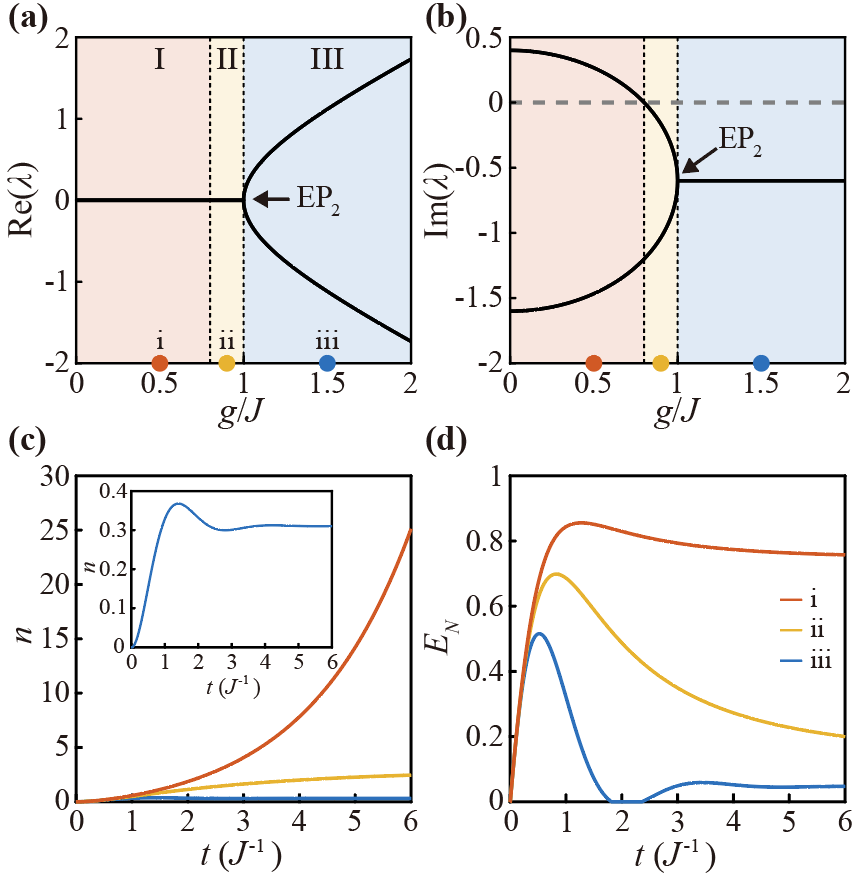}
    	\caption{Dynamical spectrum and time evolution of mode populations and entanglement under local dissipation. 
(a),(b) Real (a) and imaginary (b) parts of the four eigenvalues of the dynamical matrix $M_1$ as functions of $g/J$. The spectrum is divided into three regions. In Region I, at least one eigenvalue has a positive imaginary part indicating dynamical instability, whereas in Regions II and III all eigenvalues have negative imaginary parts, which are separated by a twofold EP$_2$. 
(c),(d) Time evolution of the equal mode population $n(t)\equiv n_1(t)=n_2(t)$ (c) and the logarithmic negativity $E_N(t)$ (d) for $g/J=0.5$ (orange), $0.9$ (yellow), and $1.5$ (blue), corresponding to the colored markers i--iii in Regions I, II, and III, respectively.
In Region I, the population diverges while the entanglement converges to a finite asymptotic value. In Regions II and III, both the population and entanglement converge to steady-state values, with Region III showing transient oscillations at short times. 
The inset in (c) shows an enlarged view of the population dynamics at point iii, resolving the transient oscillations. 
Here, we take $\Gamma_{1,2}/J=1.2$ and $\kappa=0$.
        }
    	\label{fig2}
    \end{figure}

    We first consider the dissipative bosonic dimer with identical independent dissipation ($\Gamma_1=\Gamma_2\equiv\Gamma$) and ignore the common bath ($\kappa=0$). In this case, the first-order dynamical matrix $M_1$ can be diagonalized analytically, and its eigenvalues are given by
    \begin{equation}
        \lambda_{1,2}=-\frac{i}{2}(\Gamma-s),
        \qquad
        \lambda_{3,4}=-\frac{i}{2}(\Gamma+s),
    \end{equation}
    where
    \begin{equation}
        s=2\sqrt{J^2-g^2}.
    \end{equation}
    The real and imaginary parts of these eigenvalues are shown in Fig.~\ref{fig2}(a) and Fig.~\ref{fig2}(b), respectively.

    The dynamical spectrum divides the parameter space into three distinct regions. In Region I, at least one eigenvalue satisfies $\mathrm{Im}(\lambda_\mu)\ge 0$, indicating dynamical instability. By contrast, Regions II and III are dynamically stable, with all eigenvalues satisfying $\mathrm{Im}(\lambda_\mu)<0$. A twofold second-order EP (EP$_2$) separates Regions II and III. The order of the EP follows from the Jordan normal form of $M_1$, as discussed in Appendix~\ref{App-B1}. The dynamical spectrum thus identifies not only stable and unstable regimes, but also two qualitatively different stable regions.

To connect the spectral classification with the observable dynamics, we consider the three representative parameter values marked in Figs.~\ref{fig2}(a) and \ref{fig2}(b). For the symmetric setting considered here and the two-mode vacuum initial state, the two mode populations remain identical, $n_1(t)=n_2(t)\equiv n(t)$. The corresponding population and entanglement dynamics are shown in Figs.~\ref{fig2}(c) and \ref{fig2}(d), respectively. 
These three spectral regions exhibit distinct dynamical behaviors. 
In Region I, the population diverges at long times, while the entanglement converges to a finite asymptotic value. In Region II, the population approaches its steady-state value monotonically, whereas the entanglement first increases and then gradually decreases toward its asymptotic value. In Region III, the population converges through damped oscillations, while the entanglement first rises to a maximum and then approaches its asymptotic value in an oscillatory manner. In this sense, the twofold EP$_2$ separating Regions II and III marks the boundary between qualitatively different transient dynamics within the stable regime.

    This correspondence can be made explicit from the analytical solutions of the population and the minimal symplectic eigenvalue $\nu_-(t)$, whose 
    full expressions are presented in Appendix~\ref{App-C1}. The population takes the form
    \begin{equation}
        n(t)
        =
        \frac{2J^2}{\Gamma^2-s^2}
        -
        \frac{J^2 e^{-2it\lambda_{1,2}}}{s(\Gamma-s)}
        +
        \frac{J^2 e^{-2it\lambda_{3,4}}}{s(\Gamma+s)}.
    \end{equation}
    Thus, the population depends on exponential factors determined by the dynamical spectrum. The signs of the imaginary parts of the eigenvalues therefore determine whether the dynamics is amplifying, monotonically relaxing, or oscillatory while relaxing, as detailed in Appendix~\ref{App-C1}.

    The logarithmic negativity is $E_N(t)=\max\{0,-\log \nu_-(t)\}$, where
    \begin{equation}
        \nu_-(t)
        =
        \sqrt{
        \xi(t)-\sqrt{\xi(t)^2-\det V(t)}
        }.
    \end{equation}
    The explicit form of $\xi(t)$ and $\det V(t)$ is given in Appendix~\ref{App-C1}. In Region I, both $\xi(t)$ and $\det V(t)$ grow exponentially with time. Nevertheless, their leading divergences cancel in the expression for $\nu_-(t)$, so that the long-time behavior remains finite. In particular, the asymptotic form is
    \begin{equation}
        \nu_-^\infty = \lim_{t\to\infty} \nu_-(t)
        \simeq
        \sqrt{\frac{\det V(t)}{2\,\xi(t)}}\bigg|_{t\to\infty}
        =
        \frac{2J\Gamma}{s(s+\Gamma)},
    \end{equation}
    as shown in Appendix~\ref{App-C1}, which yields a finite asymptotic logarithmic negativity
    \begin{equation}
        E_N^\infty  = \lim_{t\to\infty} E_N(t) = \max\{0,-\log \left(\frac{2J\Gamma}{s(s+\Gamma)}\right)\}.
    \end{equation}
    Therefore, in the present open bosonic system, dynamical instability does not necessarily lead to unbounded entanglement growth.

\section{Finite Asymptotic Entanglement Diagram}  
    \label{sec4}  

    \begin{figure}
    	\centering
    	\includegraphics[width=\linewidth]{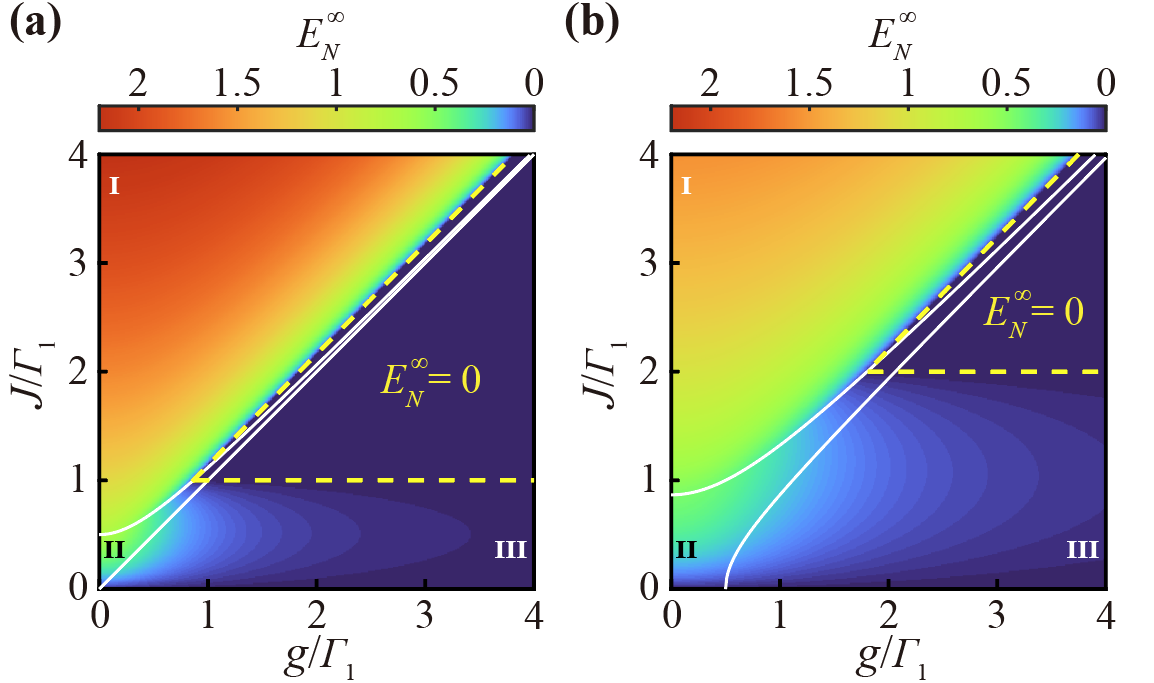}
    	\caption{
Asymptotic logarithmic negativity $E_N^\infty$ in the absence of a common bath ($\kappa=0$), shown as a function of $g/\Gamma_1$ and $J/\Gamma_1$.
(a) Symmetric local dissipation, $\Gamma_2=\Gamma_1$.
(b) Asymmetric local dissipation, $\Gamma_2=3\Gamma_1$.
The color scale indicates $E_N^\infty$. The white solid lines delineate the dynamical regions I, II, and III, while the yellow dashed lines mark the boundary between vanishing ($E_N^\infty = 0$) and nonzero ($E_N^\infty > 0$) asymptotic entanglement. Compared with the symmetric case, asymmetric local dissipation reduces the dynamically unstable region and enlarges the parameter regime supporting nonzero asymptotic entanglement.
        }
    	\label{fig3}
    \end{figure}
    
    We now focus on the long-time limit and characterize the asymptotic entanglement $E_N^\infty$ in Fig.~\ref{fig3}. 
    The resulting diagram is present in the $(g/\Gamma_1,J/\Gamma_1)$ plane for two representative dissipation settings: symmetric local dissipation with $\Gamma_1=\Gamma_2$ [Fig.~\ref{fig3}(a)] and asymmetric local dissipation with $\Gamma_2=3\Gamma_1$ [Fig.~\ref{fig3}(b)].

    In addition to the dynamical boundaries separating Regions I, II, and III, the diagram exhibits another distinct boundary between vanishing and nonzero asymptotic entanglement. These two boundaries encode different physical information. The white solid lines are determined by the spectrum of the dynamical matrix and mark transitions among different types of dynamical regimes, whereas the yellow dashed lines mark the boundary where the asymptotic logarithmic negativity changes from finite to zero.

    The analytical expressions for $E_N^\infty$ in the three dynamical regions follow from the long-time second-order moments summarized in Appendices~\ref{App-C1} and \ref{App-C2}. Qualitatively, Fig.~\ref{fig3} reveals several robust features. For symmetric dissipation, nonzero asymptotic entanglement occupies a broad portion of Region I and, within the stable regime, the region below the yellow dashed boundary, while $E_N^\infty$ vanishes in the upper-right part of the diagram. The comparison between Figs.~\ref{fig3}(a) and \ref{fig3}(b) shows directly that dissipation asymmetry shifts both the dynamical boundary and the boundary between vanishing and nonzero asymptotic entanglement. In particular, the dynamically unstable region is reduced, while the boundary between vanishing and nonzero asymptotic entanglement shifts, enlarging the parameter regime with nonzero asymptotic entanglement.

    \begin{figure}
        \centering
        \includegraphics[width=7cm]{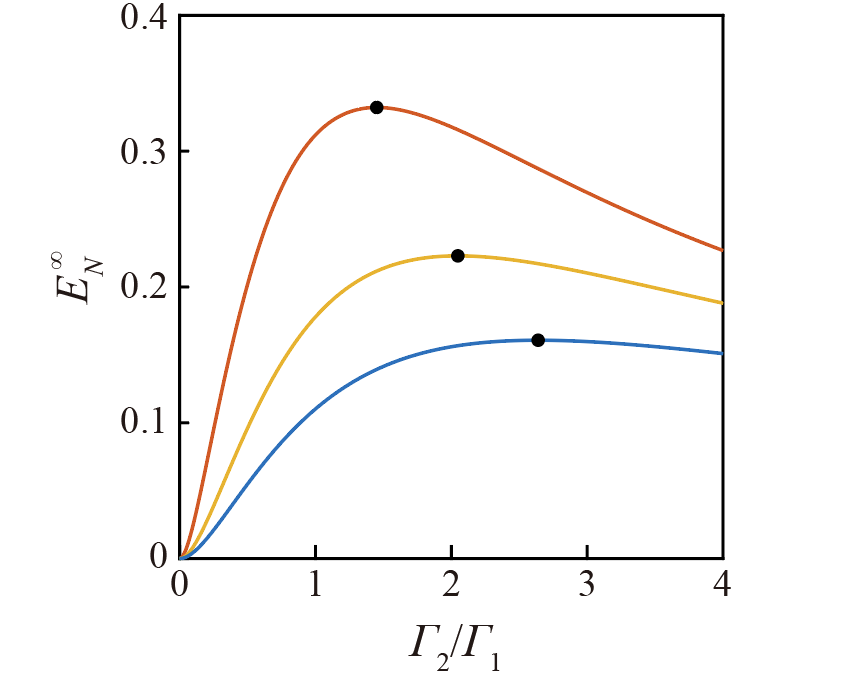}
        \caption{
Asymptotic logarithmic negativity $E_N^\infty$ as a function of the local-dissipation ratio $\Gamma_2/\Gamma_1$ for $g/\Gamma_1=0.5$ (orange), $0.75$ (yellow), and $1.0$ (blue). Here, $J/\Gamma_1=0.5$ and $\kappa=0$, and the system remains dynamically stable along all three parameter variations. The black dots indicate the maxima of $E_N^\infty$, located at $\Gamma_2/\Gamma_1\simeq1.45$, $2.05$, and $2.64$ for $g/\Gamma_1=0.5$, $0.75$, and $1.0$, respectively. The asymptotic entanglement therefore depends nonmonotonically on the dissipation asymmetry and reaches a maximum at a finite value of $\Gamma_2/\Gamma_1$ for fixed $g/\Gamma_1$ and $J/\Gamma_1$.
        }
        \label{fig4}
    \end{figure}

    The magnitude of the asymptotic entanglement is also modified in a nonuniform manner. Comparing Fig.~\ref{fig3}(a) and Fig.~\ref{fig3}(b), one finds that $E_N^\infty$ is generally reduced in the unstable regime and generally enhanced in the stable regimes. 
    More specifically, the enhancement in the stable regime is not strictly monotonic along arbitrary parameter cuts, as illustrated by the fixed-parameter slice shown in Fig.~\ref{fig4}. The three curves correspond to different choices of $g/\Gamma_1$ within the dynamically stable regime, while $J/\Gamma_1$ is held fixed. Although asymmetric local dissipation generally enlarges the finite-entanglement region in the two-dimensional diagram, the asymptotic entanglement along a fixed cut is not monotonic in $\Gamma_2/\Gamma_1$. Instead, each stable-branch curve exhibits a maximum at an intermediate dissipation asymmetry, illustrating the existence of an optimal local-loss imbalance for enhancing $E_N^\infty$. The detailed dependence on the dissipation asymmetry is given by the analytical expressions collected in Appendix~\ref{App-C2}. The asymptotic entanglement diagram accordingly provides a unified view of both the spectral organization of the dynamics and the long-time entanglement resource available in each parameter regime.

\section{Common-Bath Engineering of Asymptotic Entanglement}
    \label{sec5}

    \begin{figure}
    	\centering
    	\includegraphics[width=\linewidth]{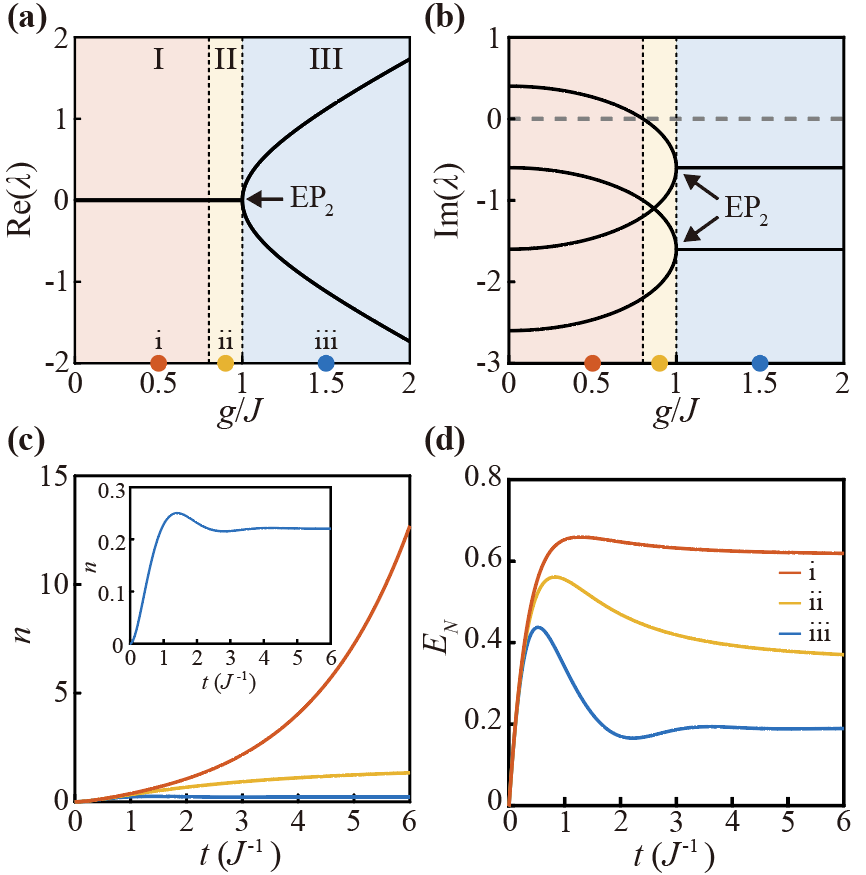}
    	\caption{
Dynamical spectrum and time evolution of population and entanglement in the presence of a common bath.
(a),(b) Real and imaginary parts, respectively, of the four eigenvalues of the dynamical matrix $M_1$ as functions of $g/J$. The vertical dotted lines mark the unchanged boundaries of the three dynamical regions I--III, while the common bath splits the twofold EP$_2$ in Fig.~\ref{fig2} into two distinct EP$_2$s.
(c),(d) Time evolution of the equal mode population $n(t)$ (c) and the logarithmic negativity $E_N(t)$ (d) for $g/J=0.5$ (orange), $0.9$ (yellow), and $1.5$ (blue), corresponding to the colored markers i--iii in Regions I, II, and III, respectively. 
The inset in (c) shows an enlarged view of the population dynamics at point iii, resolving the transient oscillations. Here, $\Gamma_1/J=\Gamma_2/J=1.2$ and $\kappa/J=1$.
    	}
    	\label{fig5}
    \end{figure}

    \begin{figure*}
    	\centering
    	\includegraphics[width=\textwidth]{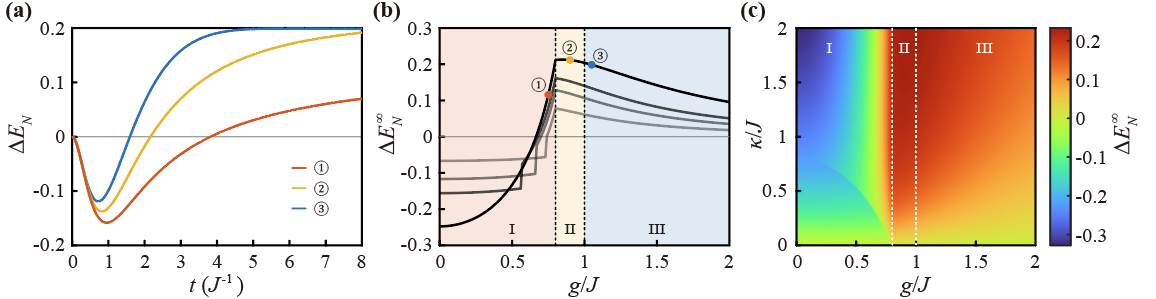}
    	\caption{
Enhancement of asymptotic entanglement induced by a common bath.
(a) Time evolution of the logarithmic-negativity difference
$\Delta E_N(t)$ at $\kappa/J=1$ for $g/J=0.75$ (orange),
$0.9$ (yellow), and $1.05$ (blue).
(b) Asymptotic entanglement difference $\Delta E_N^\infty$ as a
function of $g/J$ for different values of $\kappa/J$. The black curves
correspond to $\kappa/J$ values from $1$ to $0.2$ in steps of $0.2$,
with the opacity decreasing as $\kappa/J$ decreases. The colored markers \ding{192}--\ding{194}
indicate the parameter values used in (a).
(c) $\Delta E_N^\infty$ as a function of $g/J$ and $\kappa/J$.
The vertical dashed lines indicate the boundaries between Regions I, II,
and III. Positive values of $\Delta E_N^\infty$ identify parameter
regimes in which the common bath enhances the asymptotic entanglement
even though it introduces an additional dissipative channel.
Here, $\Gamma_1/J=\Gamma_2/J=1.2$.
    	}
    	\label{fig6}
    \end{figure*}
    
    Having established the spectral-entanglement structure in the absence of a common bath, we now ask how it is modified when a common bath is introduced. Besides being a standard ingredient in reservoir engineering, a common bath qualitatively reshapes the dissipative structure and can therefore alter both the non-Hermitian spectrum and the resulting asymptotic entanglement. We address this question first for identical local dissipation, $\Gamma_1=\Gamma_2\equiv\Gamma$. In this case, the eigenvalues of $M_1$ become
    \begin{equation}
    \begin{aligned}
        \lambda_1=-\frac{i}{2}(\Gamma-s),\qquad
        \lambda_2=-\frac{i}{2}(\Gamma+2\kappa-s),\\
        \lambda_3=-\frac{i}{2}(\Gamma+s),\qquad
        \lambda_4=-\frac{i}{2}(\Gamma+2\kappa+s).
    \end{aligned}
    \end{equation}
    Figures~\ref{fig5}(a) and \ref{fig5}(b) show that the common bath lifts the degeneracy of the $\kappa=0$ spectrum by shifting one branch of the imaginary spectrum downward. The former twofold EP$_2$ structure is correspondingly split into two distinct EP$_2$s (see Appendix~\ref{App-B3} for details). The boundaries between Regions I, II, and III, however, remain unchanged. In particular, Region I is still dynamically unstable, whereas Regions II and III remain dynamically stable.

    The corresponding population and entanglement dynamics are shown in Figs.~\ref{fig5}(c) and \ref{fig5}(d). Qualitatively, the classification established in Sec.~\ref{sec3} remains valid: Region I still retains its hallmark of a divergent population with finite asymptotic entanglement, while Regions II and III exhibit convergent evolution toward asymptotic or stable states. Moreover, compared with Fig.~\ref{fig2}, the population is reduced in magnitude [Fig.~\ref{fig5}(c)], and for the same representative parameters the common bath enhances the asymptotic entanglement, such as the blue and yellow curves in Fig.~\ref{fig5}(d). The principal effect of the common bath thus lies not in changing the region boundaries, but in modifying the dynamics within them. 
    
    To characterize the common-bath contribution, we introduce
    \begin{equation}
        \Delta E_N(t;\kappa) = E_N(t;\kappa)-E_N(t;0),
    \end{equation}
    together with its long-time limit
    \begin{equation}
        \Delta E_N^\infty(\kappa) = E_N^\infty(\kappa)-E_N^\infty(0).
    \end{equation}
    Figure~\ref{fig6} illustrates the effect of the common bath on the asymptotic entanglement. Fig.~\ref{fig6}(a) shows the time evolution of $\Delta E_N(t;\kappa)$ for three representative parameter sets from Regions I, II, and III. For the stable-region parameters (yellow and blue curves), the introduction of the common bath initially lowers the logarithmic negativity, as expected from the additional dissipative channel. At longer times, however, $\Delta E_N(t;\kappa)$ becomes positive, showing that the common bath ultimately enhances the asymptotic entanglement. The orange curve further shows that this enhancement is not restricted to the stable regimes: even inside the dynamically unstable Region I, $\Delta E_N(t;\kappa)$ becomes positive after approximately $Jt\approx 4$ for the parameter set shown. This long-time enhancement can be understood as arising from a quasi-BIC-type protection induced by the common dissipative channel, as we discuss in the next section.

    Figure~\ref{fig6}(b) provides a more systematic view by plotting $\Delta E_N^\infty$ versus $g/J$ for different $\kappa/J$. The black curves correspond to $\kappa/J$ values from $1$ to $0.2$, with opacity decreasing accordingly, and the colored markers indicate the three parameter values used for the time traces in Fig.~\ref{fig6}(a). The curves show directly that broad enhancement windows appear in the stable regimes, while a finite parameter window with $\Delta E_N^\infty>0$ also survives inside Region I. The full heatmap in Fig.~\ref{fig6}(c) gives the global picture of how the common bath modifies the asymptotic entanglement. It is shown that the asymptotic entanglement can be enhanced over a broad parameter range in Regions II and III. By contrast, most of Region I exhibits reduced asymptotic entanglement, while a narrow enhancement window also exists inside Region I.

    The discontinuous features visible in both Figs.~\ref{fig6}(b) and \ref{fig6}(c) originate from the change of the leading asymptotic contribution across $s-\Gamma-\kappa=0$, as discussed in Appendix~\ref{App-C3}. The corresponding analytical expressions 
    are summarized there. These results show that the common bath acts as a structured dissipative resource: rather than simply suppressing quantum correlations, it can selectively reduce or enhance the asymptotic entanglement depending on where the parameter point lies in the diagram.

\section{BIC-Type Protection in the common-bath-only Limit}

    To gain further insight into the long-time enhancement induced by a common bath, we consider the limiting case without independent dissipation, namely $\Gamma_{1,2}=0$ and $\kappa\neq0$. In this common-bath setting, a BIC-type perspective is particularly useful. For systems with only exchange coupling and a shared dissipative environment, a common perspective is to identify a bright mode that  remains coupled to the bath and a dark mode decoupled. In the present model, the additional two-mode-squeezing interaction does not invalidate this picture, but instead admits a natural generalization.

    Concretely, in the common-bath-only limit the master equation reduces to
    \begin{equation}
        \dot{\rho}
        =
        -i[\hat{H},\rho]
        + \kappa \mathcal{D}[\hat{a}_1+\hat{a}_2]\rho.
    \end{equation}
    Defining the orthogonal modes
    \begin{equation}
        \hat{a}_{\pm}
        =
        \frac{\hat{a}_1\pm\hat{a}_2}{\sqrt2},
    \end{equation}
    one finds
    \begin{equation}
        \dot{\rho}
        =
        -i[\hat{H}_+-\hat{H}_-,\rho]
        + \kappa \mathcal{D}[\sqrt2\,\hat{a}_+]\rho,
    \end{equation}
    where
    \begin{equation}
        \begin{aligned}
        \hat{H}_+-\hat{H}_-
        &=
        g \hat{a}_+^\dagger \hat{a}_+
        +
        \frac{J}{2}\left(\hat{a}_+^2+\hat{a}_+^{\dagger 2}\right)
        \\
        &\quad
        -
        g \hat{a}_-^\dagger \hat{a}_-
        -
        \frac{J}{2}\left(\hat{a}_-^2+\hat{a}_-^{\dagger 2}\right).
        \end{aligned}
    \end{equation}
    The dissipator therefore acts only on the symmetric mode $\hat{a}_+$, while the antisymmetric mode $\hat{a}_-$ remains exactly decoupled from the bath. Meanwhile, the two-mode-squeezing interaction is converted into independent single-mode squeezing terms for the two orthogonal modes and therefore preserves this decoupling. As a result, the antisymmetric mode supports undamped squeezing dynamics, which in the original modal basis gives rise to a protected channel for entanglement generation.

    \begin{figure}[t]
        \centering
        \includegraphics[width=8cm]{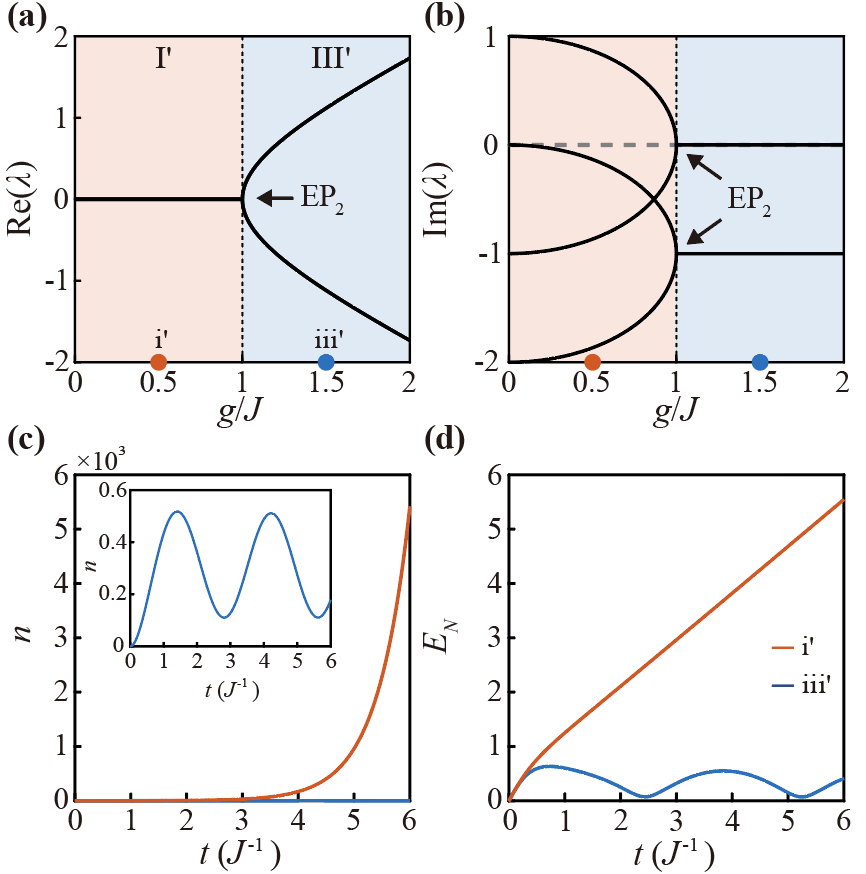}
        \caption{
Dynamical spectrum and time evolution of population and entanglement in the common-bath-only limit.
(a),(b) Real and imaginary parts, respectively, of the four eigenvalues of the dynamical matrix $M_1$ as functions of $g/J$. At $g/J=1$, two distinct EP$_2$s separate Regions I$'$ and III$'$. In Region I$'$, all eigenvalues are purely imaginary, whereas in Region III$'$, two eigenvalues are purely real and the other two are complex. The largest imaginary part is nonnegative in both regions, precluding asymptotic stability. Region I$'$ contains an amplifying mode, whereas Region III$'$ contains undamped modes.
(c),(d) Time evolution of the equal mode population $n(t)$ (c) and the logarithmic negativity $E_N(t)$ (d) for $g/J=0.5$ (orange) and $1.5$ (blue), corresponding to the colored markers i$'$ and iii$'$ in Regions I$'$ and III$'$, respectively. In Region I$'$, both the population and entanglement grow without bound, whereas in Region III$'$ they exhibit persistent oscillations. The inset in (c) shows an enlarged view of the population dynamics at point iii$'$, resolving the persistent oscillations. These behaviors are qualitatively identical to those in the corresponding regions of the dissipation-free system. Here, $\Gamma_1=\Gamma_2=0$ and $\kappa/J=1$.
    }
        \label{fig7}
    \end{figure}

    The consequences of this mode structure are reflected in the dynamical spectrum and in the corresponding population and entanglement dynamics. In this limit, the first-order dynamical matrix has the eigenvalues
    \begin{equation}
    \begin{aligned}
        \lambda_{1}=\frac{i s}{2},\qquad
        \lambda_{2}=-\frac{i}{2}(2\kappa-s),\\
        \lambda_{3}=-\frac{i s}{2},\qquad
        \lambda_{4}=-\frac{i}{2}(2\kappa+s).
    \end{aligned}
    \end{equation}
    The corresponding spectrum is shown in Fig.~\ref{fig7}(a) and Fig.~\ref{fig7}(b). Since at least one eigenvalue always has a nonnegative imaginary part in this limit, no strictly stable region exists. Instead, the spectrum is divided by a pair of EP$_2$s at $g=J$ into two dynamically unstable regions, denoted I$'$ and III$'$.
    In Region I$'$, all eigenvalues are purely imaginary, whereas in Region III$'$, two eigenvalues are purely real while the other two remain complex.

    The lower panels of Fig.~\ref{fig7} show how this spectral structure is reflected in the population and entanglement dynamics. Figure~\ref{fig7}(c) shows that the long-time population behaves as
    \begin{align}
        \text{Region I$'$ ($s>0$):}\qquad
        P(t)&\overset{t\gg 0}{\approx}\frac{J^2 e^{s t}}{2 s^2},
        \\
        \text{Region III$'$:}\qquad
        P(t)&\overset{t\gg 0}{\approx}
        \frac{2 J^2 \left(2 \kappa ^2+\omega ^2\right)}{4 \kappa ^2 \omega ^2+\omega ^4}
        \notag\\
        &\quad
        -\frac{J^2 \cos (t \omega )}{\omega ^2},
    \end{align}
    where $\omega=-is=2\sqrt{g^2-J^2}$ in Region III$'$. Thus, Region I$'$ exhibits exponential growth, whereas Region III$'$ retains long-time oscillations.

    Now we will focus on the entanglement dynamics, whose analytical discussion is provided in Appendix~\ref{App-C4}.
    In Region I$'$, the long-time asymptotics satisfy $\xi(t)\gg\det V(t)$, so that $\nu_-(t)\to0$ and the logarithmic negativity is unbounded. In Region III$'$, the long-time forms of $\xi(t)$ and $\det V(t)$ retain an explicit oscillatory dependence, 
\begin{equation}
            \xi(t) \overset{t\gg0}{\approx}
            -\frac{8 \kappa J^2 \left(2 \kappa \cos (t \omega )-\omega \sin (t \omega )\right)}{4 \kappa ^2 \omega ^2+\omega ^4}
            +\frac{4 J^2}{\omega ^2}+1,
\end{equation}
\begin{equation}
            \det V(t) \overset{t\gg0}{\approx}
            \frac{4 \kappa ^2+4 J^2+\omega ^2}{4 \kappa ^2+\omega ^2},
\end{equation}
    which leads to the persistent oscillations of $E_N(t)$ displayed in Fig.~\ref{fig7}(d). These behaviors are qualitatively identical to those in the corresponding regions of the ideal system without dissipation~\cite{Yu2025Exceptional}. In this sense, the common-bath-only limit provides a physically transparent picture of the protection mechanism induced by the shared dissipative channel. Once independent dissipation is restored, this exact protection is reduced to a quasi-BIC-type remnant, manifested as the common-bath-induced enhancement of the asymptotic entanglement discussed above.

\section{Conclusion}

Our results establish a direct connection between the non-Hermitian dynamical spectrum of an open bosonic dimer and its population and entanglement behavior. The spectrum of the first-order dynamical matrix divides the parameter space into distinct dynamical regimes, which in turn organize the population dynamics and the evolution of the logarithmic negativity. Most notably, dynamical instability does not necessarily imply unbounded entanglement growth: even in the dynamically unstable regime, the bosonic population can diverge while the asymptotic entanglement remains finite.

We further show that dissipation reshapes the asymptotic entanglement landscape in two distinct ways. Asymmetric local dissipation shifts both the dynamical stability boundary and the boundary between vanishing and nonzero asymptotic entanglement. In particular, for identical local dissipation, a common bath leaves the dynamical-region boundaries unchanged but qualitatively reshapes the asymptotic entanglement landscape, yielding broad enhancement windows in the stable regimes together with a narrow enhancement window even inside the unstable regime.

The time-domain analysis further shows that the common bath can initially reduce the entanglement, as expected from the additional dissipative channel, while nevertheless enhancing the long-time asymptotic value. In the common-bath-only limit, this behavior can be traced to a dark-mode or BIC-type protection mechanism in the orthogonal modal basis. Once independent dissipation is restored, the exact protection is reduced to a quasi-BIC-type remnant, reflected in the common-bath-induced enhancement of the asymptotic entanglement.

Taken together, these results show that the non-Hermitian dynamical spectrum of an open quadratic bosonic system provides a unified framework for characterizing dynamical stability, nonequilibrium evolution, and long-time entanglement.
These results are directly relevant to bosonic platforms in which beam-splitter, squeezing, and dissipative couplings can be tailored, including photonic~\cite{wang2025large, jia2025continuous}, optomechanical~\cite{Del2022,Wanjura2023}, and superconducting implementations~\cite{Busnaina2024}. More broadly, asymptotic entanglement provides a useful diagnostic of how spectral structures are reflected in open-system dynamics.
Go further, it would be interesting to extend this study from a quantum dimer to multimode bosonic systems and explore how dynamical spectra~\cite{wanjura2020a,tian2023a} affect the multimode entanglement~\cite{vanloock2003,Gao2026}.
Going beyond the quadratic Gaussian regime, it would also be valuable to clarify how interaction-modified spectra govern the emergence and long-time behavior of non-Gaussian CV entanglement~\cite{venkatraman2023,tian2022c,tian2025b}.

\begin{acknowledgements}
    This work is supported by the Quantum Science and Technology-National Science and Technology Major Project (Grants No. 2024ZD0302401, No. 2025ZD0301000 and No. 2021ZD0301500), the National Natural Science Foundation of China (Grants No.~12125402, No.~12534016, No.~12474256), and the Beijing Natural Science Foundation (Grant No. Z240007). F.-X. S. acknowledges support from the Fundamental Research Funds for the Central Universities (Grant No. 539926010).
\end{acknowledgements}

\begin{appendix}
\section{Dynamical matrices, entanglement characterization, and symmetry analysis}
\label{App-A}

In this appendix, we derive the dynamical matrices employed in the main text for the dissipative bosonic dimer, introduce the logarithmic negativity used to characterize bipartite entanglement, and summarize the symmetry properties of the first-order dynamical matrix in the absence and presence of dissipation.

        \subsection{Master equation and moment dynamics}\label{App-A1}

        The coherent Hamiltonian is
        \begin{equation}\label{eqA1}
            \hat{H}
            =
            g \hat{a}_1^\dagger \hat{a}_2
            +
            J \hat{a}_1^\dagger \hat{a}_2^\dagger
            + \mathrm{h.c.},
        \end{equation}
        which is Eq.~(\ref{eq1}) of the main text. The open-system dynamics are described by the Markovian master equation
        \begin{equation}\label{eqA2}
            \dot{\rho}
            =
            -i[\hat{H},\rho]
            + \Gamma_1 \mathcal{D}[\hat{a}_1]\rho
            + \Gamma_2 \mathcal{D}[\hat{a}_2]\rho
            + \kappa \mathcal{D}[\hat{a}_1+\hat{a}_2]\rho,
        \end{equation}
        where
        \begin{equation}
            \mathcal{D}[\hat{o}]\rho
            =
            \hat{o}\rho \hat{o}^\dagger
            - \frac{1}{2}\{\hat{o}^\dagger \hat{o},\rho\}.
        \end{equation}
        For an arbitrary system operator $\hat{O}$, the master equation leads to
        \begin{equation}\label{eqA3}
            \begin{aligned}
            i\frac{d}{dt}\langle \hat{O}\rangle
            &=
            \langle[\hat{O},\hat{H}]\rangle 
            + \sum_{\mu}i \gamma_\mu
            \left\langle
            \hat{L}_\mu^\dagger \hat{O}\hat{L}_\mu
            - \frac{1}{2}\{\hat{L}_\mu^\dagger \hat{L}_\mu,\hat{O}\}
            \right\rangle \\
            &=
            \left\langle[\hat{O},\hat{H}] 
            + \sum_{\mu}\frac{i\gamma_\mu}{2}(
            [\hat{L}_\mu^\dagger, \hat{O}]\hat{L}_\mu
            + \hat{L}_\mu^\dagger [\hat{O}, \hat{L}_\mu])
            \right\rangle ,
            \end{aligned}
        \end{equation}
        where the jump operators are
        \begin{equation}
            \hat{L}_1=\hat{a}_1,\qquad
            \hat{L}_2=\hat{a}_2,\qquad
            \hat{L}_c=\hat{a}_1+\hat{a}_2,
        \end{equation}
        with the corresponding rates
        $\gamma_1=\Gamma_1$, $\gamma_2=\Gamma_2$, and $\gamma_c=\kappa$.

        We define the operator vector
        \begin{equation}
            \Psi
            =
            \left(
            \hat{a}_1,
            \hat{a}_1^\dagger,
            \hat{a}_2,
            \hat{a}_2^\dagger
            \right)^{\mathrm T},
        \end{equation}
        the first-order moment vector
        \begin{equation}
            v_1 = \langle \Psi \rangle,
        \end{equation}
        and the second-order moment vector
        \begin{equation}
            v_2 = \mathrm{vec}\!\left(\langle \Psi \Psi^{\mathrm T}\rangle\right).
        \end{equation}

        It is convenient to introduce the linear map $\mathscr{L}(\hat{O})$ through
        \begin{equation}
            \mathscr{L}(\hat{O})
            =[\hat{O},\hat{H}] 
            + \sum_{\mu}\frac{i\gamma_\mu}{2}(
            [\hat{L}_\mu^\dagger, \hat{O}]\hat{L}_\mu
            + \hat{L}_\mu^\dagger [\hat{O}, \hat{L}_\mu])
            .
        \end{equation}
        The operator basis $\Psi$ is closed under the action of the linear map $\mathscr{L}$, so that
\begin{equation}
    \mathscr{L}(\Psi_i)
    =
    \sum_{k=1}^{4}(M_1)_{ik}\Psi_k,
\end{equation}
where $M_1$ is the matrix representation of $\mathscr{L}$ in the basis $\Psi$. Consequently, the first-order moments satisfy
\begin{equation}
    i\frac{d}{dt}v_1=M_1v_1,
\end{equation}
with
        \begin{equation}\label{eqA5a}
            M_1=
            \begin{bmatrix}
                -\frac{i(\Gamma_1+\kappa)}{2} & 0 & g-\frac{i\kappa}{2} & J \\
                0 & -\frac{i(\Gamma_1+\kappa)}{2} & -J & -g-\frac{i\kappa}{2} \\
                g-\frac{i\kappa}{2} & J & -\frac{i(\Gamma_2+\kappa)}{2} & 0 \\
                -J & -g-\frac{i\kappa}{2} & 0 & -\frac{i(\Gamma_2+\kappa)}{2}
            \end{bmatrix}.
        \end{equation}

        To derive the second-order dynamics, we define the operator matrix
        \begin{equation}
            S=\Psi\Psi^{\mathrm T},
            \qquad
            S_{ij}=\Psi_i\Psi_j.
        \end{equation}
        Applying $\mathscr{L}$ to $S_{ij}$ yields
        \begin{widetext}
        \begin{equation}
            \begin{aligned}
            \left\langle\mathscr{L}(S_{ij}) \right\rangle
			=&\left\langle[\Psi_i\Psi_j,\hat{H}] 
            + \sum_{\mu}\frac{i\gamma_\mu}{2}(
            [\hat{L}_\mu^\dagger, \Psi_i\Psi_j]\hat{L}_\mu
            + \hat{L}_\mu^\dagger [\Psi_i\Psi_j, \hat{L}_\mu])
            \right\rangle\\
			=&\left\langle[\Psi_i,\hat{H}]\Psi_j 
            + \sum_{\mu}\frac{i\gamma_\mu}{2}(
            [\hat{L}_\mu^\dagger, \Psi_i]\hat{L}_\mu \Psi_j
            + \hat{L}_\mu^\dagger [\Psi_i, \hat{L}_\mu]\Psi_j)
            \right\rangle\\
			&+\left\langle\Psi_i[\Psi_j,\hat{H}] 
            + \sum_{\mu}\frac{i\gamma_\mu}{2}(
            \Psi_i[\hat{L}_\mu^\dagger, \Psi_j]\hat{L}_\mu
            + \Psi_i\hat{L}_\mu^\dagger [\Psi_j, \hat{L}_\mu])
            \right\rangle\\
			&+\left\langle
			\sum_{\mu}\frac{i\gamma_\mu}{2}(
            [\hat{L}_\mu^\dagger, \Psi_i]\Psi_j \hat{L}_\mu - [\hat{L}_\mu^\dagger, \Psi_i]\hat{L}_\mu \Psi_j)
            \right\rangle
			+\left\langle
			\sum_{\mu}\frac{i\gamma_\mu}{2}(
            \hat{L}_\mu^\dagger\Psi_i[\Psi_j, \hat{L}_\mu]  - \Psi_i\hat{L}_\mu^\dagger[\Psi_j, \hat{L}_\mu])
            \right\rangle\\
			=&\left\langle\mathscr{L}(\Psi_i) \Psi_j\right\rangle + \left\langle \Psi_i\mathscr{L}(\Psi_j) \right\rangle
			+ \left\langle
			\sum_{\mu}i\gamma_\mu(
            [\hat{L}_\mu^\dagger, \Psi_i][\Psi_j, \hat{L}_\mu])
            \right\rangle\\
			=&\sum_k (M_1)_{ik}\left\langle\Psi_k \Psi_j\right\rangle + \sum_k (M_1)_{jk}\left\langle\Psi_i \Psi_k\right\rangle
			+ (D_S)_{ij}\\
			=&\sum_k (M_1)_{ik}\left\langle S_{kj}\right\rangle + \sum_k \left\langle S_{ik} \right\rangle (M_1^\mathrm{T})_{kj}
			+ (D_S)_{ij}
            \end{aligned}
        \end{equation}
        \end{widetext}
        where $(D_S)_{ij} = \left\langle \sum_{\mu}i\gamma_\mu ([\hat{L}_\mu^\dagger, \Psi_i][\Psi_j,\hat{L}_\mu]) \right\rangle$ is a constant inhomogeneous term generated by the Lindblad dissipators through the bosonic commutation relations. In matrix form, this gives
        \begin{equation}\label{eqA4}
            i\frac{d}{dt}\langle S\rangle
            =
            M_1\langle S\rangle
            +
            \langle S\rangle M_1^{\mathrm T}
            +
            D_S,
        \end{equation}
Vectorizing Eq.~(\ref{eqA4}) using
$\mathrm{vec}(AXB)=(B^{\mathrm T}\otimes A)\mathrm{vec}(X)$
gives
\begin{equation}
    i\frac{d}{dt}v_2 = M_2 v_2 + D,
\end{equation}
where
\begin{equation}
    M_2 = I\otimes M_1 + M_1\otimes I,
\end{equation}
with $I$ the $4\times4$ identity matrix, and
    $D=\mathrm{vec}(D_S)
    =
    [0,0,0,0,i(\Gamma_1+\kappa),0,i\kappa,0,
    0,0,0,0,i\kappa,0,i(\Gamma_2+\kappa),0]^{\mathrm T}.$

In addition, the second-order moment matrix $\langle S\rangle$ takes the form
\begin{equation}
    \langle S\rangle
    =
    \left\langle
    \begin{bmatrix}
    \hat{a}_1^2 & \hat{a}_1\hat{a}_1^\dagger & \hat{a}_1\hat{a}_2 & \hat{a}_1\hat{a}_2^\dagger \\
    \hat{a}_1^\dagger\hat{a}_1 & (\hat{a}_1^\dagger)^2 & \hat{a}_1^\dagger\hat{a}_2 & \hat{a}_1^\dagger\hat{a}_2^\dagger \\
    \hat{a}_2\hat{a}_1 & \hat{a}_2\hat{a}_1^\dagger & \hat{a}_2^2 & \hat{a}_2\hat{a}_2^\dagger \\
    \hat{a}_2^\dagger\hat{a}_1 & \hat{a}_2^\dagger\hat{a}_1^\dagger & \hat{a}_2^\dagger\hat{a}_2 & (\hat{a}_2^\dagger)^2
    \end{bmatrix}
    \right\rangle .
\end{equation}
For convenience, we label the 16 entries according to
\begin{equation}
    \begin{bmatrix}
    1 & 5 & 9 & 13 \\
    2 & 6 & 10 & 14 \\
    3 & 7 & 11 & 15 \\
    4 & 8 & 12 & 16
    \end{bmatrix}.
\end{equation}
Among its $16$ entries, the following relations hold:
\begin{align}
    \langle S\rangle_{6(16)}&=\langle S\rangle_{1(11)}^\ast, \\
    \langle S\rangle_{5(15)}&=\langle S\rangle_{2(12)}+1, \\
    \langle S\rangle_{3}&=\langle S\rangle_{9}, \qquad
    \langle S\rangle_{4}=\langle S\rangle_{13}, \\
    \langle S\rangle_{7}&=\langle S\rangle_{10}, \qquad
    \langle S\rangle_{8}=\langle S\rangle_{14}, \\
    \langle S\rangle_{8(7)}&=\langle S\rangle_{3(4)}^\ast .
\end{align}
Therefore, only six independent second-order moments are required to reconstruct the full matrix $\langle S\rangle$, which simplifies the subsequent calculations.

\subsection{Initial state and entanglement characterization}
\label{App-A2}

Throughout the time-domain analysis, we take the initial state to be the two-mode vacuum. The first-order moments evolve according to
\begin{equation}
    i\frac{d}{dt}v_1=M_1v_1.
\end{equation}
Since $v_1(0)=0$, it follows that $v_1(t)\equiv0$ at all times. The Gaussian dynamics are therefore fully characterized by the covariance matrix (CM), from which we quantify bipartite entanglement using the logarithmic negativity.

We define the quadrature vector
\begin{equation}
    R=(\hat{X}_1,\hat{P}_1,\hat{X}_2,\hat{P}_2)^{\mathrm T},
\end{equation}
where
\begin{equation}
    \hat{X}_j=\frac{\hat{a}_j+\hat{a}_j^\dagger}{\sqrt{2}},
    \qquad
    \hat{P}_j=\frac{\hat{a}_j-\hat{a}_j^\dagger}{i\sqrt{2}}.
\end{equation}
Since $v_1(t)\equiv0$, the CM elements reduce to
\begin{equation}
    V_{mn}
    =
    \left\langle
    \hat{R}_m\hat{R}_n+\hat{R}_n\hat{R}_m
    \right\rangle .
\end{equation}
The CM takes the block form
\begin{equation}
    V\equiv
    \begin{pmatrix}
        \sigma_A & \gamma_{AB} \\
        \gamma_{AB}^{\mathrm T} & \sigma_B
    \end{pmatrix}.
\end{equation}

The bipartite entanglement is quantified by the logarithmic negativity,
\begin{equation}\label{eq:EN}
    E_N(t)=\max\{0,-\log\nu_-(t)\},
\end{equation}
where $\nu_-(t)$ is the smallest symplectic eigenvalue of the partially transposed CM, obtained via the transformation $\hat{P}_B\rightarrow-\hat{P}_B$. Explicitly,
\begin{equation}\label{eq:numin}
    \nu_-(t)
    =
    \sqrt{
    \xi(t)-\sqrt{\xi^2(t)-\det V(t)}
    },
\end{equation}
with
\begin{equation}
    \xi(t)
    =
    \frac{
        \det\sigma_A
        +
        \det\sigma_B
        -
        2\det\gamma_{AB}
    }{2}.
\end{equation}

        \subsection{Symmetry properties of the dynamical matrix}\label{App-A3}

        To describe the relevant spectral symmetries, we define the chiral operator
\begin{equation}
    \Pi=I_2\otimes\sigma_x,
\end{equation}
and the antiunitary operator
\begin{equation}
    \mathcal{PT}=(I_2\otimes\sigma_x)\mathcal{K},
\end{equation}
where $\mathcal{K}$ denotes complex conjugation.

        In the dissipationless limit $\Gamma_1=\Gamma_2=\kappa=0$, one verifies that
        \begin{equation}
            \Pi M_1 \Pi = -M_1,
        \end{equation}
        so that $M_1$ possesses chiral symmetry. At the same time,
        \begin{equation}
            \mathcal{PT}\, M_1 \,\mathcal{PT} = -M_1.
        \end{equation}
corresponding to anti-$\mathcal{PT}$ symmetry. 
        These relations imply that the spectrum is symmetric with respect to both the real and imaginary axes, or equivalently centrosymmetric about the origin.

Once dissipation is introduced, the chiral symmetry is broken, whereas the anti-$\mathcal{PT}$ symmetry remains preserved for the dissipative models considered in this work. Consequently, the dynamical spectrum retains the reflection symmetry about the imaginary axis discussed in the main text: if $\lambda$ is an eigenvalue, then so is $-\lambda^*$.

    \section{Spectral structure and EPs of the dissipative bosonic dimer}\label{App-B}

In this appendix, we summarize the analytical spectral structure of the first-order dynamical matrix $M_1$ and the EPs relevant to the main-text discussion of Fig.~\ref{fig2}, Fig.~\ref{fig5} and Fig.~\ref{fig7}.

        Although the second-order moments and the entanglement dynamics are directly governed by $M_2$, the relation $M_2=I\otimes M_1+M_1\otimes I$ implies that the spectrum of $M_2$ is built from pairwise sums of the eigenvalues of $M_1$. Therefore, the spectral boundaries and EP conditions relevant to the main-text dynamical regimes can be identified from the simpler first-order matrix $M_1$.

        \subsection{Identical independent dissipation}\label{App-B1}

        For identical independent dissipation $\Gamma_1=\Gamma_2\equiv\Gamma$ and in the absence of a common bath ($\kappa=0$), the first-order dynamical matrix is
        \begin{equation}
            M_1 =
            \begin{pmatrix}
                -\frac{i\Gamma}{2} & 0 & g & J \\
                0 & -\frac{i\Gamma}{2} & -J & -g \\
                g & J & -\frac{i\Gamma}{2} & 0 \\
                -J & -g & 0 & -\frac{i\Gamma}{2}
            \end{pmatrix}.
        \end{equation}
        Its eigenvalues can be written analytically as
        \begin{equation}\label{eq:eig_id}
            \lambda_{1,2}=-\frac{i}{2}(\Gamma-s),
            \qquad
            \lambda_{3,4}=-\frac{i}{2}(\Gamma+s),
        \end{equation}
        where
        \begin{equation}\label{eq:s}
            s=2\sqrt{J^2-g^2}.
        \end{equation}

        Equation~(\ref{eq:eig_id}) yields the three dynamical regions discussed in the main text. Region I is characterized by at least one eigenvalue with nonnegative imaginary part, while in Regions II and III all eigenvalues have strictly negative imaginary parts. The boundary between Regions II and III is determined by a twofold EP$_2$.

        The corresponding EP condition is $g=J$.
        At this point, $M_1$ is brought to its Jordan form
        \begin{equation}\label{eq:id_EP}
            \mathcal{J}_{\Gamma}
            =
            V_{\Gamma}^{-1}M_1\big|_{g=J}V_{\Gamma}
            =
            \begin{pmatrix}
                -\frac{i\Gamma}{2} & 1 & 0 & 0 \\
                0 & -\frac{i\Gamma}{2} & 0 & 0 \\
                0 & 0 & -\frac{i\Gamma}{2} & 1 \\
                0 & 0 & 0 & -\frac{i\Gamma}{2}
            \end{pmatrix},
        \end{equation}
        \begin{equation}
            V_{\Gamma}
            =
            \begin{pmatrix}
                0 & -\frac{1}{J} & -1 & 0 \\
                0 & 0 & 1 & 0 \\
                -1 & 0 & 0 & -\frac{1}{J} \\
                1 & 0 & 0 & 0
            \end{pmatrix}.
        \end{equation}
        Away from the EP, the matrix can be diagonalized by
        \begin{equation}
            V'_\Gamma=
            \begin{pmatrix}
                -\frac{J}{\sqrt{g^2-J^2}} & -\frac{g}{\sqrt{g^2-J^2}} & \frac{J}{\sqrt{g^2-J^2}} & \frac{g}{\sqrt{g^2-J^2}} \\
                \frac{g}{\sqrt{g^2-J^2}} & \frac{J}{\sqrt{g^2-J^2}} & -\frac{g}{\sqrt{g^2-J^2}} & -\frac{J}{\sqrt{g^2-J^2}} \\
                0 & 1 & 0 & 1 \\
                1 & 0 & 1 & 0
            \end{pmatrix},
        \end{equation}
        so that no EP exists for $g\neq J$.

        \subsection{Asymmetric independent dissipation}\label{App-B2}

        For asymmetric independent dissipation, it is convenient to write $\Gamma_1=\Gamma+\delta\Gamma$ and $\Gamma_2=\Gamma-\delta\Gamma$. The first-order dynamical matrix becomes
        \begin{equation}
            M_1 =
            \begin{pmatrix}
                -\frac{i(\Gamma+\delta\Gamma)}{2} & 0 & g & J \\
                0 & -\frac{i(\Gamma+\delta\Gamma)}{2} & -J & -g \\
                g & J & -\frac{i(\Gamma-\delta\Gamma)}{2} & 0 \\
                -J & -g & 0 & -\frac{i(\Gamma-\delta\Gamma)}{2}
            \end{pmatrix},
        \end{equation}
        with eigenvalues
        \begin{equation}\label{eq:eig_asym}
            \lambda_{1,2}=-\frac{i}{2}(\Gamma-\tilde{s}),
            \qquad
            \lambda_{3,4}=-\frac{i}{2}(\Gamma+\tilde{s}),
        \end{equation}
        where
        \begin{equation}
            \tilde{s}=\sqrt{4J^2-4g^2+\delta\Gamma^2}.
        \end{equation}
        The EP condition is shifted to
        \begin{equation}
            g=\sqrt{J^2+\frac{\delta\Gamma^2}{4}}.
        \end{equation}
        At this point, the Jordan form is
        \begin{equation}
            \begin{aligned}
                \mathcal{J}_{\delta\Gamma}
                &=
                V_{\delta\Gamma}^{-1}
                M_1\Big|_{g=\sqrt{J^2+\delta\Gamma^2/4}}
                V_{\delta\Gamma}
                \\
                &=
                \begin{pmatrix}
                    -\frac{i\Gamma}{2} & 1 & 0 & 0 \\
                    0 & -\frac{i\Gamma}{2} & 0 & 0 \\
                    0 & 0 & -\frac{i\Gamma}{2} & 1 \\
                    0 & 0 & 0 & -\frac{i\Gamma}{2}
                \end{pmatrix},
            \end{aligned}
        \end{equation}
        where
        \begin{widetext}
        \begin{equation}
            V_{\delta\Gamma}
            =
            \begin{pmatrix}
                -\frac{2 i J}{\delta\Gamma} & \frac{4 J}{\delta\Gamma^2} & -\frac{i \sqrt{\delta\Gamma^2+4 J^2}}{\delta\Gamma} & \frac{2 \sqrt{\delta\Gamma^2+4 J^2}}{\delta\Gamma^2} \\
                \frac{i \sqrt{\delta\Gamma^2+4 J^2}}{\delta\Gamma} & -\frac{2 \sqrt{\delta\Gamma^2+4 J^2}}{\delta\Gamma^2} & \frac{2 i J}{\delta\Gamma} & -\frac{4 J}{\delta\Gamma^2} \\
                0 & 0 & 1 & 0 \\
                1 & 0 & 0 & 0
            \end{pmatrix}.
        \end{equation}
        Away from the EP, one may diagonalize $M_1$ using
        \begin{equation}
            V_1=
            \begin{pmatrix}
                \frac{J \left(\sqrt{-\delta\Gamma^2+4 g^2-4 J^2}+i \delta\Gamma \right)}{2 \left(J^2-g^2\right)} &
                -\frac{g \left(\sqrt{-\delta\Gamma^2+4 g^2-4 J^2}+i \delta\Gamma \right)}{2 \left(g^2-J^2\right)} &
                \frac{J \left(\sqrt{-\delta\Gamma^2+4 g^2-4 J^2}-i \delta\Gamma \right)}{2 \left(g^2-J^2\right)} &
                \frac{g \left(\sqrt{-\delta\Gamma^2+4 g^2-4 J^2}-i \delta\Gamma \right)}{2 \left(g^2-J^2\right)} \\
                \frac{g \left(\sqrt{-\delta\Gamma^2+4 g^2-4 J^2}+i \delta\Gamma \right)}{2 \left(g^2-J^2\right)} &
                \frac{J \left(\sqrt{-\delta\Gamma^2+4 g^2-4 J^2}+i \delta\Gamma \right)}{2 \left(g^2-J^2\right)} &
                -\frac{g \left(\sqrt{-\delta\Gamma^2+4 g^2-4 J^2}-i \delta\Gamma \right)}{2 \left(g^2-J^2\right)} &
                \frac{J \left(\sqrt{-\delta\Gamma^2+4 g^2-4 J^2}-i \delta\Gamma \right)}{2 \left(J^2-g^2\right)} \\
                0 & 1 & 0 & 1 \\
                1 & 0 & 1 & 0
            \end{pmatrix},
        \end{equation}
        \end{widetext}
        for $g\neq J$, and
        \begin{equation}
            V_2=
            \begin{pmatrix}
                \frac{i\delta\Gamma}{J} & -1 & -\frac{iJ}{\delta\Gamma} & -\frac{iJ}{\delta\Gamma} \\
                0 & 1 & \frac{iJ}{\delta\Gamma} & \frac{iJ}{\delta\Gamma} \\
                -1 & 0 & 0 & 1 \\
                1 & 0 & 1 & 0
            \end{pmatrix},
        \end{equation}
        for $g=J$. Thus no EP exists away from $g=\sqrt{J^2+\delta\Gamma^2/4}$.

        \subsection{Identical independent dissipation with a common bath}\label{App-B3}

        For $\Gamma_1=\Gamma_2\equiv\Gamma$ and $\kappa\neq0$, the first-order dynamical matrix is
        \begin{equation}
            M_1 =
            \begin{pmatrix}
                -\frac{i\Gamma}{2}-\frac{i\kappa}{2} & 0 & g-\frac{i\kappa}{2} & J \\
                0 & -\frac{i\Gamma}{2}-\frac{i\kappa}{2} & -J & -g-\frac{i\kappa}{2} \\
                g-\frac{i\kappa}{2} & J & -\frac{i\Gamma}{2}-\frac{i\kappa}{2} & 0 \\
                -J & -g-\frac{i\kappa}{2} & 0 & -\frac{i\Gamma}{2}-\frac{i\kappa}{2}
            \end{pmatrix}.
        \end{equation}
        Its eigenvalues are
        \begin{equation}\label{eq:eig_id_kap}
        \begin{aligned}
            \lambda_1=-\frac{i}{2}(\Gamma-s),\qquad
            \lambda_2=-\frac{i}{2}(\Gamma+2\kappa-s),\\
            \lambda_3=-\frac{i}{2}(\Gamma+s),\qquad
            \lambda_4=-\frac{i}{2}(\Gamma+2\kappa+s).
        \end{aligned}
        \end{equation}
        where $s$ is defined in Eq.~(\ref{eq:s}).
        The common bath modifies the spectrum quantitatively while leaving the dynamical boundaries discussed in the main text unchanged.

        The EP condition remains $g=J$. 
        At this point, the Jordan form becomes
        \begin{equation}
            \begin{aligned}
                \mathcal{J}_{\kappa}
                &=
                V_{\kappa}^{-1}M_1\big|_{g=J}V_{\kappa}
                \\
                &=
                \begin{pmatrix}
                    -\frac{i\Gamma}{2} & 1 & 0 & 0 \\
                    0 & -\frac{i\Gamma}{2} & 0 & 0 \\
                    0 & 0 & -\frac{i}{2}(\Gamma+2\kappa) & 1 \\
                    0 & 0 & 0 & -\frac{i}{2}(\Gamma+2\kappa)
                \end{pmatrix},
            \end{aligned}
        \end{equation}
        with
        \begin{equation}\label{eq:V_kappa}
            V_{\kappa}
            =
            \begin{pmatrix}
                1 & -\frac{1}{J} & -1 & -\frac{1}{J} \\
                -1 & 0 & 1 & 0 \\
                -1 & \frac{1}{J} & -1 & -\frac{1}{J} \\
                1 & 0 & 1 & 0
            \end{pmatrix}.
        \end{equation}
Comparison with Eq.~(\ref{eq:id_EP}) shows that the twofold degeneracy of the EP$_2$ is lifted, yielding two distinct EP$_2$s, as discussed in the main text.
        Away from the EP, $M_1$ is diagonalized by
        \begin{widetext}
        \begin{equation}\label{eq:V_kappa_p}
            V'_{\kappa}=
            \begin{pmatrix}
                \frac{\sqrt{g^2-J^2}+g}{J} & \frac{g-\sqrt{g^2-J^2}}{J} & \frac{\sqrt{g^2-J^2}-g}{J} & -\frac{\sqrt{g^2-J^2}+g}{J} \\
                -1 & -1 & 1 & 1 \\
                -\frac{\sqrt{g^2-J^2}+g}{J} & \frac{\sqrt{g^2-J^2}-g}{J} & \frac{\sqrt{g^2-J^2}-g}{J} & -\frac{\sqrt{g^2-J^2}+g}{J} \\
                1 & 1 & 1 & 1
            \end{pmatrix},
        \end{equation}
        \end{widetext}
        so that no EP exists for $g\neq J$.

For the common-bath-only limit, the spectral and EP analysis follows directly by setting $\Gamma=0$ in the above results. The eigenvalues of the first-order dynamical matrix become
\begin{equation}\label{eq:eig_kap}
\begin{aligned}
    \lambda_1=\frac{is}{2},\qquad
    \lambda_2=-\frac{i}{2}(2\kappa-s),\\
    \lambda_3=-\frac{is}{2},\qquad
    \lambda_4=-\frac{i}{2}(2\kappa+s),
\end{aligned}
\end{equation}
where $s$ is defined in Eq.~(\ref{eq:s}). The EP condition remains $g=J$. The corresponding Jordan structure and the diagonalization away from EP follow from the same $V_\kappa$ and $V'_\kappa$ given in Eqs.~(\ref{eq:V_kappa}) and~(\ref{eq:V_kappa_p}), respectively.

    \section{Analytical entanglement dynamics and asymptotic formulas}\label{App-C}

In this appendix, we collect the analytical expressions underlying the results shown in Figs.~\ref{fig2}--\ref{fig7}, including the population dynamics, second-order moments, logarithmic negativity, and finite asymptotic entanglement diagrams.

        \subsection{Identical independent dissipation}\label{App-C1}

        We start with the first main-text setting $\Gamma_1=\Gamma_2\equiv\Gamma$ and $\kappa=0$. Solving the second-order equation
        \begin{equation}
            i\frac{d}{dt}v_2=M_2v_2+D
        \end{equation}
        yields closed analytical expressions for all second-order moments. 
        With $s$ defined in Eq.~(\ref{eq:s}), the nonzero independent moments are
        \begin{widetext}
        \begin{equation}\label{eq:id_dis_v2}
        \begin{aligned}
            \langle \hat{a}_1^2\rangle(t)
            &=
            \langle \hat{a}_2^2\rangle(t)
            =
            \frac{-2gJ}{\Gamma^2-s^2}
            +
            \frac{gJ e^{-2it\lambda_{1,2}}}{s(\Gamma-s)}
            -
            \frac{gJ e^{-2it\lambda_{3,4}}}{s(\Gamma+s)}, \\
            \langle \hat{a}_1^\dagger \hat{a}_1\rangle(t)
            &=
            \langle \hat{a}_2^\dagger \hat{a}_2\rangle(t)
            =
            \frac{2J^2}{\Gamma^2-s^2}
            -
            \frac{J^2 e^{-2it\lambda_{1,2}}}{s(\Gamma-s)}
            +
            \frac{J^2 e^{-2it\lambda_{3,4}}}{s(\Gamma+s)}, \\
            \langle \hat{a}_1\hat{a}_2\rangle(t)
            &=
            \frac{-iJ\Gamma}{\Gamma^2-s^2}
            +
            \frac{iJ e^{-2it\lambda_{1,2}}}{2(\Gamma-s)}
            +
            \frac{iJ e^{-2it\lambda_{3,4}}}{2(\Gamma+s)}.
        \end{aligned}
        \end{equation}
        \end{widetext}
where the eigenvalues $\lambda_j$ are given in Eq.~(\ref{eq:eig_id}). The exponential time dependence of the second-order moments is therefore directly inherited from the dynamical spectrum. In particular, the signs of the imaginary parts of the eigenvalues determine whether the moments exhibit amplification or relax to a stationary value, while the real parts determine whether this approach is monotonic or oscillatory.

Using the CM convention and entanglement characterization introduced in Appendix~\ref{App-A2}, the logarithmic negativity is determined by the minimal symplectic eigenvalue $\nu_-(t)$ given in Eq.~(\ref{eq:numin}). For the symmetric-dissipation setting considered here, the corresponding quantities $\xi(t)$ and $\det V(t)$ take the explicit forms
        \begin{widetext}
        \begin{align}
            \xi(t)
            &=
            \frac{e^{-2 t (s + \Gamma)}}{s (s - \Gamma)^{2} (s + \Gamma)^{2}}
            \Big[
            e^{2 t (s + \Gamma)} s (s^{2} - \Gamma^{2})^{2}
            \notag\\
            &\qquad\qquad
            + 2 J^{2}
            \big(
            (1 + e^{4 s t} - 2 e^{2 t (s + \Gamma)}) s^{3}
            \notag\\
            &\qquad\qquad\quad
            + 2 (-1 + e^{4 s t} + e^{t (s + \Gamma)} - e^{t (3 s + \Gamma)}) s^{2} \Gamma
            \notag\\
            &\qquad\qquad\quad
            + (1 + e^{4 s t} - 4 e^{t (s + \Gamma)} + 6 e^{2 t (s + \Gamma)} - 4 e^{t (3 s + \Gamma)}) s \Gamma^{2}
            \notag\\
            &\qquad\qquad\quad
            - 2 e^{t (s + \Gamma)} (-1 + e^{2 s t}) \Gamma^{3}
            \big)
            \Big],
        \end{align}
        and
        \begin{align}
            \det V(t)
            &=
            \frac{e^{-4 t (s + \Gamma)}}{s^{2} (s - \Gamma)^{2} (s + \Gamma)^{2}}
            \Big(
            16 e^{4 s t} J^{4} s^{2}
            - 32 e^{t (3 s + \Gamma)} J^{4} s \Gamma
            \notag\\
            &\qquad\qquad
            + 32 e^{t (5 s + \Gamma)} J^{4} s \Gamma
            + 16 e^{2 t (s + \Gamma)} J^{4} \Gamma^{2}
            + 16 e^{2 t (3 s + \Gamma)} J^{4} \Gamma^{2}
            \notag\\
            &\qquad\qquad
            - 8 e^{3 t (s + \Gamma)} J^{2} s \Gamma \bigl(-4 J^{2} + s^{2} - \Gamma^{2}\bigr)
            + 8 e^{5 s t + 3 t \Gamma} J^{2} s \Gamma \bigl(-4 J^{2} + s^{2} - \Gamma^{2}\bigr)
            \notag\\
            &\qquad\qquad
            + e^{4 t (s + \Gamma)} s^{2} \bigl(4 J^{2} - s^{2} + \Gamma^{2}\bigr)^{2}
            \notag\\
            &\qquad\qquad
            - 8 e^{2 t (2 s + \Gamma)} J^{2} \bigl(-s^{4} + s^{2} \Gamma^{2} + 4 J^{2} (s^{2} + \Gamma^{2})\bigr)
            \Big).
        \end{align}
        \end{widetext}

\subsubsection{Asymptotic analysis in the three dynamical regimes}

For identical independent dissipation in the absence of a common bath, we denote the equal mode populations by
        \begin{equation}
            \begin{aligned}
            n(t)
            &=
            \langle \hat{a}_1^\dagger \hat{a}_1\rangle(t)
            =
            \langle \hat{a}_2^\dagger \hat{a}_2\rangle(t)
            \\
            &=
            \frac{2J^2}{\Gamma^2-s^2}
            -
            \frac{J^2 e^{-t(\Gamma-s)}}{s(\Gamma-s)}
            +
            \frac{J^2 e^{-t(\Gamma+s)}}{s(\Gamma+s)},
            \end{aligned}
            \label{eqC1}
        \end{equation}
The population dynamics in the three dynamical regimes and at their boundaries can then be characterized as follows:
        \begin{widetext}
        \begin{equation}
        \begin{aligned}
            \text{Region I:}\qquad
            n(t) &\overset{t\gg 0}{\approx}
            \frac{J^2 e^{-t(\Gamma-s)}}{-s(\Gamma-s)}, \\
            \text{I--II boundary }(\Gamma=s):\qquad
            n(t) &=
            \frac{J^2}{2\Gamma^2}
            \left(e^{-2\Gamma t}+2\Gamma t-1\right), \\
            \text{Region II:}\qquad
            n(t) &\overset{t\gg 0}{\approx}
            \frac{2J^2}{\Gamma^2-s^2}, \\
            \text{II--III boundary }(s=0):\qquad
            n(t) &=
            \frac{2J^2}{\Gamma^2}\left[1-e^{-\Gamma t}(\Gamma t+1)\right], \\
            \text{Region III:}\qquad
            n(t) &=
            \frac{2 J^2}{\Gamma^2 + \omega^2}
            -
            \frac{2J^2 e^{-t\Gamma}}{\omega \sqrt{\Gamma^2 + \omega^2}}
            \sin(\omega t+\phi),
            \qquad
            \tan\phi=\frac{\omega}{\Gamma}.
        \end{aligned}
        \end{equation}
        \end{widetext}
        where $\omega=-is$ in Region III. 
These expressions explicitly capture the unbounded population growth in Region I, the monotonic approach to a stationary value in Region II, and the oscillatory approach to a stationary value in Region III.

Next, we analyze the asymptotic entanglement through $\xi(t)$ and $\det V(t)$. In Region I,
        \begin{align}
            \xi(t) &\overset{t\gg 0}{\approx}
            \frac{2J^2 e^{2t(s-\Gamma)}}{(\Gamma-s)^2}, \\
            \det V(t) &\overset{t\gg 0}{\approx}
            \frac{16\Gamma^2J^4 e^{2t(s-\Gamma)}}{s^2(\Gamma^2-s^2)^2},
        \end{align}
Although both quantities grow exponentially, the common leading exponential factors cancel in evaluating $\nu_-(t)$, yielding
        \begin{equation}\label{eq:nu_inf_id_R1}
            \nu_-^\infty \simeq \sqrt{\frac{\det V(t)}{2\,\xi(t)}}\bigg|_{t\to\infty}
            = \frac{2J\Gamma}{s(s+\Gamma)} ,
        \end{equation}
        and therefore
        \begin{equation}
            E_N^\infty = \max\{0,-\log \left(\frac{2J\Gamma}{s(s+\Gamma)}\right)\}.
            \label{eqC4}
        \end{equation}
At the I--II boundary $\Gamma=s$, one finds
        \begin{widetext}
        \begin{align}
            \xi(t)
            &=
            1 + \frac{e^{-4 t \Gamma} J^{2} \bigl(1 + 2 e^{2 t \Gamma} + e^{4 t \Gamma} \bigl(-3 + 8 t \Gamma + 4 t^{2} \Gamma^{2}\bigr)\bigr)}{2 \Gamma^{2}}, \\
            \det V(t)
            &=
            \frac{e^{-4 t \Gamma} \bigl(2 J^{2} (\Gamma t + 1) + e^{2 t \Gamma} \bigl(2 J^{2} (\Gamma t - 1) + \Gamma^{2}\bigr)\bigr)^{2}}{\Gamma^{4}},
        \end{align}
        \end{widetext}
        and therefore
        \begin{equation}
            \nu_-^\infty \simeq \sqrt{\frac{\det V(t)}{2\,\xi(t)}}\bigg|_{t\to\infty} = \frac{J}{\Gamma},
        \end{equation}
which continuously matches the Region-I result in Eq.~(\ref{eq:nu_inf_id_R1}) at $\Gamma=s$.
        In Region II,
        \begin{align}
            \xi(t) &\overset{t\gg 0}{\approx}
            1 - \frac{4 J^{2} \bigl(s^{2} - 3 \Gamma^{2}\bigr)}{\bigl(s^{2} - \Gamma^{2}\bigr)^{2}}, \\
            \det V(t) &\overset{t\gg 0}{\approx}
            \frac{\bigl(4 J^{2} - s^{2} + \Gamma^{2}\bigr)^{2}}{\bigl(s^{2} - \Gamma^{2}\bigr)^{2}},
        \end{align}
        which yields
        \begin{widetext}
        \begin{equation}\label{eq:nu_inf_id_R2}
            \nu_-^\infty
            =
            \frac{
            \sqrt{
            -4 J^{2} \bigl(s^{2} - 3 \Gamma^{2}\bigr)
            + \bigl(s^{2} - \Gamma^{2}\bigr)^{2}
            - 4 J \Gamma \sqrt{-4 J^{2} \bigl(s^{2} - 2 \Gamma^{2}\bigr) + \bigl(s^{2} - \Gamma^{2}\bigr)^{2}}
            }
            }{-s^{2} + \Gamma^{2}}.
        \end{equation}
At the II--III boundary $s=0$, one obtains
        \begin{align}
            \xi(t)
            &=
            1 + \frac{4 e^{-2 t \Gamma} J^{2} \bigl(1 + 3 e^{2 t \Gamma} - 2 e^{t \Gamma} (2 + t \Gamma)\bigr)}{\Gamma^{2}}, \\
            \det V(t)
            &=
            \frac{e^{-4 t \Gamma} \bigl(e^{2 t \Gamma} \Gamma^{2} + 4 J^{2} \bigl(-1 + e^{2 t \Gamma} - 2 e^{t \Gamma} t \Gamma\bigr)\bigr)^{2}}{\Gamma^{4}},
        \end{align}
        and
        \begin{equation}
            \nu_-^\infty
            =
            \frac{\sqrt{12 J^{2} + \Gamma^{2} - 4 J \sqrt{8 J^{2} + \Gamma^{2}}}}{\Gamma}.
        \end{equation}
In Region III, substituting $s=i\omega$ into the general expressions gives
        \begin{align}
            \xi(t)
            &=
            \frac{e^{-2 t \Gamma}}{\omega \bigl(\Gamma^{2} + \omega^{2}\bigr)^{2}}
            \Big(
            e^{2 t \Gamma} \omega \bigl(12 J^{2} \Gamma^{2} + \Gamma^{4} + 4 J^{2} \omega^{2} + 2 \Gamma^{2} \omega^{2} + \omega^{4}\bigr)
            \notag\\
            &\qquad\qquad
            - 4 J^{2} \omega \bigl(-\Gamma^{2} \cos(2 t \omega) + \omega^{2} \cos(2 t \omega) + 2 \Gamma \omega \sin(2 t \omega)\bigr)
            \notag\\
            &\qquad\qquad
            - 8 J^{2} \Gamma e^{t \Gamma} \bigl(2 \Gamma \omega \cos(t \omega) + \Gamma^{2} \sin(t \omega) - \omega^{2} \sin(t \omega)\bigr)
            \Big), \\
            \det V(t)
            &=
            \frac{e^{-4 t \Gamma} \bigl(4 \bigl(-1 + e^{2 t \Gamma}\bigr) J^{2} \omega + e^{2 t \Gamma} \omega \bigl(\Gamma^{2} + \omega^{2}\bigr) - 8 e^{t \Gamma} J^{2} \Gamma \sin(t \omega)\bigr)^{2}}{\omega^{2} \bigl(\Gamma^{2} + \omega^{2}\bigr)^{2}}.
        \end{align}
The appearance of the trigonometric terms reflects the oscillatory dynamics characteristic of Region III, consistent with the oscillatory approach toward the stationary state shown in the main text.
        In the long-time limit,
        \begin{align}
            \xi(t) &\overset{t\gg 0}{\approx}
            1 + \frac{4 J^{2} \bigl(3 \Gamma^{2} + \omega^{2}\bigr)}{\bigl(\Gamma^{2} + \omega^{2}\bigr)^{2}}, \\
            \det V(t) &\overset{t\gg 0}{\approx}
            \frac{\bigl(4 J^{2} + \Gamma^{2} + \omega^{2}\bigr)^{2}}{\bigl(\Gamma^{2} + \omega^{2}\bigr)^{2}},
        \end{align}
        and hence
        \begin{equation}\label{eq:nu_inf_id_R3}
            \nu_-^\infty
            =
            \frac{
            \sqrt{
            \bigl(\Gamma^{2} + \omega^{2}\bigr)^{2}
            + 4 J^{2} \bigl(3 \Gamma^{2} + \omega^{2}\bigr)
            - 4 J \Gamma \sqrt{\bigl(\Gamma^{2} + \omega^{2}\bigr)^{2} + 4 J^{2} \bigl(2 \Gamma^{2} + \omega^{2}\bigr)}
            }
            }{\Gamma^{2} + \omega^{2}},
        \end{equation}
        \end{widetext}
The asymptotic expressions in Regions II and III share the same analytical structure under the substitution $s=i\omega$. Despite the qualitatively different transient entanglement dynamics in the two stable regions, the asymptotic entanglement admits a unified analytical description.

These asymptotic results also allow the condition for nonzero asymptotic entanglement, $\nu_-^\infty<1$, to be obtained analytically. In Region I, Eq.~(\ref{eq:nu_inf_id_R1}) gives
\begin{equation}
    2J\Gamma-s(s+\Gamma)<0.
\end{equation}
In Regions II and III, Eq.~(\ref{eq:nu_inf_id_R2}) and Eq.~(\ref{eq:nu_inf_id_R3}) give the same condition
\begin{equation}
    J-\Gamma<0.
\end{equation}
The corresponding equalities determine the boundary between vanishing and nonzero asymptotic entanglement shown by the yellow dashed lines in Fig.~\ref{fig3}(a).

        \subsection{Asymmetric independent dissipation}\label{App-C2}
We now consider asymmetric independent dissipation, $\Gamma_1\neq\Gamma_2$, with $\kappa=0$. Using the notation introduced in Appendix~\ref{App-B2}, the nonzero independent second-order moments are
\begin{widetext}
        \begin{equation}
        \begin{aligned}
            \langle \hat{a}_1^2\rangle(t)
            &=
            \frac{-2gJ}{\Gamma^2-\tilde{s}^2}\left(1-\frac{\delta\Gamma}{\Gamma}\right)
            +
            \frac{gJ e^{-2it\lambda_{1,2}}}{\tilde{s}(\Gamma-\tilde{s})}\left(1-\frac{\delta\Gamma}{\tilde{s}}\right)
            -
            \frac{gJ e^{-2it\lambda_{3,4}}}{\tilde{s}(\Gamma+\tilde{s})}\left(1+\frac{\delta\Gamma}{\tilde{s}}\right)
            +
            \frac{2gJ\delta\Gamma}{\tilde{s}^2\Gamma}e^{-\Gamma t},
            \\
            \langle \hat{a}_2^2\rangle(t)
            &=
            \frac{-2gJ}{\Gamma^2-\tilde{s}^2}\left(1+\frac{\delta\Gamma}{\Gamma}\right)
            +
            \frac{gJ e^{-2it\lambda_{1,2}}}{\tilde{s}(\Gamma-\tilde{s})}\left(1+\frac{\delta\Gamma}{\tilde{s}}\right)
            -
            \frac{gJ e^{-2it\lambda_{3,4}}}{\tilde{s}(\Gamma+\tilde{s})}\left(1-\frac{\delta\Gamma}{\tilde{s}}\right)
            -
            \frac{2gJ\delta\Gamma}{\tilde{s}^2\Gamma}e^{-\Gamma t}, \\
            \langle \hat{a}_1^\dagger \hat{a}_1\rangle(t)
            &=
            \frac{2J^2}{\Gamma^2-\tilde{s}^2}\left(1-\frac{\delta\Gamma}{\Gamma}\right)
            -
            \frac{J^2 e^{-2it\lambda_{1,2}}}{\tilde{s}(\Gamma-\tilde{s})}\left(1-\frac{\delta\Gamma}{\tilde{s}}\right)
            +
            \frac{J^2 e^{-2it\lambda_{3,4}}}{\tilde{s}(\Gamma+\tilde{s})}\left(1+\frac{\delta\Gamma}{\tilde{s}}\right)
            -
            \frac{2J^2\delta\Gamma}{\tilde{s}^2\Gamma}e^{-\Gamma t},
            \\
            \langle \hat{a}_2^\dagger \hat{a}_2\rangle(t)
            &=
            \frac{2J^2}{\Gamma^2-\tilde{s}^2}\left(1+\frac{\delta\Gamma}{\Gamma}\right)
            -
            \frac{J^2 e^{-2it\lambda_{1,2}}}{\tilde{s}(\Gamma-\tilde{s})}\left(1+\frac{\delta\Gamma}{\tilde{s}}\right)
            +
            \frac{J^2 e^{-2it\lambda_{3,4}}}{\tilde{s}(\Gamma+\tilde{s})}\left(1-\frac{\delta\Gamma}{\tilde{s}}\right)
            +
            \frac{2J^2\delta\Gamma}{\tilde{s}^2\Gamma}e^{-\Gamma t}, \\
            \langle \hat{a}_1\hat{a}_2\rangle(t)
            &=
            \frac{-iJ\Gamma}{\Gamma^2-\tilde{s}^2}\left(1-\frac{\delta\Gamma^2}{\Gamma^2}\right)
            +
            \frac{iJ e^{-2it\lambda_{1,2}}}{2(\Gamma-\tilde{s})}\left(1-\frac{\delta\Gamma^2}{\tilde{s}^2}\right)
            +
            \frac{iJ e^{-2it\lambda_{3,4}}}{2(\Gamma+\tilde{s})}\left(1-\frac{\delta\Gamma^2}{\tilde{s}^2}\right)
            +
            \frac{iJ\delta\Gamma^2}{\tilde{s}^2\Gamma}e^{-\Gamma t}.
        \end{aligned}
        \end{equation}
        \end{widetext}
where the eigenvalues $\lambda_j$ are given in Eq.~(\ref{eq:eig_asym}).
As in the symmetric-dissipation case, the exponential time dependence of the second-order moments is governed by the dynamical spectrum. 
For $\delta\Gamma=0$, one has $\tilde{s}=s$, and the above expressions reduce to the symmetric-dissipation results in Eq.~(\ref{eq:id_dis_v2}).

From these second-order moments, the corresponding CM and entanglement dynamics can be obtained following Appendix~\ref{App-A2}. For the asymmetric-dissipation setting considered here, the quantities $\xi(t)$ and $\det V(t)$ entering the minimal symplectic eigenvalue $\nu_-(t)$ can be organized as finite sums of exponential terms,
\begin{equation}
    \xi(t) = \sum_{n=1}^{7} \tilde{C}_n e^{\tilde{a}_n t},
    \qquad
    \det V(t) = \sum_{n=1}^{13} \tilde{B}_n e^{\tilde{d}_n t}.
\end{equation}
The corresponding exponents and coefficients are listed below.
\begin{widetext}
\begingroup
\renewcommand{\arraystretch}{2.0}
\setlength{\extrarowheight}{2pt}
\begin{equation*}
\begin{array}{c c c}
\hline
n & \tilde{a}_n & \tilde{C}_n \\
\hline
1 &
2(\tilde{s}-\Gamma) &
\dfrac{2J^2\left(\tilde{s}^2-\delta\Gamma^2\right)}
      {\tilde{s}^2(\tilde{s}-\Gamma)^2}
\\
2 &
\tilde{s}-\Gamma &
-\dfrac{4J^2\left(\Gamma \tilde{s}-\delta\Gamma^2\right)}
       {\tilde{s}^2(\tilde{s}-\Gamma)^2}
\\
3 &
0 &
\dfrac{
\left(\Gamma^3-\Gamma \tilde{s}^2\right)^2
-4J^2\left(\Gamma^2-\delta\Gamma^2\right)
\left(\tilde{s}^2-3\Gamma^2\right)}
{\Gamma^2(\tilde{s}-\Gamma)^2(\Gamma+\tilde{s})^2}
\\
4 &
-\Gamma &
-\dfrac{8\delta\Gamma^2J^2}
       {\Gamma^2 \tilde{s}^2}
\\
5 &
-2\Gamma &
\dfrac{4\delta\Gamma^2J^2}
      {\Gamma^2 \tilde{s}^2}
\\
6 &
-\Gamma-\tilde{s} &
\dfrac{4J^2\left(\delta\Gamma^2+\Gamma \tilde{s}\right)}
      {\tilde{s}^2(\Gamma+\tilde{s})^2}
\\
7 &
-2(\Gamma+\tilde{s}) &
\dfrac{2J^2\left(\tilde{s}^2-\delta\Gamma^2\right)}
      {\tilde{s}^2(\Gamma+\tilde{s})^2}
\\
\hline
\end{array}
\end{equation*}
\begin{equation*}
\begin{array}{c c c}
\hline
n & \tilde{d}_n & \tilde{B}_n \\
\hline
1 & 2(\tilde{s}-\Gamma) &
\dfrac{4J^2\left(\tilde{s}^2-\delta\Gamma^2\right)\left(4J^2\left(\Gamma^2-\delta\Gamma^2\right)+\delta\Gamma^2(\Gamma+\tilde{s})^2\right)}
{\tilde{s}^4(\tilde{s}-\Gamma)^2(\Gamma+\tilde{s})^2}
\\
2 & \tilde{s}-\Gamma &
\begin{array}{c}
\dfrac{8J^2}{\Gamma^2 \tilde{s}^2 (\tilde{s}-\Gamma)^2 (\Gamma+\tilde{s})^2}
\left[
4J^2\left(\Gamma^2-\delta\Gamma^2\right)\left(\delta\Gamma^2-\Gamma \tilde{s}\right)
\right.
\\[2pt]
\left.
-\Gamma\left(
\Gamma\delta\Gamma^2(\Gamma^2-2\delta\Gamma^2)
-\tilde{s}^3(\Gamma^2-2\delta\Gamma^2)
+\Gamma\delta\Gamma^2 \tilde{s}^2
+\tilde{s}(\Gamma^4-2\delta\Gamma^4)
\right)
\right]
\end{array}
\\
3 & \tilde{s}-2\Gamma &
\dfrac{16\delta\Gamma^2J^2\left(
-\Gamma^2\delta\Gamma^2
-4J^2(\Gamma^2-\delta\Gamma^2)
+\Gamma \tilde{s}^3+\Gamma^2 \tilde{s}^2-\Gamma\delta\Gamma^2 \tilde{s}
\right)}
{\Gamma^2 \tilde{s}^4 (\tilde{s}-\Gamma)(\Gamma+\tilde{s})}
\\
4 & \tilde{s}-3\Gamma &
\dfrac{32J^4(\Gamma^2-\delta\Gamma^2)(\delta\Gamma^2+\Gamma \tilde{s})}
{\Gamma^2 \tilde{s}^2 (\tilde{s}-\Gamma)^2 (\Gamma+\tilde{s})^2}
\\
5 & 0 &
\dfrac{
16J^4(\Gamma^2-\delta\Gamma^2)^2
+8\Gamma^2J^2(\Gamma^2-\delta\Gamma^2)(\Gamma^2+2\delta\Gamma^2-\tilde{s}^2)
+\Gamma^4(\tilde{s}^2-\Gamma^2)^2}
{\Gamma^4 (\tilde{s}-\Gamma)^2 (\Gamma+\tilde{s})^2}
\\
6 & -\Gamma &
\dfrac{16\delta\Gamma^2J^2\left(
\Gamma^2(\Gamma^2-2\delta\Gamma^2+\tilde{s}^2)
-4J^2(\Gamma^2-\delta\Gamma^2)\right)}
{\Gamma^4 \tilde{s}^2 (\Gamma-\tilde{s})(\Gamma+\tilde{s})}
\\
7 & -2\Gamma &
\begin{array}{c}
\dfrac{8J^2}{\Gamma^4 \tilde{s}^4 (\tilde{s}-\Gamma)^2 (\Gamma+\tilde{s})^2}
\bigl[
\Gamma^2\bigl(
-3\Gamma^4\delta\Gamma^4
+\tilde{s}^6(\Gamma^2+\delta\Gamma^2)
-\tilde{s}^4(\Gamma^4+4\Gamma^2\delta\Gamma^2+2\delta\Gamma^4)
\bigr)
\\[2pt]
\Gamma^2 \tilde{s}^2(3\Gamma^4\delta\Gamma^2+5\Gamma^2\delta\Gamma^4)
-4J^2(\Gamma^2-\delta\Gamma^2)\bigl(
3\Gamma^4\delta\Gamma^2
+\tilde{s}^4(\Gamma^2+3\delta\Gamma^2)
\bigr)
\\[2pt]
-4J^2(\Gamma^2-\delta\Gamma^2)\bigl(
+\tilde{s}^2(\Gamma^4-4\Gamma^2\delta\Gamma^2)
\bigr)
\bigr]
\end{array}
\\
8 & -3\Gamma &
\dfrac{64\delta\Gamma^2J^4(\Gamma^2-\delta\Gamma^2)}
{\Gamma^4 \tilde{s}^2 (\tilde{s}-\Gamma)(\Gamma+\tilde{s})}
\\
9 & -4\Gamma &
\dfrac{16J^4(\Gamma^2-\delta\Gamma^2)^2}
{\Gamma^4 (\tilde{s}-\Gamma)^2 (\Gamma+\tilde{s})^2}
\\
10 & -\Gamma-\tilde{s} &
\begin{array}{c}
\dfrac{8\Gamma J^2}{\Gamma^2 \tilde{s}^2 (\tilde{s}-\Gamma)^2 (\Gamma+\tilde{s})^2}
\bigl[
4J^2(\Gamma^2-\delta\Gamma^2)(\delta\Gamma^2+\Gamma \tilde{s})
-\Gamma^2\delta\Gamma^2(\Gamma^2-2\delta\Gamma^2)
\\[2pt]
-\Gamma \tilde{s}^3(\Gamma^2-2\delta\Gamma^2)
-\Gamma^2\delta\Gamma^2 \tilde{s}^2
+\Gamma \tilde{s}(\Gamma^4-2\delta\Gamma^4)
\bigr]
\end{array}
\\
11 & -2\Gamma-\tilde{s} &
\dfrac{16\delta\Gamma^2J^2\left(
-\Gamma^2\delta\Gamma^2
-4J^2(\Gamma^2-\delta\Gamma^2)
-\Gamma \tilde{s}^3+\Gamma^2 \tilde{s}^2+\Gamma\delta\Gamma^2 \tilde{s}
\right)}
{\Gamma^2 \tilde{s}^4 (\tilde{s}-\Gamma)(\Gamma+\tilde{s})}
\\
12 & -3\Gamma-\tilde{s} &
-\dfrac{32J^4(\Gamma^2-\delta\Gamma^2)(\Gamma \tilde{s}-\delta\Gamma^2)}
{\Gamma^2 \tilde{s}^2 (\tilde{s}-\Gamma)^2 (\Gamma+\tilde{s})^2}
\\
13 & -2(\Gamma+\tilde{s}) &
\dfrac{4J^2\left(\tilde{s}^2-\delta\Gamma^2\right)\left(
4J^2(\Gamma^2-\delta\Gamma^2)
+\delta\Gamma^2(\tilde{s}-\Gamma)^2
\right)}
{\tilde{s}^4 (\tilde{s}-\Gamma)^2 (\Gamma+\tilde{s})^2}
\\
\hline
\end{array}
\end{equation*}
\endgroup
\end{widetext}

        \subsubsection{Asymptotic analysis in the three dynamical regimes}
We first analyze the population dynamics. We denote the two mode populations by
\begin{equation}
    n_j(t)
    =
    \langle \hat{a}_j^\dagger \hat{a}_j\rangle(t),
    \qquad j=1,2.
\end{equation}
Using the exact second-order moments given above, the population dynamics in the three dynamical regimes and at their boundaries can be characterized as follows:
\begin{widetext}
\begin{equation}
\begin{aligned}
    \text{Region I:}\qquad
    n_1(t)
    &\overset{t\gg 0}{\approx}
    \frac{J^2 e^{-t(\Gamma-\tilde{s})}}{-\tilde{s} (\Gamma - \tilde{s})}\left(1-\frac{\delta\Gamma}{\tilde{s}}\right),
    \\
    n_2(t)
    &\overset{t\gg 0}{\approx}
    \frac{J^2 e^{-t(\Gamma-\tilde{s})}}{-\tilde{s} (\Gamma - \tilde{s})}\left(1+\frac{\delta\Gamma}{\tilde{s}}\right),
    \\
    \text{I--II boundary }(\Gamma=\tilde{s}):
    n_1(t)
    &=
    \frac{J^2}{2\Gamma^3}\left[ \big(e^{-2\Gamma t} + 2\Gamma t - 1\big)\Gamma + \big(e^{-2\Gamma t} - 4e^{-\Gamma t} - 2\Gamma t + 3\big)\delta\Gamma \right],
    \\
    n_2(t)
    &=
    \frac{J^2}{2\Gamma^3}\left[ \big(e^{-2\Gamma t} + 2\Gamma t - 1\big)\Gamma - \big(e^{-2\Gamma t} - 4e^{-\Gamma t} - 2\Gamma t + 3\big)\delta\Gamma \right],
    \\
    \text{Region II:}\qquad
    n_1(t)
    &\overset{t\gg 0}{\approx}
    \frac{2 J^2}{\Gamma^2 - \tilde{s}^2}\left(1-\frac{\delta\Gamma}{\Gamma}\right),
    \\
    n_2(t)
    &\overset{t\gg 0}{\approx}
    \frac{2 J^2}{\Gamma^2 - \tilde{s}^2}\left(1+\frac{\delta\Gamma}{\Gamma}\right),
    \\
    \text{II--III boundary }(\tilde{s}=0):
    n_1(t)
    &=
    \frac{J^2}{\Gamma^3}\left[ 2\Gamma\bigl(1 - e^{-\Gamma t}(\Gamma t + 1)\bigr) + \bigl(e^{-\Gamma t}(\Gamma^2 t^2 + 2\Gamma t + 2) - 2\bigr)\delta\Gamma \right],
    \\
    n_2(t)
    &=
    \frac{J^2}{\Gamma^3}\left[ 2\Gamma\bigl(1 - e^{-\Gamma t}(\Gamma t + 1)\bigr) - \bigl(e^{-\Gamma t}(\Gamma^2 t^2 + 2\Gamma t + 2) - 2\bigr)\delta\Gamma \right],
    \\
    \text{Region III:}\qquad
    n_1(t)
    &=
    \frac{2 J^2}{\Gamma^2 + \tilde{\omega}^2}\left(1-\frac{\delta\Gamma}{\Gamma}\right)
    - \frac{2J^2 e^{-t\Gamma}}{\tilde{\omega} \sqrt{\Gamma^2 + \tilde{\omega}^2}}
    \sin(\tilde{\omega} t+\phi)
    \\
    &\quad
    - \frac{2J^2 \delta\Gamma  e^{-\Gamma t}}{\tilde{\omega}^2 \sqrt{\Gamma^2 + \tilde{\omega}^2}}
    \cos(\tilde{\omega} t+\phi)
    +\frac{2J^2\delta\Gamma}{\tilde{\omega}^2\Gamma}e^{-\Gamma t},
    \\
    n_2(t)
    &=
    \frac{2 J^2}{\Gamma^2 + \tilde{\omega}^2}\left(1+\frac{\delta\Gamma}{\Gamma}\right)
    - \frac{2J^2 e^{-t\Gamma}}{\tilde{\omega} \sqrt{\Gamma^2 + \tilde{\omega}^2}}
    \sin(\tilde{\omega} t+\phi)
    \\
    &\quad
    + \frac{2J^2 \delta\Gamma  e^{-\Gamma t}}{\tilde{\omega}^2 \sqrt{\Gamma^2 + \tilde{\omega}^2}}
    \cos(\tilde{\omega} t+\phi)
    -\frac{2J^2\delta\Gamma}{\tilde{\omega}^2\Gamma}e^{-\Gamma t},
\end{aligned}
\end{equation}
\end{widetext}
where $\tilde{\omega}=-i\tilde{s}=\sqrt{4g^2-4J^2-\delta\Gamma^2}$ in Region III and $\tan\phi=\tilde{\omega}/\Gamma$. These expressions retain the characteristic population dynamics of the three regions found for identical dissipation, while the dissipation asymmetry leads to unequal populations of the two modes and shifts the corresponding dynamical boundaries.

We next analyze the asymptotic entanglement through $\xi(t)$ and $\det V(t)$. In Region I,
\begin{widetext}
\begin{equation}
    \xi(t)\overset{t\gg 0}{\approx}
    \frac{2J^2\left(\tilde{s}^2-\delta\Gamma^2\right)e^{2t(\tilde{s}-\Gamma)}}
    {\tilde{s}^2(\tilde{s}-\Gamma)^2},
\end{equation}
\begin{equation}
    \det V(t)\overset{t\gg 0}{\approx}
    \frac{4J^2\left(\tilde{s}^2-\delta\Gamma^2\right)e^{2t(\tilde{s}-\Gamma)}
    \left(4J^2\left(\Gamma^2-\delta\Gamma^2\right)+\delta\Gamma^2(\Gamma+\tilde{s})^2\right)}
    {\tilde{s}^4(\tilde{s}-\Gamma)^2(\Gamma+\tilde{s})^2}.
\end{equation}
As in the symmetric-dissipation case, both quantities grow exponentially, while their leading divergent contributions cancel in evaluating $\nu_-(t)$, yielding
        \begin{equation}\label{eq:nu_inf_asym_R1}
            \nu_-^\infty \simeq \sqrt{\frac{\det V(t)}{2\,\xi(t)}}\bigg|_{t\to\infty}
            =
            \frac{\sqrt{\delta \Gamma ^2 (\Gamma +\tilde{s})^2+4 J^2 \left(\Gamma ^2-\delta \Gamma ^2\right)}}{\tilde{s}(\Gamma +\tilde{s})} .
        \end{equation}
\end{widetext}
At the I--II boundary $\Gamma=\tilde{s}$, one finds
        \begin{widetext}
            \begin{equation}
    \xi(t)=
    1+\frac{J^2}{2\Gamma^4}
    \left(7\delta\Gamma^2+4\Gamma^4t^2-\Gamma^2\left(4\delta\Gamma^2t^2+3\right)+8\Gamma^3t\right)
    -\frac{8\delta\Gamma^2J^2}{\Gamma^4}e^{-\Gamma t}
    +\frac{J^2\left(\Gamma^2-\delta\Gamma^2\right)}{2\Gamma^4}e^{-4\Gamma t}
    +\frac{J^2\left(\Gamma^2+5\delta\Gamma^2\right)}{\Gamma^4}e^{-2\Gamma t}.
\end{equation}
\begin{align}
    \det V(t)&=
    \frac{4J^4\left(\Gamma^2-\delta\Gamma^2\right)
    \left(-4\delta\Gamma^2+\Gamma^4t^2+\Gamma^2\left(1-\delta\Gamma^2t^2\right)-2\Gamma^3t+4\Gamma\delta\Gamma^2t\right)}{\Gamma^8}
    \notag\\
    &\quad
    +\frac{J^2\left(\Gamma^2-\delta\Gamma^2\right)e^{-4\Gamma t}
    \left(\Gamma^2\delta\Gamma^2+4J^2\left(-4\delta\Gamma^2+\Gamma^4t^2+\Gamma^2\left(1-\delta\Gamma^2t^2\right)+2\Gamma^3t-4\Gamma\delta\Gamma^2t\right)\right)}{\Gamma^8}
    \notag\\
    &\quad
    +\frac{J^2\left(-9\delta\Gamma^4+4\Gamma^4\left(\delta\Gamma^2t^2-1\right)+\Gamma^2\left(\delta\Gamma^2-4\delta\Gamma^4t^2\right)+4\Gamma^5t-8\Gamma^3\delta\Gamma^2t+12\Gamma\delta\Gamma^4t\right)}{\Gamma^6}
    \notag\\
    &\quad
    +\frac{8\delta\Gamma^2J^2\left(\Gamma^2-\delta\Gamma^2\right)e^{-3\Gamma t}}{\Gamma^8}
    \left[4J^2(\Gamma t+2)-\Gamma^2\right]
    \notag\\
    &\quad
    -\frac{8\delta\Gamma^2J^2e^{-\Gamma t}}{\Gamma^8}
    \left[4J^2\left(\Gamma^2-\delta\Gamma^2\right)(\Gamma t-2)+\Gamma^2\left(\Gamma^2-3\delta\Gamma^2-2\Gamma^3t+2\Gamma\delta\Gamma^2t\right)\right]
    \notag\\
    &\quad
    +\frac{e^{-2\Gamma t}}{\Gamma^8}
    \Bigl[
    8J^4\left(\Gamma^2-\delta\Gamma^2\right)\left(-12\delta\Gamma^2+\Gamma^4t^2+\Gamma^2\left(\delta\Gamma^2t^2-1\right)\right)
    \notag\\
    &\hspace{3.5em}
    +2\Gamma^2J^2\left(2\Gamma^4+7\Gamma^2\delta\Gamma^2-11\delta\Gamma^4+2\Gamma^5t-4\Gamma^3\delta\Gamma^2t+2\Gamma\delta\Gamma^4t\right)
    \Bigr]
    +1.
\end{align}
        \end{widetext}
In the long-time limit, the minimal symplectic eigenvalue reduces to
        \begin{equation}
            \nu_-^\infty \simeq \sqrt{\frac{\det V(t)}{2\,\xi(t)}}\bigg|_{t\to\infty}
            =
            \frac{\sqrt{\Gamma ^2 \delta \Gamma ^2+J^2 \left(\Gamma ^2-\delta \Gamma ^2\right)}}{\Gamma ^2}.
        \end{equation}
This result continuously matches the Region-I expression in Eq.~(\ref{eq:nu_inf_asym_R1}) at $\Gamma=\tilde{s}$. 
In Region II one obtains
        \begin{widetext}
        \begin{equation}
    \xi(t) \overset{t\gg 0}{\approx}
    \frac{\left(\Gamma ^3-\Gamma  \tilde{s}^2\right)^2-4 J^2 \left(\Gamma ^2-\delta \Gamma ^2\right) \left(\tilde{s}^2-3 \Gamma ^2\right)}{\Gamma ^2 (\tilde{s}-\Gamma )^2 (\Gamma +\tilde{s})^2},
    \\
\end{equation}
\begin{equation}
    \det V(t) \overset{t\gg 0}{\approx}
    \frac{
    16 J^4 \left(\Gamma ^2-\delta \Gamma ^2\right)^2
    +8 \Gamma ^2 J^2 \left(\Gamma ^2-\delta \Gamma ^2\right) \left(\Gamma ^2+2 \delta \Gamma ^2-\tilde{s}^2\right)
    +\Gamma ^4 \left(\tilde{s}^2-\Gamma ^2\right)^2
    }{\Gamma ^4 (\tilde{s}-\Gamma )^2 (\Gamma +\tilde{s})^2}.
\end{equation}
which yields
        \begin{equation}\label{eq:nu_inf_asym_R2}
            \nu_-^\infty
            =
            \sqrt{\frac{-4 J^2 \left(\Gamma ^2-\delta \Gamma ^2\right) \left(\tilde{s}^2-3 \Gamma ^2\right)-4 \Gamma  J \sqrt{\left(\Gamma ^2-\delta \Gamma ^2\right)^2 \left(\left(\tilde{s}^2-\Gamma ^2\right)^2-4 J^2 \left(\tilde{s}^2-2 \Gamma ^2\right)\right)}+\left(\Gamma ^3-\Gamma  \tilde{s}^2\right)^2}{\Gamma ^2 (\tilde{s}-\Gamma )^2 (\Gamma +\tilde{s})^2}} .
        \end{equation}
At the II--III boundary $\tilde{s}=0$, 
\begin{align}
    \xi(t)
    &=
    1+\frac{12J^2\left(\Gamma^2-\delta\Gamma^2\right)}{\Gamma^4}
    -\frac{4J^2e^{-2\Gamma t}}{\Gamma^4}
    \left(3\delta\Gamma^2+\Gamma^2\left(2\delta\Gamma^2t^2-1\right)+4\Gamma\delta\Gamma^2t\right)
    \notag\\
    &\quad
    +\frac{4J^2e^{-\Gamma t}}{\Gamma^4}
    \left(6\delta\Gamma^2+\Gamma^2\left(\delta\Gamma^2t^2-4\right)-2\Gamma^3t+4\Gamma\delta\Gamma^2t\right).
\end{align}
\begin{align}
    \det V(t)&=
    \frac{32J^4\left(\Gamma^2-\delta\Gamma^2\right)e^{-3\Gamma t}}{\Gamma^8}
    \left(2\delta\Gamma^2+\Gamma^2\delta\Gamma^2t^2+2\Gamma^3t\right)
    +\frac{16J^4\left(\Gamma^2-\delta\Gamma^2\right)^2e^{-4\Gamma t}}{\Gamma^8}
    \notag\\
    &\quad
    +\frac{\Gamma^8+16J^4\left(\Gamma^2-\delta\Gamma^2\right)^2+8J^2\left(\Gamma^6+\Gamma^4\delta\Gamma^2-2\Gamma^2\delta\Gamma^4\right)}{\Gamma^8}
    \notag\\
    &\quad
    +\frac{8e^{-\Gamma t}}{\Gamma^8}
    \Bigl[
    4J^4\left(\Gamma^2-\delta\Gamma^2\right)\left(2\delta\Gamma^2+\Gamma^2\delta\Gamma^2t^2-2\Gamma^3t\right)
    \notag\\
    &\hspace{3.5em}
    +\Gamma^2J^2\left(-2\Gamma^2\delta\Gamma^2+4\delta\Gamma^4+t^2\left(2\Gamma^2\delta\Gamma^4-\Gamma^4\delta\Gamma^2\right)-2t\left(\Gamma^5-2\Gamma\delta\Gamma^4\right)\right)
    \Bigr]
    \notag\\
    &\quad
    -\frac{4e^{-2\Gamma t}}{\Gamma^8}
    \Bigl[
    4J^4\left(\Gamma^2-\delta\Gamma^2\right)\left(6\delta\Gamma^2+\Gamma^4t^2\left(\delta\Gamma^2t^2-4\right)+\Gamma^2\left(4\delta\Gamma^2t^2+2\right)\right)
    \notag\\
    &\hspace{3.5em}
    +\Gamma^2J^2\left(4\delta\Gamma^4+\Gamma^4\left(\delta\Gamma^4t^4+2\right)+4\Gamma^3\delta\Gamma^4t^3+\Gamma^2\left(8\delta\Gamma^4t^2-2\delta\Gamma^2\right)+8\Gamma\delta\Gamma^4t\right)
    \Bigr].
\end{align}
In the long-time limit, the minimal symplectic eigenvalue becomes
        \begin{equation}
            \nu_-^\infty
            =
            \frac{\sqrt{\Gamma ^4+12 J^2 \left(\Gamma ^2-\delta \Gamma ^2\right)+4 J \left(\delta \Gamma ^2-\Gamma ^2\right) \sqrt{\Gamma ^2+8 J^2}}}{\Gamma ^2}.
        \end{equation}
        \end{widetext}
In Region III, the time-dependent expressions for $\xi(t)$ and $\det V(t)$ can be recast under the substitution $\tilde{s}=i\tilde{\omega}$ into forms containing explicit oscillatory terms, consistent with the transient oscillatory entanglement dynamics discussed in the main text. Moreover, the asymptotic solution shares the same analytical structure as the Region-II result in Eq.~(\ref{eq:nu_inf_asym_R2}) under the same substitution.

The condition for nonzero asymptotic entanglement, $\nu_-^\infty<1$, can then be obtained analytically. In Region I, Eq.~(\ref{eq:nu_inf_asym_R1}) gives
\begin{equation}
    (\delta\Gamma^2-\tilde{s}^2)(\Gamma+\tilde{s})^2
    +4J^2(\Gamma^2-\delta\Gamma^2)<0.
\end{equation}
In Regions II and III, for $\Gamma>|\delta\Gamma|$, the condition reduces to
\begin{equation}
    J-\Gamma<0.
\end{equation}
The limiting case $\Gamma=|\delta\Gamma|$, corresponding to a vanishing local dissipation rate for one of the two modes, does not support nonzero asymptotic entanglement in stable regions. The boundaries between vanishing and nonzero asymptotic entanglement are therefore obtained by saturating these inequalities and are shown by the yellow dashed lines in Fig.~\ref{fig3}(b).

        \subsection{Identical independent dissipation with a common bath}\label{App-C3}

We next consider identical local dissipation,
$\Gamma_1=\Gamma_2\equiv\Gamma$,
in the presence of a common bath with collective dissipation rate $\kappa$.
Using the notation introduced in Appendix~\ref{App-B3}, the exact second-order moments are
\begin{widetext}
\begin{equation}
\begin{aligned}
    \langle a_1^2\rangle (t)= \langle a_2^2\rangle (t)
    &=
    -\frac{2 g J (\Gamma ^2-s^2) + 4 g \kappa  J (\Gamma +\kappa )-i \kappa  J (\Gamma ^2+2 \Gamma  \kappa +s^2)}{(\Gamma ^2-s^2) (\Gamma +2 \kappa -s) (\Gamma +2 \kappa +s)}
    \\
    &\quad
    +\frac{(2 g J-i J s) e^{-2it\lambda_{1}}}{4 s (\Gamma -s)}
    +\frac{(2 g J+i J s) e^{-2it\lambda_{2}}}{4 s (\Gamma +2 \kappa -s)}
    \\
    &\quad
    -\frac{(2 g J+i J s) e^{-2it\lambda_{3}}}{4 s (\Gamma +s)}
    -\frac{(2 g J-i J s) e^{-2it\lambda_{4}}}{4 s (\Gamma +2 \kappa +s)},
    \\
    \langle a_1^\dagger a_1 \rangle (t)= \langle a_2^\dagger a_2 \rangle (t)
    &=
    \frac{2 J^2 (\Gamma ^2-s^2) +4 \kappa  J^2 (\Gamma +\kappa )}{(\Gamma ^2-s^2) (\Gamma +2 \kappa -s) (\Gamma +2 \kappa +s)}
    \\
    &\quad
    -\frac{J^2 e^{-2it\lambda_{1}}}{2 s (\Gamma -s)}
    -\frac{J^2 e^{-2it\lambda_{2}}}{2 s (\Gamma +2 \kappa -s)}
    +\frac{J^2 e^{-2it\lambda_{3}}}{2 s (\Gamma +s)}
    +\frac{J^2 e^{-2it\lambda_{4}}}{2 s (\Gamma +2 \kappa +s)},
    \\
    \langle a_1 a_2\rangle(t)
    &=
    \frac{-i J (\Gamma +\kappa ) (\Gamma ^2-s^2) -i 2 \Gamma  \kappa  J (\Gamma +\kappa ) + 4 g \kappa  J (\Gamma +\kappa )}{(\Gamma ^2-s^2) (\Gamma +2 \kappa -s) (\Gamma +2 \kappa +s)}
    \\
    &\quad
    +\frac{(-2 g J+i J s) e^{-2it\lambda_{1}}}{4 s (\Gamma -s)}
    +\frac{(2 g J+i J s) e^{-2it\lambda_{2}}}{4 s (\Gamma +2 \kappa -s)}
    \\
    &\quad
    +\frac{(2 g J+i J s) e^{-2it\lambda_{3}}}{4 s (\Gamma +s)}
    +\frac{(-2 g J+i J s) e^{-2it\lambda_{4}}}{4 s (\Gamma +2 \kappa +s)},
    \\
    \langle a_1 a_2^\dagger\rangle(t)
    &=
    -\frac{4 \kappa  J^2 (\Gamma +\kappa )}{(\Gamma ^2-s^2) (\Gamma +2 \kappa -s) (\Gamma +2 \kappa +s)}
    +\frac{J^2 e^{-2it\lambda_{1}}}{2 s (\Gamma -s)}
    -\frac{J^2 e^{-2it\lambda_{2}}}{2 s (\Gamma +2 \kappa -s)}
    \\
    &\quad
    -\frac{J^2 e^{-2it\lambda_{3}}}{2 s (\Gamma +s)}
    +\frac{J^2 e^{-2it\lambda_{4}}}{2 s (\Gamma +2 \kappa +s)} .
\end{aligned}
\end{equation}
\end{widetext}
where the eigenvalues $\lambda_j$ are given in Eq.~(\ref{eq:eig_id_kap}). As in the previous cases, the exponential time dependence remains governed by the dynamical spectrum.

The entanglement dynamics follows from the CM constructed from the above second-order moments. For the common-bath setting, the quantities $\xi(t)$ and $\det V(t)$ entering the minimal symplectic eigenvalue $\nu_-(t)$ admit a finite exponential decomposition,
\begin{align}
    \xi(t)&=\sum_{n=1}^{7} C_n e^{a_n t},
    \\
    \det V(t)&=\sum_{n=1}^{15} B_n e^{d_n t},
\end{align}
with coefficients listed below.
\begin{widetext}
\begingroup
\renewcommand{\arraystretch}{2.0}
\setlength{\extrarowheight}{2pt}
\begin{equation*}
\begin{array}{c c c}
\hline
n & a_n & C_n \\
\hline
1 & 2(-\Gamma-\kappa+s) &
\dfrac{2J^2}
{(s-\Gamma)(-\Gamma-2\kappa+s)}
\\
2 & s-\Gamma &
-\dfrac{2J^2(\Gamma+2\kappa)}
{s(s-\Gamma)(-\Gamma-2\kappa+s)}
\\
3 & -\Gamma-2\kappa+s &
-\dfrac{2\Gamma J^2}
{s(s-\Gamma)(-\Gamma-2\kappa+s)}
\\
4 & 0 &
\dfrac{
\left(s^2-\Gamma^2\right)\left(s^2-(\Gamma+2\kappa)^2\right)
-4J^2\left(-3\Gamma^2-6\Gamma\kappa-4\kappa^2+s^2\right)}
{(s-\Gamma)(\Gamma+s)(-\Gamma-2\kappa+s)(\Gamma+2\kappa+s)}
\\
5 & -\Gamma-s &
\dfrac{2J^2(\Gamma+2\kappa)}
{s(\Gamma+s)(\Gamma+2\kappa+s)}
\\
6 & -\Gamma-2\kappa-s &
\dfrac{2\Gamma J^2}
{s(\Gamma+s)(\Gamma+2\kappa+s)}
\\
7 & -2(\Gamma+\kappa+s) &
\dfrac{2J^2}
{(\Gamma+s)(\Gamma+2\kappa+s)}
\\
\hline
\end{array}
\end{equation*}
\begin{equation*}
\begin{array}{c c c}
\hline
n & d_n & B_n \\
\hline
1 & 2(-\Gamma-\kappa+s) &
\dfrac{16\Gamma J^4(\Gamma+2\kappa)}
{s^2(s-\Gamma)(\Gamma+s)(-\Gamma-2\kappa+s)(\Gamma+2\kappa+s)}
\\
2 & s-\Gamma &
-\dfrac{4\Gamma J^2\left((\Gamma+2\kappa)^2+4J^2-s^2\right)}
{s(s-\Gamma)(\Gamma+s)(-\Gamma-2\kappa+s)(\Gamma+2\kappa+s)}
\\
3 & -\Gamma-2\kappa+s &
-\dfrac{4J^2(\Gamma+2\kappa)\left(\Gamma^2+4J^2-s^2\right)}
{s(s-\Gamma)(\Gamma+s)(-\Gamma-2\kappa+s)(\Gamma+2\kappa+s)}
\\
4 & -3\Gamma-2\kappa+s &
\dfrac{16J^4(\Gamma+2\kappa)}
{s(s-\Gamma)(\Gamma+s)(-\Gamma-2\kappa+s)(\Gamma+2\kappa+s)}
\\
5 & -3\Gamma-4\kappa+s &
\dfrac{16\Gamma J^4}
{s\left(s^2-\Gamma^2\right)\left(s^2-(\Gamma+2\kappa)^2\right)}
\\
6 & 0 &
\dfrac{
\left(\Gamma^2+4J^2-s^2\right)
\left((\Gamma+2\kappa)^2+4J^2-s^2\right)}
{(s-\Gamma)(\Gamma+s)(-\Gamma-2\kappa+s)(\Gamma+2\kappa+s)}
\\
7 & -2\Gamma &
-\dfrac{4J^2\left((\Gamma+2\kappa)^2+4J^2-s^2\right)}
{(s-\Gamma)(\Gamma+s)(-\Gamma-2\kappa+s)(\Gamma+2\kappa+s)}
\\
8 & -2(\Gamma+\kappa) &
-\dfrac{32\Gamma J^4(\Gamma+2\kappa)}
{s^2(s-\Gamma)(\Gamma+s)(-\Gamma-2\kappa+s)(\Gamma+2\kappa+s)}
\\
9 & -2(\Gamma+2\kappa) &
-\dfrac{4J^2\left(\Gamma^2+4J^2-s^2\right)}
{(s-\Gamma)(\Gamma+s)(-\Gamma-2\kappa+s)(\Gamma+2\kappa+s)}
\\
10 & -4(\Gamma+\kappa) &
\dfrac{16J^4}
{(s-\Gamma)(\Gamma+s)(-\Gamma-2\kappa+s)(\Gamma+2\kappa+s)}
\\
11 & -\Gamma-s &
\dfrac{4\Gamma J^2\left((\Gamma+2\kappa)^2+4J^2-s^2\right)}
{s(s-\Gamma)(\Gamma+s)(-\Gamma-2\kappa+s)(\Gamma+2\kappa+s)}
\\
12 & -\Gamma-2\kappa-s &
\dfrac{4J^2(\Gamma+2\kappa)\left(\Gamma^2+4J^2-s^2\right)}
{s(s-\Gamma)(\Gamma+s)(-\Gamma-2\kappa+s)(\Gamma+2\kappa+s)}
\\
13 & -3\Gamma-2\kappa-s &
-\dfrac{16J^4(\Gamma+2\kappa)}
{s(s-\Gamma)(\Gamma+s)(-\Gamma-2\kappa+s)(\Gamma+2\kappa+s)}
\\
14 & -3\Gamma-4\kappa-s &
-\dfrac{16\Gamma J^4}
{s\left(s^2-\Gamma^2\right)\left(s^2-(\Gamma+2\kappa)^2\right)}
\\
15 & -2(\Gamma+\kappa+s) &
\dfrac{16\Gamma J^4(\Gamma+2\kappa)}
{s^2(s-\Gamma)(\Gamma+s)(-\Gamma-2\kappa+s)(\Gamma+2\kappa+s)}
\\
\hline
\end{array}
\end{equation*}
\endgroup
\end{widetext}

\subsubsection{Asymptotic analysis in the three dynamical regimes}

Using the exact second-order moments given above, we first analyze the population dynamics. Since the two mode populations remain identical,
\begin{equation}
    n(t)
    =
    \langle \hat a_1^\dagger\hat a_1\rangle(t)
    =
    \langle \hat a_2^\dagger\hat a_2\rangle(t),
\end{equation}
the population dynamics in the three dynamical regimes and at their boundaries can be characterized as follows:
\begin{widetext}
\begin{equation}
\begin{aligned}
    \text{Region I:}\qquad
    n(t)&\overset{t\gg 0}{\approx}
    \frac{J^2 e^{-t (\Gamma -s)}}{-2 s (\Gamma -s)},
    \\
\text{I--II boundary }(\Gamma=s):\qquad
    n(t)&=
    \frac{J^2 \left(-\kappa ^2+\Gamma ^2 (2 \kappa  t+1)+\Gamma  \kappa  (2 \kappa  t-1)\right)}{4 \Gamma ^2 \kappa  (\Gamma +\kappa )}+\frac{J^2 e^{-2 \Gamma  t}}{4 \Gamma ^2}
    -\frac{J^2 e^{-2 \kappa  t}}{4 \Gamma  \kappa }+\frac{J^2 e^{-2 t (\Gamma +\kappa )}}{4 \Gamma  (\Gamma +\kappa )},
    \\
    \text{Region II:}\qquad
    n(t)&\overset{t\gg 0}{\approx}
    \frac{2 J^2 \left(\Gamma ^2-s^2\right) +4 \kappa  J^2 (\Gamma +\kappa )}{\left(\Gamma ^2-s^2\right) (\Gamma +2 \kappa -s) (\Gamma +2 \kappa +s)},
    \\
\text{II--III boundary }(s=0):\qquad
    n(t)&=
    \frac{2 J^2 \left(\Gamma ^2+2 \Gamma  \kappa +2 \kappa ^2\right)}{\Gamma ^2 (\Gamma +2 \kappa )^2}-\frac{J^2 e^{-\Gamma t} (\Gamma  t+1)}{\Gamma ^2}
    -\frac{J^2 e^{-t (\Gamma +2 \kappa )} (t (\Gamma +2 \kappa )+1)}{(\Gamma +2 \kappa )^2},
    \\
    \text{Region III:}\qquad
    n(t)&=
    \frac{2 J^2 \left(\Gamma ^2+\omega ^2\right)+4 \kappa  J^2 (\Gamma +\kappa )}{\left(\Gamma ^2+\omega ^2\right) \left((\Gamma +2 \kappa )^2+\omega ^2\right)}
    -\frac{J^2 e^{-\Gamma t} \sin (\omega t +\phi )}{\omega  \sqrt{\Gamma ^2+\omega ^2}}
    \\
    &\quad
    -\frac{J^2 e^{-t (\Gamma +2 \kappa )} \left(\sqrt{\Gamma ^2+\omega ^2} \sin (\omega t +\phi )+2 \kappa  \sin (\omega t )\right)}{\omega  \left((\Gamma +2 \kappa )^2+\omega ^2\right)},
\end{aligned}
\end{equation}
\end{widetext}
where $\omega=-is$ in Region III and $\tan\phi=\omega/\Gamma$. These expressions retain the characteristic population dynamics of the three regions found without a common bath, while the common-bath coupling modifies both the transient evolution and the stationary population.
        
We next analyze the asymptotic entanglement through $\xi(t)$ and $\det V(t)$. Unlike the previous cases, the dominant long-time contribution in Region I changes depending on the sign of $s-\Gamma-\kappa$, giving rise to two distinct asymptotic subregions.
For $s-\Gamma-\kappa>0$, the leading exponential contributions to both $\xi(t)$ and $\det V(t)$ scale as $e^{2t(s-\Gamma-\kappa)}$, so
\begin{equation}
    \xi(t)\overset{t\gg 0}{\approx}
    \frac{2 J^2 e^{2 t (-\Gamma -\kappa +s)}}{(s-\Gamma ) (-\Gamma -2 \kappa +s)}.
\end{equation}
\begin{equation}
    \det V(t)\overset{t\gg 0}{\approx}
    \frac{16 \Gamma  J^4 (\Gamma +2 \kappa ) e^{2 t (-\Gamma -\kappa +s)}}{s^2 (s-\Gamma ) (\Gamma +s) (-\Gamma -2 \kappa +s) (\Gamma +2 \kappa +s)}.
\end{equation}
As in the previous cases, the leading-order contributions cancel in evaluating $\nu_-(t)$, yielding
\begin{equation}\label{eq:nu_inf_id_kap_R1_01}
    \nu_-^\infty \simeq \sqrt{\frac{\det V(t)}{2\,\xi(t)}}\bigg|_{t\to\infty}
    =
    2 \sqrt{\frac{\Gamma  J^2 (\Gamma +2 \kappa )}{s^2 (\Gamma +s) (\Gamma +2 \kappa +s)}} .
\end{equation}
For $s-\Gamma>0$ and $s-\Gamma-\kappa\le0$, the leading exponential contributions change to $e^{t(s-\Gamma)}$, yielding
\begin{equation}
    \xi(t)\overset{t\gg 0}{\approx}
    -\frac{2 J^2 (\Gamma +2 \kappa ) e^{t (s-\Gamma )}}{s (s-\Gamma ) (-\Gamma -2 \kappa +s)}.
\end{equation}
\begin{equation}
    \det V(t)\overset{t\gg 0}{\approx}
    -\frac{4 \Gamma  J^2 e^{t (s-\Gamma )} \left((\Gamma +2 \kappa )^2+4 J^2-s^2\right)}{s (s-\Gamma ) (\Gamma +s) (-\Gamma -2 \kappa +s) (\Gamma +2 \kappa +s)}.
\end{equation}
The same cancellation mechanism gives
\begin{equation}\label{eq:nu_inf_id_kap_R1_02}
    \nu_-^\infty \simeq \sqrt{\frac{\det V(t)}{2\,\xi(t)}}\bigg|_{t\to\infty}
    =
    \sqrt{\frac{\Gamma  \left((\Gamma +2 \kappa )^2+4 J^2-s^2\right)}{(\Gamma +2 \kappa ) (\Gamma +s) (\Gamma +2 \kappa +s)}} .
\end{equation}
Thus, Region I is further divided by $s-\Gamma-\kappa=0$ into two asymptotic subregions with different dominant long-time contributions. The resulting asymptotic expressions for $\nu_-^\infty$ are discontinuous across this boundary, as their limiting values differ on the two sides. This jump in $\nu_-^\infty$ leads to the discontinuity of the finite asymptotic entanglement observed in Fig.~\ref{fig6}(b) and Fig.~\ref{fig6}(c) of the main text.

At the I--II boundary $\Gamma=s$, one finds
\begin{widetext}
\begin{align}
    \xi(t)
    &=
    \frac{J^2 (\Gamma +2 \kappa ) e^{-2 \Gamma  t}}{2 \Gamma ^2 (\Gamma +\kappa )}
    +\frac{J^2 \left(-2 \kappa ^2+2 \Gamma ^3 t+\Gamma ^2 (6 \kappa  t+2)+\Gamma  \kappa  (4 \kappa  t-1)\right)}{2 \Gamma ^2 \kappa  (\Gamma +\kappa )}
    \notag\\
    &\quad
    -\frac{J^2 (\Gamma  t+1) e^{-2 \kappa  t}}{\Gamma  \kappa }
    +\frac{J^2 e^{-2 t (\Gamma +\kappa )}}{2 \Gamma  (\Gamma +\kappa )}
    +\frac{J^2 e^{-2 t (2 \Gamma +\kappa )}}{2 \Gamma  (\Gamma +\kappa )}+1,
    \\
    \det V(t)
    &=
    \frac{2 J^4 (\Gamma +2 \kappa ) (\Gamma  t+1) e^{-2 t (2 \Gamma +\kappa )}}{\Gamma ^3 \kappa  (\Gamma +\kappa )}
    -\frac{2 J^4 (\Gamma  t+1) e^{-4 t (\Gamma +\kappa )}}{\Gamma ^2 \kappa  (\Gamma +\kappa )}
    \notag\\
    &\quad
    +\frac{2 J^2 e^{-2 \Gamma  t} (\Gamma  t+1) \left(\kappa  (\Gamma +\kappa )+J^2\right)}{\Gamma ^2 \kappa  (\Gamma +\kappa )}
    -\frac{J^2 \left(\Gamma ^2+2 J^2 (\Gamma  t-1)\right) e^{-2 t (\Gamma +2 \kappa )}}{\Gamma ^2 \kappa  (\Gamma +\kappa )}
    \notag\\
    &\quad
    +\frac{\left(\kappa  (\Gamma +\kappa )+J^2\right) \left(\Gamma ^2+2 J^2 (\Gamma  t-1)\right)}{\Gamma ^2 \kappa  (\Gamma +\kappa )}
    -\frac{J^2 (\Gamma +2 \kappa ) \left(4 J^2-\Gamma ^2\right) e^{-2 t (\Gamma +\kappa )}}{\Gamma ^3 \kappa  (\Gamma +\kappa )}
    \notag\\
    &\quad
    -\frac{J^2 (\Gamma +2 \kappa ) e^{-2 \kappa  t} \left(\Gamma ^2+2 J^2 (\Gamma  t-1)\right)}{\Gamma ^3 \kappa  (\Gamma +\kappa )}.
\end{align}
\end{widetext}
Thus, the long-time minimal symplectic eigenvalue reduces to
\begin{equation}
    \nu_-^\infty \simeq \sqrt{\frac{\det V(t)}{2\,\xi(t)}}\bigg|_{t\to\infty}
    =
    \sqrt{\frac{\kappa  (\Gamma +\kappa )+J^2}{\Gamma ^2+3 \Gamma  \kappa +2 \kappa ^2}},
\end{equation}
which continuously matches the second Region-I asymptotic expression at $\Gamma=s$.
In Region II,
\begin{widetext}
\begin{align}
    \xi(t) \overset{t\gg 0}{\approx}
    \frac{\left(s^2-\Gamma ^2\right) \left(s^2-(\Gamma +2 \kappa )^2\right)-4 J^2 \left(-3 \Gamma ^2-6 \Gamma  \kappa -4 \kappa ^2+s^2\right)}{(s-\Gamma ) (\Gamma +s) (-\Gamma -2 \kappa +s) (\Gamma +2 \kappa +s)},
    \\
    \det V(t) \overset{t\gg 0}{\approx}
    \frac{\left(\Gamma ^2+4 J^2-s^2\right) \left((\Gamma +2 \kappa )^2+4 J^2-s^2\right)}{(s-\Gamma ) (\Gamma +s) (-\Gamma -2 \kappa +s) (\Gamma +2 \kappa +s)}.
\end{align}
Thus, the long-time minimal symplectic eigenvalue reduces to
\begin{equation}\label{eq:nu_inf_id_kap_R2}
\begin{aligned}
    \nu_-^\infty
    &=
    \sqrt{\frac{
    -4 J^2 \left(-3 \Gamma ^2-6 \Gamma  \kappa -4 \kappa ^2+s^2\right)
    +\left(s^2-\Gamma ^2\right) \left(s^2-(\Gamma +2 \kappa )^2\right)
    }{(s-\Gamma ) (\Gamma +s) (-\Gamma -2 \kappa +s) (\Gamma +2 \kappa +s)}} \\
    &\quad\times
    \sqrt{1
    -\frac{
    4 J (\Gamma +\kappa ) \sqrt{\left(s^2-\Gamma ^2\right) \left(s^2-(\Gamma +2 \kappa )^2\right)-4 J^2 \left(s^2-2 \left(\Gamma ^2+2 \Gamma  \kappa +2 \kappa ^2\right)\right)}
    }{
    -4 J^2 \left(-3 \Gamma ^2-6 \Gamma  \kappa -4 \kappa ^2+s^2\right)
    +\left(s^2-\Gamma ^2\right) \left(s^2-(\Gamma +2 \kappa )^2\right)
    }}.
\end{aligned}
\end{equation}
At the II--III boundary $s=0$, one finds
\begin{align}
    \xi(t)
    &=
    \frac{4 J^2 \left(3 \Gamma ^2+6 \Gamma  \kappa +4 \kappa ^2\right)}{\Gamma ^2 (\Gamma +2 \kappa )^2}
    -\frac{4 J^2 e^{-\Gamma t} \left(2 \kappa +\Gamma ^2 t+2 \Gamma  (\kappa  t+1)\right)}{\Gamma ^2 (\Gamma +2 \kappa )}
    \notag\\
    &\quad
    -\frac{4 J^2 e^{-t (\Gamma +2 \kappa )} \left(2 \kappa +\Gamma ^2 t+2 \Gamma  (\kappa  t+1)\right)}{\Gamma  (\Gamma +2 \kappa )^2}
    +\frac{4 J^2 e^{-2 t (\Gamma +\kappa )}}{\Gamma ^2+2 \Gamma  \kappa }+1,
    \\
    \det V(t)
    &=
    \frac{64 J^4 t^2 e^{-2 t (\Gamma +\kappa )}}{\Gamma ^2+2 \Gamma  \kappa }
    +\frac{32 J^4 t e^{t (-3 \Gamma -2 \kappa )}}{\Gamma ^2 (\Gamma +2 \kappa )}
    +\frac{16 J^4 e^{-4 t (\Gamma +\kappa )}}{\Gamma ^2 (\Gamma +2 \kappa )^2}
    +\frac{32 J^4 t e^{t (-3 \Gamma -4 \kappa )}}{\Gamma  (\Gamma +2 \kappa )^2}
    \notag\\
    &\quad
    +\frac{\left(\Gamma ^2+4 J^2\right) \left((\Gamma +2 \kappa )^2+4 J^2\right)}{\Gamma ^2 (\Gamma +2 \kappa )^2}
    +\frac{4 J^2 e^{-2 \Gamma  t} \left(-\frac{4 J^2}{(\Gamma +2 \kappa )^2}-1\right)}{\Gamma ^2}
    \notag\\
    &\quad
    -\frac{8 J^2 t \left(\Gamma ^2+4 J^2\right) e^{t (-\Gamma -2 \kappa )}}{\Gamma ^2 (\Gamma +2 \kappa )}
    -\frac{4 J^2 \left(\Gamma ^2+4 J^2\right) e^{-2 t (\Gamma +2 \kappa )}}{\Gamma ^2 (\Gamma +2 \kappa )^2}
    -\frac{8 J^2 t e^{-\Gamma t} \left((\Gamma +2 \kappa )^2+4 J^2\right)}{\Gamma  (\Gamma +2 \kappa )^2}.
\end{align}
Accordingly, the long-time minimal symplectic eigenvalue is
\begin{equation}
    \nu_-^\infty
    =
    \frac{\sqrt{\Gamma ^2 (\Gamma +2 \kappa )^2+4 J^2 \left(3 \Gamma ^2+6 \Gamma  \kappa +4 \kappa ^2\right)-4 J (\Gamma +\kappa ) \sqrt{\Gamma ^2 (\Gamma +2 \kappa )^2+8 J^2 \left(\Gamma ^2+2 \Gamma  \kappa +2 \kappa ^2\right)}}}{\Gamma  (\Gamma +2 \kappa )}.
\end{equation}
\end{widetext} 
In Region III, substituting $s=i\omega$ into the time-dependent expressions naturally recovers the oscillatory transient behavior discussed in the main text. The corresponding asymptotic solution follows from the same substitution applied to the Region-II result in Eq.~(\ref{eq:nu_inf_id_kap_R2}).

The condition for nonzero asymptotic entanglement, $\nu_-^\infty<1$, can likewise be obtained analytically. For the first asymptotic subregion of Region I, Eq.~(\ref{eq:nu_inf_id_kap_R1_01}) gives
\begin{equation}
    4 \Gamma J^2 (\Gamma +2 \kappa )-s^2 (\Gamma +s) (\Gamma +2 \kappa +s)<0
\end{equation}
For the second asymptotic subregion of Region I, Eq.~(\ref{eq:nu_inf_id_kap_R1_02}) gives
\begin{equation}
    \Gamma \left((\Gamma +2 \kappa )^2+4 J^2-s^2\right)-(\Gamma +2 \kappa ) (\Gamma +s) (\Gamma +2 \kappa +s)<0
\end{equation}
In Regions II and III, Eq.~(\ref{eq:nu_inf_id_kap_R2}) and its analytic continuation give the common condition
\begin{equation}
    J-(\Gamma +\kappa )<0
\end{equation}

        \begin{figure}
        \centering
        \includegraphics[width=8cm]{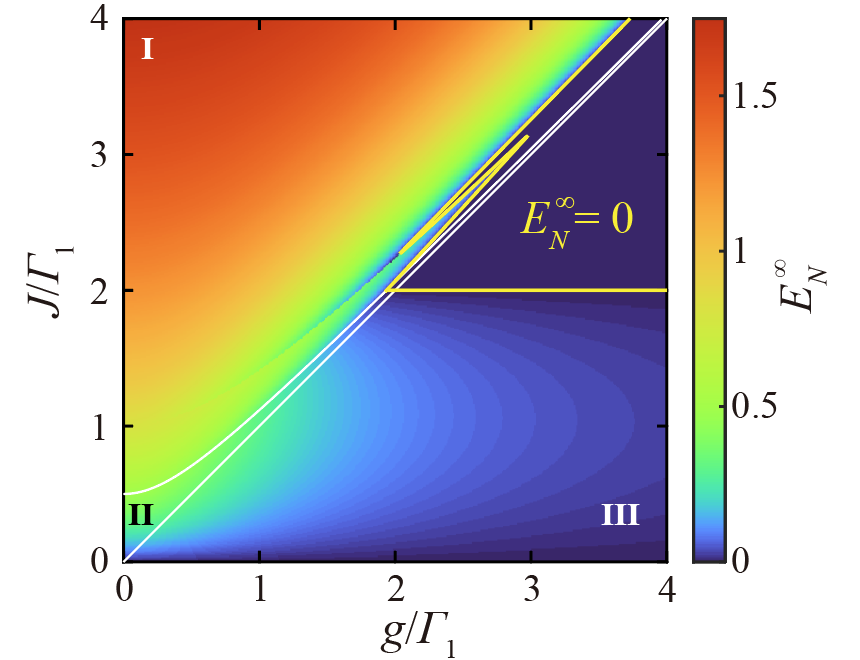}
        \caption{
Asymptotic logarithmic negativity $E_N^\infty$ of the dissipative quadratic bosonic dimer with a common bath, shown as a function of $g/\Gamma_1$ and $J/\Gamma_1$. Here, $\kappa/\Gamma_1=1$ and $\Gamma_2/\Gamma_1=1$. The white and yellow solid lines mark the dynamical-region boundaries and the boundary of the $E_N^\infty=0$ region, respectively. Within Region I, $E_N^\infty$ remains finite but changes discontinuously upon crossing the line $s-\Gamma-\kappa=0$. This discontinuity results from a change in the dominant asymptotic contribution, yielding two distinct analytical branches.
        }
        \label{figSf1}
        \end{figure}

Based on the analytical expressions for $\nu_-^\infty$ derived above together with Eq.~(\ref{eq:EN}), the full asymptotic-entanglement landscape can be constructed, as shown in Fig.~\ref{figSf1}. The white lines mark the dynamical boundaries separating Regions I, II, and III, while the yellow lines indicate the boundaries between vanishing and nonzero asymptotic entanglement. 
Consistent with the analytical results derived above, the discontinuity of the finite asymptotic entanglement inside Region I reflects the transition between the two dominant asymptotic contributions across $s-\Gamma-\kappa=0$. 
Together with the $\Delta E_N^\infty$ map in the main text, this figure distinguishes the full asymptotic entanglement landscape from the common-bath-induced enhancement or suppression.

        \subsection{Common bath without independent dissipation}\label{App-C4}

We finally consider the special case in which only a common bath is present, namely $\Gamma_{1,2}=0$ and $\kappa\neq0$. 
With the same notation introduced in Appendix~\ref{App-B3}, the corresponding second-order moments are
\begin{widetext}
\begin{equation}
\begin{aligned}
    \langle a_1^2\rangle (t)= \langle a_2^2\rangle (t)
    &=
    \frac{J \left(-4 g \kappa ^2+s^2 (2 g+i \kappa )\right)}{s^2 (s-2 \kappa ) (2 \kappa +s)}
    -\frac{J (2 g-i s) e^{-2it\lambda_1}}{4 s^2}
    -\frac{J (2 g+i s) e^{-2it\lambda_2}}{4 s (s-2 \kappa )}
    \\
    &\quad
    -\frac{J (2 g+i s) e^{-2it\lambda_3}}{4 s^2}
    +\frac{(-2 g J+i J s) e^{-2it\lambda_4}}{4 s (2 \kappa +s)},
    \\
    \langle a_1^\dagger a_1 \rangle (t) = \langle a_2^\dagger a_2 \rangle (t)
    &=
    -\frac{2 J^2 \left(s^2-2 \kappa ^2\right)}{s^2 (s-2 \kappa ) (2 \kappa +s)}
    +\frac{J^2 e^{-2it\lambda_1}}{2 s^2}
    +\frac{J^2 e^{-2it\lambda_2}}{2 s (s-2 \kappa )}
    +\frac{J^2 e^{-2it\lambda_3}}{2 s^2}
    +\frac{J^2 e^{-2it\lambda_4}}{2 s (2 \kappa +s)},
    \\
    \langle a_1 a_2 \rangle (t)
    &=
    \frac{\kappa  J \left(4 g \kappa +i s^2\right)}{s^2 (s-2 \kappa ) (2 \kappa +s)}
    +\frac{J (2 g-i s) e^{-2it\lambda_1}}{4 s^2}
    -\frac{J (2 g+i s) e^{-2it\lambda_2}}{4 s (s-2 \kappa )}
    \\
    &\quad
    +\frac{J (2 g+i s) e^{-2it\lambda_3}}{4 s^2}
    +\frac{(-2 g J+i J s) e^{-2it\lambda_4}}{4 s (2 \kappa +s)},
    \\
    \langle a_1 a_2^\dagger \rangle (t)
    &=
    -\frac{4 \kappa ^2 J^2}{s^2 (s-2 \kappa ) (2 \kappa +s)}
    -\frac{J^2 e^{-2it\lambda_1}}{2 s^2}
    +\frac{J^2 e^{-2it\lambda_2}}{2 s (s-2 \kappa )}
    -\frac{J^2 e^{-2it\lambda_3}}{2 s^2}
    +\frac{J^2 e^{-2it\lambda_4}}{2 s (2 \kappa +s)}.
\end{aligned}
\end{equation}
\end{widetext}
where the eigenvalues $\lambda_j$ are given in Eq.~(\ref{eq:eig_kap}). Although the dynamical regimes in this limit differ from those with independent dissipation, the exponential time dependence of the second-order moments remains governed by the dynamical spectrum.

Using the above second-order moments, the entanglement dynamics can be obtained from the corresponding CM. The quantities $\xi(t)$ and $\det V(t)$ entering the minimal symplectic eigenvalue $\nu_-(t)$ take the explicit forms
\begin{widetext}
\begin{equation}
    \xi(t) =
    \frac{4 \kappa  J^2 e^{-s t}}{s^2 (2 \kappa +s)}
    -\frac{4 \kappa  J^2 e^{s t}}{s^2 (s-2 \kappa )}
    -\frac{4 J^2}{s^2}
    +\frac{2 J^2 e^{-2 t (\kappa +s)}}{s (2 \kappa +s)}
    +\frac{2 J^2 e^{-2 t (\kappa -s)}}{s (s-2 \kappa )}+1,
\end{equation}
and
\begin{equation}
    \det V(t) =
    \frac{-4 \kappa ^2-4 J^2+s^2}{(s-2 \kappa ) (2 \kappa +s)}
    +\frac{4 J^2 e^{-4 \kappa  t}}{(s-2 \kappa ) (2 \kappa +s)}
    +\frac{8 \kappa  J^2 e^{-t (2 \kappa -s)}}{s (s-2 \kappa ) (2 \kappa +s)}
    -\frac{8 \kappa  J^2 e^{-t (2 \kappa +s)}}{s (s-2 \kappa ) (2 \kappa +s)}.
\end{equation}
\end{widetext}

\subsubsection{Asymptotic analysis in the two dynamical regimes}

Based on the above second-order moments, the identical mode populations are given by
\begin{equation}
    n(t)
    =
    \langle \hat a_1^\dagger\hat a_1\rangle(t)
    =
    \langle \hat a_2^\dagger\hat a_2\rangle(t).
\end{equation}
The population dynamics in the two spectral regimes and at the EP separating them are summarized as follows:
\begin{widetext}
\begin{equation}
\begin{aligned}
    \text{Region I$'$ ($s>0$):}\qquad
    n(t)&\overset{t\gg 0}{\approx}
    \frac{J^2 e^{s t}}{2 s^2},
    \\
    \text{I$'$--III$'$ boundary }(s=0):\qquad
    \lim_{s\rightarrow 0} n(t)
    &=
    \frac{J^2 \left(2 \kappa ^2 t^2+1\right)}{4 \kappa ^2}
    -\frac{J^2 e^{-2 \kappa  t} (2 \kappa  t+1)}{4 \kappa ^2} 
    \overset{t\gg 0}{\approx} \frac{J^2 \left(2 \kappa ^2 t^2+1\right)}{4 \kappa ^2},
    \\
    \text{Region III$'$:}\qquad
    n(t)
    &=
    \frac{2 J^2 \left(2 \kappa ^2+\omega ^2\right)}{4 \kappa ^2 \omega ^2+\omega ^4}
    -\frac{J^2 e^{-2 \kappa  t} (2 \kappa  \sin (t \omega )+\omega  \cos (t \omega ))}{4 \kappa ^2 \omega +\omega ^3}
    -\frac{J^2 \cos (t \omega )}{\omega ^2}
    \notag\\
    &\overset{t\gg 0}{\approx}
    \frac{2 J^2 \left(2 \kappa ^2+\omega ^2\right)}{4 \kappa ^2 \omega ^2+\omega ^4}
    -\frac{J^2 \cos (t \omega )}{\omega ^2}.
\end{aligned}
\end{equation}
\end{widetext}
where $\omega=-is$ in Region III$'$. These expressions reveal the distinct population dynamics induced by the spectral structure: exponential amplification in Region I$'$, polynomial growth at the EP, and persistent oscillations in Region III$'$.

Next, we examine the asymptotic entanglement dynamics through the quantities $\xi(t)$ and $\det V(t)$. 
In Region I$'$, the dominant long-time contributions depend on the relative magnitude of $s$ and $2\kappa$, resulting in two distinct asymptotic forms for $\xi(t)$ and $\det V(t)$.
For $s>2\kappa$,
\begin{equation}
    \xi(t)\overset{t\gg 0}{\approx}
    \frac{2 J^2 e^{2 t (s-\kappa )}}{s (s-2 \kappa )},
\end{equation}
\begin{equation}
    \det V(t)\overset{t\gg 0}{\approx}
    \frac{8 \kappa  J^2 e^{t (s-2 \kappa )}}{s (s-2 \kappa ) (2 \kappa +s)},
\end{equation}
Since $\xi(t)$ becomes dominant over $\det V(t)$ for $t\gg0$,
\begin{equation}
    \nu_-(t)\overset{t\gg 0}{\approx}
    \sqrt{\xi(t)-\sqrt{\xi^2(t)}} \overset{t\gg 0}{\approx} 0.
\end{equation}
For $s>0$ and $s\le 2 \kappa$,
\begin{equation}
    \xi(t)\overset{t\gg 0}{\approx}
    -\frac{4 \kappa  J^2 e^{s t}}{s^2 (s-2 \kappa )},
\end{equation}

\begin{equation}
    \det V(t)\overset{t\gg 0}{\approx}
    \frac{-4 \kappa ^2-4 J^2+s^2}{(s-2 \kappa ) (2 \kappa +s)},
\end{equation}
Likewise, $\xi(t)$ dominates over $\det V(t)$ in the long-time limit, yielding
\begin{equation}
    \nu_-(t)\overset{t\gg 0}{\longrightarrow}0 .
\end{equation}
At the internal boundary $s=2\kappa$
\begin{widetext}
\begin{equation}
    \lim_{s\rightarrow 2\kappa}\xi(t)=
    -\frac{J^2}{\kappa ^2}
    +\frac{J^2 e^{2 \kappa  t} (2 \kappa  t+1)}{2 \kappa ^2}
    +\frac{J^2 e^{-6 \kappa  t}}{4 \kappa ^2}
    +\frac{J^2 e^{-2 \kappa  t}}{4 \kappa ^2}+1
    \overset{t\gg 0}{\approx}
    \frac{J^2 e^{2 \kappa  t} (2 \kappa  t+1)}{2 \kappa ^2},
\end{equation}
\begin{equation}
    \lim_{s\rightarrow 2\kappa}\det V(t)=
    \frac{J^2 (2 \kappa  t-1)}{2 \kappa ^2}
    +\frac{J^2 e^{-4 \kappa  t} (2 \kappa  t+1)}{2 \kappa ^2}+1
    \overset{t\gg 0}{\approx}
    \frac{J^2t}{\kappa},
\end{equation}
\end{widetext}
At this boundary, $\xi(t)$ remains dominant over $\det V(t)$ in the long-time limit, and therefore
\begin{equation}
    \nu_-(t)\overset{t\gg0}{\longrightarrow}0 .
\end{equation}
Accordingly, throughout Region I$'$, $\xi(t)$ dominates over $\det V(t)$ in the long-time limit, leading to $\nu_-(t)\rightarrow0$ and hence to unbounded entanglement growth, as discussed in the main text.

At the I$'$--III$'$ boundary $s=0$, one finds
\begin{widetext}
\begin{equation}
    \lim_{s\rightarrow 0}\xi(t)=
    J^2 \left(\frac{1}{\kappa ^2}+2 t^2+\frac{2 t}{\kappa }\right)
    -\frac{J^2 e^{-2 \kappa  t} (4 \kappa  t+1)}{\kappa ^2}+1
    \overset{t\gg 0}{\approx}
    2 J^2 t^2,
\end{equation}
\begin{equation}
    \lim_{s\rightarrow 0}\det V(t)=
    \frac{J^2}{\kappa ^2}
    -\frac{J^2 e^{-4 \kappa  t}}{\kappa ^2}
    -\frac{4 J^2 t e^{-2 \kappa  t}}{\kappa }+1
    \overset{t\gg 0}{\approx}
    \frac{J^2}{\kappa ^2}+1.
\end{equation}
\end{widetext}
Thus, the exponential growth of $\xi(t)$ in Region I$'$ is replaced by polynomial growth at the EP, while $\det V(t)$ approaches a finite value in this limit, resulting in a slower unbounded growth of entanglement.
        
        In Region III$'$, by contrast, 
\begin{widetext}
\begin{equation}
    \xi(t)=
    -\frac{4 J^2 e^{-2 \kappa  t} (2 \kappa  \sin (2 t \omega )+\omega  \cos (2 t \omega ))}{4 \kappa ^2 \omega +\omega ^3}
    -\frac{8 \kappa  J^2 (2 \kappa  \cos (t \omega )-\omega  \sin (t \omega ))}{4 \kappa ^2 \omega ^2+\omega ^4}
    +\frac{4 J^2}{\omega ^2}+1,
\end{equation}
\begin{equation}
    \det V(t)=
    \frac{4 \kappa ^2+4 J^2+\omega ^2}{4 \kappa ^2+\omega ^2}
    -\frac{16 \kappa  J^2 e^{-2 \kappa  t} \sin (t \omega )}{4 \kappa ^2 \omega +\omega ^3}
    -\frac{4 J^2 e^{-4 \kappa  t}}{4 \kappa ^2+\omega ^2}.
\end{equation}
\end{widetext}
Then the long-time behavior is governed by
        \begin{equation}
            \xi(t) \overset{t\gg0}{\approx}
            -\frac{8 \kappa J^2 \left(2 \kappa \cos (t \omega )-\omega \sin (t \omega )\right)}{4 \kappa ^2 \omega ^2+\omega ^4}
            +\frac{4 J^2}{\omega ^2}+1,
       \end{equation}
       \begin{equation}
            \det V(t) \overset{t\gg0}{\approx}
            \frac{4 \kappa ^2+4 J^2+\omega ^2}{4 \kappa ^2+\omega ^2},
        \end{equation}
which makes the persistent oscillatory entanglement dynamics explicit.

These results show that, although the common-bath-only case contains no strictly stable region, its spectral structure still organizes two qualitatively distinct types of unstable dynamics: amplification in Region I$'$ and persistent oscillations in Region III$'$. These dynamical behaviors are qualitatively analogous to those in the corresponding regions of the ideal dissipation-free system.

\end{appendix}

\bibliography{refs}

@article{Busnaina2024,
  title={Quantum simulation of the bosonic Kitaev chain},
  author={Busnaina, Jamal H and Shi, Zheng and McDonald, Alexander and Dubyna, Dmytro and Nsanzineza, Ibrahim and Hung, Jimmy SC and Chang, CW Sandbo and Clerk, Aashish A and Wilson, Christopher M},
  journal={Nat. Commun.},
  volume={15},
  number={1},
  pages={3065},
  year={2024},
  publisher={Nature Publishing Group UK London},
  doi = {10.1038/s41467-024-47186-8}
}

@article{wanjura2020a,
  title = {Topological Framework for Directional Amplification in Driven-Dissipative Cavity Arrays},
  author = {Wanjura, Clara C. and Brunelli, Matteo and Nunnenkamp, Andreas},
  year = 2020,
  month = jun,
  journal = {Nat. Commun.},
  volume = {11},
  number = {1},
  pages = {3149},
  publisher = {Nature Publishing Group},
  issn = {2041-1723},
  doi = {10.1038/s41467-020-16863-9},
}

@article{tian2023a,
  title = {Nonreciprocal Amplification Transition in a Topological Photonic Network},
  author = {Tian, Mingsheng and Sun, Fengxiao and Shi, Kaiye and Xu, Haitan and He, Qiongyi and Zhang, Wei},
  year = 2023,
  month = may,
  journal = {Photon. Res.},
  volume = {11},
  number = {5},
  pages = {852--857},
  publisher = {Optica Publishing Group},
  issn = {2327-9125},
  doi = {10.1364/PRJ.485595},
  copyright = {\copyright{} 2023 Chinese Laser Press}
}

@misc{venkatraman2023,
      title={A driven quantum superconducting circuit with multiple tunable degeneracies}, 
      author={Jayameenakshi Venkatraman and Rodrigo G. Cortinas and Nicholas E. Frattini and Xu Xiao and Michel H. Devoret},
      year={2023},
      eprint={2211.04605},
      archivePrefix={arXiv},
      primaryClass={quant-ph}, 
}

@article{vanloock2003,
  title = {Detecting Genuine Multipartite Continuous-Variable Entanglement},
  author = {{van Loock}, Peter and Furusawa, Akira},
  year = 2003,
  month = may,
  journal = {Phys. Rev. A},
  volume = {67},
  number = {5},
  pages = {052315},
  publisher = {American Physical Society},
  doi = {10.1103/PhysRevA.67.052315}
}

@article{Gao2026,
  title = {Classifying multipartite continuous-variable entanglement structures through data-augmented neural networks},
  author = {Gao, Xiaoting and Tian, Mingsheng and Sun, Feng-Xiao and Wu, Ya-Dong and Xiang, Yu and He, Qiongyi},
  year = {2026},
  journal = {Nat. Mach. Intell.},
  publisher = {Nature Publishing Group},
  issn = {2522-5839},
  url = {https://doi.org/10.1038/s42256-026-01284-y},
  doi = {10.1038/s42256-026-01284-y}
}

@article{tian2022c,
  title = {Characterizing {{Multipartite}} Non-{{Gaussian Entanglement}} for a {{Three-Mode Spontaneous Parametric Down-Conversion Process}}},
  author = {Tian, Mingsheng and Xiang, Yu and Sun, Feng-Xiao and Fadel, Matteo and He, Qiongyi},
  year = 2022,
  month = aug,
  journal = {Phys. Rev. Appl.},
  volume = {18},
  number = {2},
  pages = {024065},
  publisher = {American Physical Society},
  doi = {10.1103/PhysRevApplied.18.024065}
}

@article{tian2025b,
  title = {Characterizing the {{Multipartite Entanglement Structure}} of {{Non-Gaussian Continuous-Variable States}} with a {{Single Evolution Operator}}},
  author = {Tian, Mingsheng and Gao, Xiaoting and Jing, Boxuan and Sun, Fengxiao and Fadel, Matteo and Gessner, Manuel and He, Qiongyi},
  year = 2025,
  month = sep,
  journal = {Phys. Rev. Lett.},
  volume = {135},
  number = {14},
  pages = {140201},
  issn = {0031-9007, 1079-7114},
  doi = {10.1103/21t1-dqn6}
}

@article{Jolin2023,
  title = {Multipartite Entanglement in a Microwave Frequency Comb},
  author = {Jolin, Shan W. and Andersson, Gustav and Hern\'andez, J. C. Rivera and Strandberg, Ingrid and Quijandr\'{\i}a, Fernando and Aumentado, Jos\'e and Borgani, Riccardo and Thol\'en, Mats O. and Haviland, David B.},
  journal = {Phys. Rev. Lett.},
  volume = {130},
  issue = {12},
  pages = {120601},
  numpages = {8},
  year = {2023},
  month = {Mar},
  publisher = {American Physical Society},
  doi = {10.1103/PhysRevLett.130.120601},
  url = {https://link.aps.org/doi/10.1103/PhysRevLett.130.120601}
}

@article{Lingua2025,
  title = {Continuous-Variable Square-Ladder Cluster States in a Microwave Frequency Comb},
  author = {Lingua, Fabio and Rivera Hern\'andez, J. C. and Cortinovis, Michele and Haviland, David B.},
  journal = {Phys. Rev. Lett.},
  volume = {134},
  issue = {18},
  pages = {183602},
  numpages = {6},
  year = {2025},
  month = {May},
  publisher = {American Physical Society},
  doi = {10.1103/PhysRevLett.134.183602},
  url = {https://link.aps.org/doi/10.1103/PhysRevLett.134.183602}
}

@article{Pogorzalek2019,
  title = {Secure Quantum Remote State Preparation of Squeezed Microwave States},
  author = {Pogorzalek, S. and Fedorov, K. G. and Xu, M. and Parra-Rodriguez, A. and Sanz, M. and Fischer, M. and Xie, E. and Inomata, K. and Nakamura, Y. and Solano, E. and others},
  journal = {Nat. Commun.},
  volume = {10},
  issue = {1},
  pages = {2604},
  year = {2019},
  month = {June},
  publisher = {Nature Publishing Group UK London},
  doi = {10.1038/s41467-019-10727-7},
  url = {https://doi.org/10.1038/s41467-019-10727-7}
}

@article{Xia2020,
  title = {Demonstration of a Reconfigurable Entangled Radio-Frequency Photonic Sensor Network},
  author = {Xia, Yi and Li, Wei and Clark, William and Hart, Darlene and Zhuang, Quntao and Zhang, Zheshen},
  journal = {Phys. Rev. Lett.},
  volume = {124},
  issue = {15},
  pages = {150502},
  numpages = {6},
  year = {2020},
  month = {Apr},
  publisher = {American Physical Society},
  doi = {10.1103/PhysRevLett.124.150502},
  url = {https://link.aps.org/doi/10.1103/PhysRevLett.124.150502}
}

@article{Kwon2022,
  title = {Quantum Metrological Power of Continuous-Variable Quantum Networks},
  author = {Kwon, Hyukgun and Lim, Youngrong and Jiang, Liang and Jeong, Hyunseok and Oh, Changhun},
  journal = {Phys. Rev. Lett.},
  volume = {128},
  issue = {18},
  pages = {180503},
  numpages = {7},
  year = {2022},
  month = {May},
  publisher = {American Physical Society},
  doi = {10.1103/PhysRevLett.128.180503},
  url = {https://link.aps.org/doi/10.1103/PhysRevLett.128.180503}
}

@article{Simon2000,
  title = {Peres-{Horodecki} Separability Criterion for Continuous Variable Systems},
  author = {Simon, R.},
  journal = {Phys. Rev. Lett.},
  volume = {84},
  issue = {12},
  pages = {2726--2729},
  numpages = {0},
  year = {2000},
  month = {Mar},
  publisher = {American Physical Society},
  doi = {10.1103/PhysRevLett.84.2726}
}

@article{Adesso2007,
	title = {Entanglement in continuous-variable systems: recent advances and current perspectives},
	volume = {40},
	issn = {1751-8113, 1751-8121},
	doi = {10.1088/1751-8113/40/28/S01},
	shorttitle = {Entanglement in continuous-variable systems},
	pages = {7821--7880},
	number = {28},
	journal={J. Phys. A Math. Theor.},
	shortjournal = {J. Phys. A: Math. Theor.},
	author = {Adesso, Gerardo and Illuminati, Fabrizio},
	urldate = {2024-07-10},
  year = {2007},
	date = {2007-07-13}
}

@article{Asavanant2024multipartite,
  title = {Multipartite continuous-variable optical quantum entanglement: Generation and application},
  author = {Asavanant, Warit and Furusawa, Akira},
  journal = {Phys. Rev. A},
  volume = {109},
  issue = {4},
  pages = {040101},
  numpages = {25},
  year = {2024},
  month = {Apr},
  publisher = {American Physical Society},
  doi = {10.1103/PhysRevA.109.040101},
  url = {https://link.aps.org/doi/10.1103/PhysRevA.109.040101}
}

@article{Su2007Experimental,
  title = {Experimental Preparation of Quadripartite Cluster and {Greenberger-Horne-Zeilinger} Entangled States for Continuous Variables},
  author = {Su, Xiaolong and Tan, Aihong and Jia, Xiaojun and Zhang, Jing and Xie, Changde and Peng, Kunchi},
  journal = {Phys. Rev. Lett.},
  volume = {98},
  issue = {7},
  pages = {070502},
  numpages = {4},
  year = {2007},
  month = {Feb},
  publisher = {American Physical Society},
  doi = {10.1103/PhysRevLett.98.070502},
  url = {https://link.aps.org/doi/10.1103/PhysRevLett.98.070502}
}

@article{yokoyama2013ultra,
  title={Ultra-large-scale continuous-variable cluster states multiplexed in the time domain},
  author={Yokoyama, Shota and Ukai, Ryuji and Armstrong, Seiji C and Sornphiphatphong, Chanond and Kaji, Toshiyuki and Suzuki, Shigenari and Yoshikawa, Jun-ichi and Yonezawa, Hidehiro and Menicucci, Nicolas C and Furusawa, Akira},
  journal={Nat. Photon.},
  volume={7},
  number={12},
  pages={982--986},
  year={2013},
  doi={10.1038/nphoton.2013.287},
  publisher={Nature Publishing Group UK London}
}

@article{Wang2020Large,
  title = {Large-Scale Quantum Network over 66 Orbital Angular Momentum Optical Modes},
  author = {Wang, Wei and Zhang, Kai and Jing, Jietai},
  journal = {Phys. Rev. Lett.},
  volume = {125},
  issue = {14},
  pages = {140501},
  numpages = {6},
  year = {2020},
  month = {Oct},
  publisher = {American Physical Society},
  doi = {10.1103/PhysRevLett.125.140501},
  url = {https://link.aps.org/doi/10.1103/PhysRevLett.125.140501}
}

@article{Warit2019Generation,
author = {Warit Asavanant  and Yu Shiozawa  and Shota Yokoyama  and Baramee Charoensombutamon  and Hiroki Emura  and Rafael N. Alexander  and Shuntaro Takeda  and Jun-ichi Yoshikawa  and Nicolas C. Menicucci  and Hidehiro Yonezawa  and others },
title = {Generation of time-domain-multiplexed two-dimensional cluster state},
journal = {Science},
volume = {366},
number = {6463},
pages = {373-376},
year = {2019},
doi = {10.1126/science.aay2645}
}

@article{Mikkel2019Deterministic,
author = {Mikkel V. Larsen  and Xueshi Guo  and Casper R. Breum  and Jonas S. Neergaard-Nielsen  and Ulrik L. Andersen },
title = {Deterministic generation of a two-dimensional cluster state},
journal = {Science},
volume = {366},
number = {6463},
pages = {369-372},
year = {2019},
doi = {10.1126/science.aay4354}
}

@article{cai2017multimode,
  title={Multimode entanglement in reconfigurable graph states using optical frequency combs},
  author={Cai, Yin and Roslund, Jonathan and Ferrini, Giulia and Arzani, Francesco and Xu, X and Fabre, Claude and Treps, Nicolas},
  journal={Nat. Commun.},
  volume={8},
  number={1},
  pages={15645},
  year={2017},
  doi={10.1038/ncomms15645},
  publisher={Nature Publishing Group UK London}
}

@article{roh2025generation,
  title={Generation of three-dimensional cluster entangled state},
  author={Roh, Chan and Gwak, Geunhee and Yoon, Young-Do and Ra, Young-Sik},
  journal={Nat. Photon.},
  volume={19},
  number={5},
  pages={526--532},
  year={2025},
  doi={10.1038/s41566-025-01631-2},
  publisher={Nature Publishing Group UK London}
}

@article{wang2025large,
  title={Large-scale cluster quantum microcombs},
  author={Wang, Ze and Li, Kangkang and Wang, Yue and Zhou, Xin and Cheng, Yinke and Jing, Boxuan and Sun, Fengxiao and Li, Jincheng and Li, Zhilin and Wu, Bingyan and others},
  journal={Light Sci. Appl.},
  volume={14},
  number={1},
  pages={164},
  year={2025},
  doi={10.1038/s41377-025-01812-2},
  publisher={Nature Publishing Group UK London}
}

@article{jia2025continuous,
  title={Continuous-variable multipartite entanglement in an integrated microcomb},
  author={Jia, Xinyu and Zhai, Chonghao and Zhu, Xuezhi and You, Chang and Cao, Yunyun and Zhang, Xuguang and Zheng, Yun and Fu, Zhaorong and Mao, Jun and Dai, Tianxiang and others},
  journal={Nature},
  volume={639},
  number={8054},
  pages={329--336},
  year={2025},
  doi={10.1038/s41586-025-08602-1},
  publisher={Nature Publishing Group UK London}
}

@article{Reid1988Quantum,
  title = {Quantum Correlations of Phase in Nondegenerate Parametric Oscillation},
  author = {Reid, M. D. and Drummond, P. D.},
  journal = {Phys. Rev. Lett.},
  volume = {60},
  issue = {26},
  pages = {2731--2733},
  numpages = {0},
  year = {1988},
  month = {Jun},
  publisher = {American Physical Society},
  doi = {10.1103/PhysRevLett.60.2731},
  url = {https://link.aps.org/doi/10.1103/PhysRevLett.60.2731}
}

@article{Villar2005Generation,
  title = {Generation of Bright Two-Color Continuous Variable Entanglement},
  author = {Villar, A. S. and Cruz, L. S. and Cassemiro, K. N. and Martinelli, M. and Nussenzveig, P.},
  journal = {Phys. Rev. Lett.},
  volume = {95},
  issue = {24},
  pages = {243603},
  numpages = {4},
  year = {2005},
  month = {Dec},
  publisher = {American Physical Society},
  doi = {10.1103/PhysRevLett.95.243603},
  url = {https://link.aps.org/doi/10.1103/PhysRevLett.95.243603}
}

@article{Coelho2009three,
author = {A. S. Coelho  and F. A. S. Barbosa  and K. N. Cassemiro  and A. S. Villar  and M. Martinelli  and P. Nussenzveig },
title = {Three-Color Entanglement},
journal = {Science},
volume = {326},
number = {5954},
pages = {823-826},
year = {2009},
doi = {10.1126/science.1178683},
URL = {https://www.science.org/doi/abs/10.1126/science.1178683}
}

@article{Weedbrook2012Gaussian,
  title = {Gaussian quantum information},
  author = {Weedbrook, Christian and Pirandola, Stefano and Garc\'{\i}a-Patr\'on, Ra\'ul and Cerf, Nicolas J. and Ralph, Timothy C. and Shapiro, Jeffrey H. and Lloyd, Seth},
  journal = {Rev. Mod. Phys.},
  volume = {84},
  issue = {2},
  pages = {621--669},
  numpages = {0},
  year = {2012},
  month = {May},
  publisher = {American Physical Society},
  doi = {10.1103/RevModPhys.84.621},
  url = {https://link.aps.org/doi/10.1103/RevModPhys.84.621}
}

@article{Djorw2014Robustness,
  title = {Robustness of continuous-variable entanglement via geometrical nonlinearity},
  author = {Djorw\'e, Philippe and Engo, S. G. Nana and Woafo, Paul},
  journal = {Phys. Rev. A},
  volume = {90},
  issue = {2},
  pages = {024303},
  numpages = {5},
  year = {2014},
  month = {Aug},
  publisher = {American Physical Society},
  doi = {10.1103/PhysRevA.90.024303},
  url = {https://link.aps.org/doi/10.1103/PhysRevA.90.024303}
}

@article{Krauter2011Entanglement,
  title = {Entanglement Generated by Dissipation and Steady State Entanglement of Two Macroscopic Objects},
  author = {Krauter, Hanna and Muschik, Christine A. and Jensen, Kasper and Wasilewski, Wojciech and Petersen, Jonas M. and Cirac, J. Ignacio and Polzik, Eugene S.},
  journal = {Phys. Rev. Lett.},
  volume = {107},
  issue = {8},
  pages = {080503},
  numpages = {5},
  year = {2011},
  month = {Aug},
  publisher = {American Physical Society},
  doi = {10.1103/PhysRevLett.107.080503},
  url = {https://link.aps.org/doi/10.1103/PhysRevLett.107.080503}
}

@article{Mamaev2018dissipative,
  doi = {10.22331/q-2018-03-27-58},
  url = {https://doi.org/10.22331/q-2018-03-27-58},
  title = {Dissipative stabilization of entangled cat states using a driven {B}ose-{H}ubbard dimer},
  author = {Mamaev, M. and Govia, L. C. G. and Clerk, A. A.},
  journal = {Quantum},
  issn = {2521-327X},
  publisher = {{Verein zur F{\"{o}}rderung des Open Access Publizierens in den Quantenwissenschaften}},
  volume = {2},
  pages = {58},
  month = mar,
  year = {2018}
}

@article{Qiu2024Genuine,
  title = {Genuine multipartite entanglement induced by a thermal acoustic reservoir},
  author = {Qiu, Qing-Yang and Lu, Zhi-Guang and He, Qiongyi and Wu, Ying and L\"u, Xin-You},
  journal = {Phys. Rev. B},
  volume = {110},
  issue = {22},
  pages = {L220301},
  numpages = {8},
  year = {2024},
  month = {Dec},
  publisher = {American Physical Society},
  doi = {10.1103/PhysRevB.110.L220301},
  url = {https://link.aps.org/doi/10.1103/PhysRevB.110.L220301}
}

@article{Li2025Programmable,
author = {Yue Li  and Yi Li  and Xu Cheng  and Lingna Wang  and Xingyu Zhao  and Waner Hou  and Kamran Rehan  and Mingdong Zhu  and Lin Yan  and Xi Qin  and others },
title = {Programmable multi-mode entanglement via dissipative engineering in vibrating trapped ions},
journal = {Sci. Adv.},
volume = {11},
number = {27},
pages = {eadv7838},
year = {2025},
doi = {10.1126/sciadv.adv7838},
URL = {https://www.science.org/doi/abs/10.1126/sciadv.adv7838}
}

@article{palomaki2013entangling,
  title={Entangling mechanical motion with microwave fields},
  author={Palomaki, TA and Teufel, JD and Simmonds, RW and Lehnert, Konrad W},
  journal={Science},
  volume={342},
  number={6159},
  pages={710--713},
  year={2013},
  doi={10.1126/science.1244563},
  publisher={American Association for the Advancement of Science}
}

@article{Hofer2011optomechanics,
  title = {Quantum entanglement and teleportation in pulsed cavity optomechanics},
  author = {Hofer, Sebastian G. and Wieczorek, Witlef and Aspelmeyer, Markus and Hammerer, Klemens},
  journal = {Phys. Rev. A},
  volume = {84},
  issue = {5},
  pages = {052327},
  numpages = {10},
  year = {2011},
  month = {Nov},
  publisher = {American Physical Society},
  doi = {10.1103/PhysRevA.84.052327},
  url = {https://link.aps.org/doi/10.1103/PhysRevA.84.052327}
}

@article{Kustura2022Mechanical,
  title = {Mechanical Squeezing via Unstable Dynamics in a Microcavity},
  author = {Kustura, Katja and Gonzalez-Ballestero, Carlos and Sommer, Andr\'es de los R\'{\i}os and Meyer, Nadine and Quidant, Romain and Romero-Isart, Oriol},
  journal = {Phys. Rev. Lett.},
  volume = {128},
  issue = {14},
  pages = {143601},
  numpages = {7},
  year = {2022},
  month = {Apr},
  publisher = {American Physical Society},
  doi = {10.1103/PhysRevLett.128.143601},
  url = {https://link.aps.org/doi/10.1103/PhysRevLett.128.143601}
}

@article{Wustmann2013Parametric,
  title = {Parametric resonance in tunable superconducting cavities},
  author = {Wustmann, Waltraut and Shumeiko, Vitaly},
  journal = {Phys. Rev. B},
  volume = {87},
  issue = {18},
  pages = {184501},
  numpages = {23},
  year = {2013},
  month = {May},
  publisher = {American Physical Society},
  doi = {10.1103/PhysRevB.87.184501},
  url = {https://link.aps.org/doi/10.1103/PhysRevB.87.184501}
}

@article{Wustmann2017Nondegenerate,
  title = {Nondegenerate Parametric Resonance in a Tunable Superconducting Cavity},
  author = {Wustmann, Waltraut and Shumeiko, Vitaly},
  journal = {Phys. Rev. Appl.},
  volume = {8},
  issue = {2},
  pages = {024018},
  numpages = {22},
  year = {2017},
  month = {Aug},
  publisher = {American Physical Society},
  doi = {10.1103/PhysRevApplied.8.024018},
  url = {https://link.aps.org/doi/10.1103/PhysRevApplied.8.024018}
}

@article{Han2021four,
author = {Kong Han and Yimin Wang and Guo-Qiang Zhang},
journal = {Opt. Express},
number = {9},
pages = {13451--13468},
publisher = {Optica Publishing Group},
title = {Enhancement of microwave squeezing via parametric down-conversion in a superconducting quantum circuit},
volume = {29},
month = {Apr},
year = {2021},
url = {https://opg.optica.org/oe/abstract.cfm?URI=oe-29-9-13451},
doi = {10.1364/OE.423373}
}

@article{Zheng2024Optomechanical,
  title = {Optomechanical cooling with simultaneous intracavity and extracavity squeezed light},
  author = {Zheng, S. S. and Sun, F. X. and Asjad, M. and Zhang, G. W. and Huo, J. and Li, J. and Zhou, J. and Ma, Z. and He, Q. Y.},
  journal = {Phys. Rev. A},
  volume = {110},
  issue = {6},
  pages = {063520},
  numpages = {14},
  year = {2024},
  month = {Dec},
  publisher = {American Physical Society},
  doi = {10.1103/PhysRevA.110.063520},
  url = {https://link.aps.org/doi/10.1103/PhysRevA.110.063520}
}

@article{Xu2026Modulation,
  title = {Modulation instability--induced multimode squeezing in quadratic frequency combs},
  author = {Xu, Haodong and Li, Nianqin and Shu, Zijun and Shen, Yang and Ji, Bo and Xie, Aiping and Yang, Feng and Yang, Dengcai and Peng, Jing and Gong, Hang and others},
  journal = {Phys. Rev. Appl.},
  volume = {25},
  issue = {2},
  pages = {024006},
  numpages = {22},
  year = {2026},
  month = {Feb},
  publisher = {American Physical Society},
  doi = {10.1103/z3nc-zq3y},
  url = {https://link.aps.org/doi/10.1103/z3nc-zq3y}
}

@article{Zheng2019Manipulation,
  title = {Manipulation and enhancement of asymmetric steering via interference effects induced by closed-loop coupling},
  author = {Zheng, Shasha and Sun, Fengxiao and Lai, Yijie and Gong, Qihuang and He, Qiongyi},
  journal = {Phys. Rev. A},
  volume = {99},
  issue = {2},
  pages = {022335},
  numpages = {9},
  year = {2019},
  month = {Feb},
  publisher = {American Physical Society},
  doi = {10.1103/PhysRevA.99.022335},
  url = {https://link.aps.org/doi/10.1103/PhysRevA.99.022335}
}

@article{sun2017phase,
  title={Phase control of entanglement and quantum steering in a three-mode optomechanical system},
  author={Sun, FX and Mao, D and Dai, YT and Ficek, Z and He, QY and Gong, QH},
  journal={New J. Phys.},
  volume={19},
  number={12},
  pages={123039},
  year={2017},
  doi={10.1088/1367-2630/aa9c9a},
  publisher={IOP Publishing}
}

@article{zheng2021enhanced,
  title={Enhanced entanglement and asymmetric {EPR} steering between magnons},
  author={Zheng, Sha-Sha and Sun, Feng-Xiao and Yuan, Huai-Yang and Ficek, Zbigniew and Gong, Qi-Huang and He, Qiong-Yi},
  journal={Sci. China Phys. Mech. Astron.},
  volume={64},
  number={1},
  pages={210311},
  year={2021},
  doi={10.1007/s11433-020-1587-5},
  publisher={Springer}
}

@article{Braunstein2005Quantum,
  title = {Quantum information with continuous variables},
  author = {Braunstein, Samuel L. and van Loock, Peter},
  journal = {Rev. Mod. Phys.},
  volume = {77},
  issue = {2},
  pages = {513--577},
  numpages = {0},
  year = {2005},
  month = {Jun},
  publisher = {American Physical Society},
  doi = {10.1103/RevModPhys.77.513},
  url = {https://link.aps.org/doi/10.1103/RevModPhys.77.513}
}

@article{Garc2009key,
  title = {Continuous-Variable Quantum Key Distribution Protocols Over Noisy Channels},
  author = {Garc\'{\i}a-Patr\'on, Ra\'ul and Cerf, Nicolas J.},
  journal = {Phys. Rev. Lett.},
  volume = {102},
  issue = {13},
  pages = {130501},
  numpages = {4},
  year = {2009},
  month = {Mar},
  publisher = {American Physical Society},
  doi = {10.1103/PhysRevLett.102.130501},
  url = {https://link.aps.org/doi/10.1103/PhysRevLett.102.130501}
}

@article{madsen2012continuous,
  title={Continuous variable quantum key distribution with modulated entangled states},
  author={Madsen, Lars S and Usenko, Vladyslav C and Lassen, Mikael and Filip, Radim and Andersen, Ulrik L},
  journal={Nat. Commun.},
  volume={3},
  number={1},
  pages={1083},
  year={2012},
  doi={10.1038/ncomms2097},
  publisher={Nature Publishing Group UK London}
}

@article{Abdi2021Continuous,
  title = {Continuous-variable multipartite vibrational entanglement},
  author = {Abdi, Mehdi},
  journal = {Phys. Rev. A},
  volume = {103},
  issue = {4},
  pages = {043520},
  numpages = {11},
  year = {2021},
  month = {Apr},
  publisher = {American Physical Society},
  doi = {10.1103/PhysRevA.103.043520},
  url = {https://link.aps.org/doi/10.1103/PhysRevA.103.043520}
}

@article{pedram2023overview,
  title={An Overview: Steady-State Quantum Entanglement via Reservoir Engineering.},
  author={Pedram, Ali and M{\"u}stecaplio{\u{g}}lu, {\"O}zg{\"u}r E},
  journal={Int. J. Unconv. Comput.},
  volume={18},
  number={1},
  pages={67--81},
  year={2023},
  url={https://www.oldcitypublishing.com/journals/ijuc-home/ijuc-issue-contents/ijuc-volume-18-number-1-2023/ijuc-18-1-p-67-81/}
}

@article{ockeloen2018stabilized,
  title={Stabilized entanglement of massive mechanical oscillators},
  author={Ockeloen-Korppi, CF and Damsk{\"a}gg, E and Pirkkalainen, J-M and Asjad, M and Clerk, AA and Massel, F and Woolley, MJ and Sillanp{\"a}{\"a}, MA},
  journal={Nature},
  volume={556},
  number={7702},
  pages={478--482},
  year={2018},
  doi={10.1038/s41586-018-0038-x},
  publisher={Nature Publishing Group UK London}
}

@article{Wang2013Reservoir,
  title = {Reservoir-Engineered Entanglement in Optomechanical Systems},
  author = {Wang, Ying-Dan and Clerk, Aashish A.},
  journal = {Phys. Rev. Lett.},
  volume = {110},
  issue = {25},
  pages = {253601},
  numpages = {5},
  year = {2013},
  month = {Jun},
  publisher = {American Physical Society},
  doi = {10.1103/PhysRevLett.110.253601},
  url = {https://link.aps.org/doi/10.1103/PhysRevLett.110.253601}
}

@article{Woolley2014squeezed,
  title = {Two-mode squeezed states in cavity optomechanics via engineering of a single reservoir},
  author = {Woolley, M. J. and Clerk, A. A.},
  journal = {Phys. Rev. A},
  volume = {89},
  issue = {6},
  pages = {063805},
  numpages = {17},
  year = {2014},
  month = {Jun},
  publisher = {American Physical Society},
  doi = {10.1103/PhysRevA.89.063805},
  url = {https://link.aps.org/doi/10.1103/PhysRevA.89.063805}
}

@article{yonezawa2004demonstration,
  title={Demonstration of a quantum teleportation network for continuous variables},
  author={Yonezawa, Hidehiro and Aoki, Takao and Furusawa, Akira},
  journal={Nature},
  volume={431},
  number={7007},
  pages={430--433},
  year={2004},
  doi={10.1038/nature02858},
  publisher={Nature Publishing Group UK London}
}

@article{Lloyd1999Computation,
  title = {Quantum Computation over Continuous Variables},
  author = {Lloyd, Seth and Braunstein, Samuel L.},
  journal = {Phys. Rev. Lett.},
  volume = {82},
  issue = {8},
  pages = {1784--1787},
  numpages = {0},
  year = {1999},
  month = {Feb},
  publisher = {American Physical Society},
  doi = {10.1103/PhysRevLett.82.1784},
  url = {https://link.aps.org/doi/10.1103/PhysRevLett.82.1784}
}

@article{armstrong2012programmable,
  title={Programmable multimode quantum networks},
  author={Armstrong, Seiji and Morizur, Jean-Fran{\c{c}}ois and Janousek, Jiri and Hage, Boris and Treps, Nicolas and Lam, Ping Koy and Bachor, Hans-A},
  journal={Nat. Commun.},
  volume={3},
  number={1},
  pages={1026},
  year={2012},
  doi={10.1038/ncomms2033},
  publisher={Nature Publishing Group UK London}
}

@article{zheng2026large,
  title={Large-scale quantum communication networks with integrated photonics},
  author={Zheng, Yun and Wang, Hanyu and Jia, Xinyu and Huang, Jiahui and Yuan, Huihong and Zhai, Chonghao and Dai, Junhao and Shi, Jingbo and Zhang, Lei and Zhang, Xuguang and others},
  journal={Nature},
  volume={651},
  number={8104},
  pages={68--75},
  year={2026},
  doi={10.1038/s41586-026-10152-z},
  publisher={Nature Publishing Group UK London}
}

@article{guo2020distributed,
  title={Distributed quantum sensing in a continuous-variable entangled network},
  author={Guo, Xueshi and Breum, Casper R and Borregaard, Johannes and Izumi, Shuro and Larsen, Mikkel V and Gehring, Tobias and Christandl, Matthias and Neergaard-Nielsen, Jonas S and Andersen, Ulrik L},
  journal={Nat. Phys.},
  volume={16},
  number={3},
  pages={281--284},
  year={2020},
  doi={10.1038/s41567-019-0743-x},
  publisher={Nature Publishing Group UK London}
}

@article{Furusawa1998teleportation,
author = {A. Furusawa  and J. L. Sørensen  and S. L. Braunstein  and C. A. Fuchs  and H. J. Kimble  and E. S. Polzik },
title = {Unconditional Quantum Teleportation},
journal = {Science},
volume = {282},
number = {5389},
pages = {706-709},
year = {1998},
doi = {10.1126/science.282.5389.706}
}

@article{Ukai2011Computation,
  title = {Demonstration of Unconditional One-Way Quantum Computations for Continuous Variables},
  author = {Ukai, Ryuji and Iwata, Noriaki and Shimokawa, Yuji and Armstrong, Seiji C. and Politi, Alberto and Yoshikawa, Jun-ichi and van Loock, Peter and Furusawa, Akira},
  journal = {Phys. Rev. Lett.},
  volume = {106},
  issue = {24},
  pages = {240504},
  numpages = {4},
  year = {2011},
  month = {Jun},
  publisher = {American Physical Society},
  doi = {10.1103/PhysRevLett.106.240504},
  url = {https://link.aps.org/doi/10.1103/PhysRevLett.106.240504}
}

@article{Larsen2021Fault,
  title = {Fault-Tolerant Continuous-Variable Measurement-based Quantum Computation Architecture},
  author = {Larsen, Mikkel V. and Chamberland, Christopher and Noh, Kyungjoo and Neergaard-Nielsen, Jonas S. and Andersen, Ulrik L.},
  journal = {PRX Quantum},
  volume = {2},
  issue = {3},
  pages = {030325},
  numpages = {19},
  year = {2021},
  month = {Aug},
  publisher = {American Physical Society},
  doi = {10.1103/PRXQuantum.2.030325},
  url = {https://link.aps.org/doi/10.1103/PRXQuantum.2.030325}
}

@article{Aspelmeyer2014optomechanics,
  title = {Cavity optomechanics},
  author = {Aspelmeyer, Markus and Kippenberg, Tobias J. and Marquardt, Florian},
  journal = {Rev. Mod. Phys.},
  volume = {86},
  issue = {4},
  pages = {1391--1452},
  numpages = {62},
  year = {2014},
  month = {Dec},
  publisher = {American Physical Society},
  doi = {10.1103/RevModPhys.86.1391},
  url = {https://link.aps.org/doi/10.1103/RevModPhys.86.1391}
}

@article{dutt2024nonlinear,
  title={Nonlinear and quantum photonics using integrated optical materials},
  author={Dutt, Avik and Mohanty, Aseema and Gaeta, Alexander L and Lipson, Michal},
  journal={Nat. Rev. Mater.},
  volume={9},
  number={5},
  pages={321--346},
  year={2024},
  doi={10.1038/s41578-024-00668-z},
  publisher={Nature Publishing Group UK London}
}

@article{wakefield2024,
  title={Non-{Hermiticity} in quantum nonlinear optics through symplectic transformations},
  author={Wakefield, Ross and Laing, Anthony and Joglekar, Yogesh N},
  journal={App. Phys. Lett.},
  volume={124},
  pages={201103},
  year={2024},
  publisher={AIP Publishing},
  doi={https://doi.org/10.1063/5.0206393},
}

@article{Ashida2020,
  title={Non-{Hermitian} physics},
  author={Ashida, Yuto and Gong, Zongping and Ueda, Masahito},
  journal={Adv. Phys.},
  volume={69},
  number={3},
  pages={249--435},
  year={2020},
  publisher={Taylor \& Francis},
  doi = {10.1080/00018732.2021.1876991}
}

@article{El2018Non,
  title={Non-{Hermitian} physics and {PT} symmetry},
  author={El-Ganainy, Ramy and Makris, Konstantinos G and Khajavikhan, Mercedeh and Musslimani, Ziad H and Rotter, Stefan and Christodoulides, Demetrios N},
  journal={Nat. Phys.},
  volume={14},
  number={1},
  pages={11--19},
  year={2018},
  doi={10.1038/nphys4323},
  publisher={Nature Publishing Group UK London}
}

@article{Del2022,
  title={Non-{Hermitian} chiral phononics through optomechanically induced squeezing},
  author={Del Pino, Javier and Slim, Jesse J and Verhagen, Ewold},
  journal={Nature},
  volume={606},
  number={7912},
  pages={82--87},
  year={2022},
  publisher={Nature Publishing Group UK London},
  doi = {10.1038/s41586-022-04609-0}
}

@article{ozdemir2019parity,
  title={Parity--time symmetry and exceptional points in photonics},
  author={{\"O}zdemir, {\c{S}}ahin Kaya and Rotter, Stefan and Nori, Franco and Yang, Lan},
  journal={Nat. Mater.},
  volume={18},
  number={8},
  pages={783--798},
  year={2019},
  doi={10.1038/s41563-019-0304-9},
  publisher={Nature Publishing Group UK London}
}

@article{Miri2019Exceptional,
author = {Mohammad-Ali Miri  and Andrea Alù },
title = {Exceptional points in optics and photonics},
journal = {Science},
volume = {363},
number = {6422},
pages = {eaar7709},
year = {2019},
doi = {10.1126/science.aar7709},
URL = {https://www.science.org/doi/abs/10.1126/science.aar7709}
}

@article{Bergholtz2021Exceptional,
  title = {Exceptional topology of non-{Hermitian} systems},
  author = {Bergholtz, Emil J. and Budich, Jan Carl and Kunst, Flore K.},
  journal = {Rev. Mod. Phys.},
  volume = {93},
  issue = {1},
  pages = {015005},
  numpages = {31},
  year = {2021},
  month = {Feb},
  publisher = {American Physical Society},
  doi = {10.1103/RevModPhys.93.015005},
  url = {https://link.aps.org/doi/10.1103/RevModPhys.93.015005}
}

@article{Minganti2019Quantum,
  title = {Quantum exceptional points of non-{Hermitian} {Hamiltonians} and {Liouvillians}: The effects of quantum jumps},
  author = {Minganti, Fabrizio and Miranowicz, Adam and Chhajlany, Ravindra W. and Nori, Franco},
  journal = {Phys. Rev. A},
  volume = {100},
  issue = {6},
  pages = {062131},
  numpages = {17},
  year = {2019},
  month = {Dec},
  publisher = {American Physical Society},
  doi = {10.1103/PhysRevA.100.062131},
  url = {https://link.aps.org/doi/10.1103/PhysRevA.100.062131}
}

@article{McDonald2018Phase,
  title = {Phase-Dependent Chiral Transport and Effective Non-{Hermitian} Dynamics in a Bosonic {Kitaev-Majorana} Chain},
  author = {McDonald, A. and Pereg-Barnea, T. and Clerk, A. A.},
  journal = {Phys. Rev. X},
  volume = {8},
  issue = {4},
  pages = {041031},
  numpages = {15},
  year = {2018},
  month = {Nov},
  publisher = {American Physical Society},
  doi = {10.1103/PhysRevX.8.041031},
  url = {https://link.aps.org/doi/10.1103/PhysRevX.8.041031}
}

@article{Lee2024Entanglement,
  title = {Entanglement Phase Transition Due to Reciprocity Breaking without Measurement or Postselection},
  author = {Lee, Gideon and Jin, Tony and Wang, Yu-Xin and McDonald, Alexander and Clerk, Aashish},
  journal = {PRX Quantum},
  volume = {5},
  issue = {1},
  pages = {010313},
  numpages = {20},
  year = {2024},
  month = {Jan},
  publisher = {American Physical Society},
  doi = {10.1103/PRXQuantum.5.010313},
  url = {https://link.aps.org/doi/10.1103/PRXQuantum.5.010313}
}

@article{Arkhipov2021Generating,
  title = {Generating high-order quantum exceptional points in synthetic dimensions},
  author = {Arkhipov, Ievgen I. and Minganti, Fabrizio and Miranowicz, Adam and Nori, Franco},
  journal = {Phys. Rev. A},
  volume = {104},
  issue = {1},
  pages = {012205},
  numpages = {13},
  year = {2021},
  month = {Jul},
  publisher = {American Physical Society},
  doi = {10.1103/PhysRevA.104.012205},
  url = {https://link.aps.org/doi/10.1103/PhysRevA.104.012205}
}

@article{Wanjura2023,
  title={Quadrature nonreciprocity in bosonic networks without breaking time-reversal symmetry},
  author={Wanjura, Clara C and Slim, Jesse J and del Pino, Javier and Brunelli, Matteo and Verhagen, Ewold and Nunnenkamp, Andreas},
  journal={Nat. Phys.},
  volume={19},
  number={10},
  pages={1429--1436},
  year={2023},
  publisher={Nature Publishing Group UK London},
  doi = {10.1038/s41567-023-02128-x}
}

@misc{yang2026spectral,
      title={From Spectral Singularities to Multipartite Entanglement Scaling at Higher-Order Exceptional Points}, 
      author={Chunlai Yang and Shuheng Liu and Xinyao Huang and Kaiye Shi and Qiongyi He},
      year={2026},
      eprint={2606.24205},
      archivePrefix={arXiv},
      primaryClass={quant-ph},
}

@article{Yu2025Exceptional,
  title={Exceptional-point-induced nonequilibrium entanglement dynamics in bosonic networks},
  author={Yu, Chenghe and Tian, Mingsheng and Kong, Ningxin and Fadel, Matteo and Huang, Xinyao and He, Qiongyi},
  journal={npj Quantum Inf.},
  volume={12},
  number={1},
  pages={14},
  year={2026},
  publisher={Nature Publishing Group UK London},
  doi={10.1038/s41534-025-01158-y}
}

@article{hsu2016bound,
  title={Bound states in the continuum},
  author={Hsu, Chia Wei and Zhen, Bo and Stone, A Douglas and Joannopoulos, John D and Solja{\v{c}}i{\'c}, Marin},
  journal={Nat. Rev. Mater.},
  volume={1},
  number={9},
  pages={16048},
  year={2016},
  doi={10.1038/natrevmats.2016.48},
  publisher={Nature Publishing Group}
}

@article{Lei2023,
  title = {Hyperparametric Oscillation via Bound States in the Continuum},
  author = {Lei, Fuchuan and Ye, Zhichao and Twayana, Krishna and Gao, Yan and Girardi, Marcello and Helgason, \'Oskar B. and Zhao, Ping and Torres-Company, Victor},
  journal = {Phys. Rev. Lett.},
  volume = {130},
  issue = {9},
  pages = {093801},
  numpages = {6},
  year = {2023},
  month = {Feb},
  publisher = {American Physical Society},
  doi = {10.1103/PhysRevLett.130.093801},
  url = {https://link.aps.org/doi/10.1103/PhysRevLett.130.093801}
}

\end{document}